\documentclass[apj]{emulateapj-rtx4}
\usepackage{calc,lscape}
\newcommand{\ew}{$pW$}
\newcommand{\pwone}{$pW$1\,(\ion{Ca}{2}\,H\&K)}
\newcommand{\pwtwo}{$pW$2\,(\ion{Si}{2}\,4130)}
\newcommand{\pwthree}{$pW$3\,(\ion{Mg}{2})}
\newcommand{\pwfour}{$pW$4\,(\ion{Fe}{2})}
\newcommand{\pwfive}{$pW$5\,(\ion{S}{2}\,W)}
\newcommand{\pwsix}{$pW$6\,(\ion{Si}{2}\,5972)}
\newcommand{\pwseven}{$pW$7\,(\ion{Si}{2}\,6355)}
\newcommand{\pweight}{$pW$8\,(\ion{Ca}{2}\,IR)}
\newcommand{\vone}{$v$\,(\ion{Ca}{2}\,H\&K)}
\newcommand{\vtwo}{$v$\,(\ion{Si}{2}\,4130)}
\newcommand{\vthree}{$v$\,(\ion{S}{2}\,5449)}
\newcommand{\vfour}{$v$\,(\ion{S}{2}\,5622)}
\newcommand{\vfive}{$v$\,(\ion{Si}{2}\,5972)}
\newcommand{\vsix}{$v$\,(\ion{Si}{2}\,6355)}
\newcommand{\vseven}{$v$\,(\ion{Ca}{2}\,IR)}
\newcommand{\dm}{$\Delta m_{15}(B)$}
\newcommand{\rsi}{$\cal{R}$(\ion{Si}{2})}
\newcommand{\deltav}{$\Delta v_{20}(\mathrm{Si})$}
\newcommand{\sndia}{SN-Ia}
\newcommand{\sneia}{SNe~Ia}
\newcommand{\omm}{\Omega_M}
\newcommand{\oml}{\Omega_\Lambda}

\shorttitle{SNe~Ia Spectroscopy by the CSP}
\shortauthors{Folatelli et al.}

\submitted{ApJ, Accepted on May 21, 2013}

\begin{document}

\title{Spectroscopy of Type Ia Supernovae by the Carnegie Supernova Project\altaffilmark{1}}

\author{%
  Gast\'on Folatelli\altaffilmark{2},
  Nidia Morrell\altaffilmark{3}, 
  Mark M. Phillips\altaffilmark{3},
  Eric Hsiao\altaffilmark{3},
  Abdo Campillay\altaffilmark{3},
  Carlos Contreras\altaffilmark{3},
  Sergio Castell\'on\altaffilmark{3},
  Mario Hamuy\altaffilmark{4},
  Wojtek Krzeminski\altaffilmark{5},
  Miguel Roth\altaffilmark{3},
  Maximilian Stritzinger\altaffilmark{6},
  Christopher R. Burns\altaffilmark{7},
  Wendy L. Freedman\altaffilmark{7},
  Barry F. Madore\altaffilmark{7,8},
  David Murphy\altaffilmark{7},
  S. E. Persson\altaffilmark{7},
  Jos\'e L. Prieto\altaffilmark{9},
  Nicholas B. Suntzeff\altaffilmark{10},
  Kevin Krisciunas\altaffilmark{10},
  Joseph P. Anderson\altaffilmark{4},
  Francisco F\"orster\altaffilmark{4},
  Jos\'e Maza\altaffilmark{4},
  Giuliano Pignata\altaffilmark{11},
  P. Andrea Rojas\altaffilmark{12},
  Luis Boldt\altaffilmark{13},
  Francisco Salgado\altaffilmark{14},
  Pamela Wyatt\altaffilmark{15}, 
  Felipe Olivares E.\altaffilmark{16},
  Avishay Gal-Yam\altaffilmark{17}, and
  Masao Sako\altaffilmark{18}
}
\altaffiltext{1}{This paper includes data gathered with the 
6.5~m Magellan Telescopes located at Las Campanas Observatory, Chile; and
  the Gemini Observatory, Cerro Pachon, Chile (Gemini Program
  GS-2008B$-$Q$-$56). Based on observations collected at the European
  Organisation for Astronomical Research in the Southern Hemisphere,
  Chile (ESO Programmes 076.A-0156, 078.D-0048, 080.A-0516, and
  082.A-0526)}
\altaffiltext{2}{Kavli Institute for the Physics and Mathematics of
  the Universe (WPI), Todai Institutes for Advanced Study, the
  University of Tokyo, Kashiwa, Japan 277-8583 (Kavli IPMU, WPI)}
\altaffiltext{3}{Las Campanas Observatory, Carnegie Observatories,
  Casilla 601, La Serena, Chile}
\altaffiltext{4}{Universidad de Chile, Departamento de Astronom\'{\i}a,
  Casilla 36-D, Santiago, Chile }
\altaffiltext{5}{N. Copernicus Astronomical Center, ul. Bartycka 18,
  00-716 Warszawa, Poland}
\altaffiltext{6}{Department of Physics and Astronomy, Aarhus University, Ny Munkegade 120, 8000 Aarhus C, Denmark}
\altaffiltext{7}{Observatories of the Carnegie Institution of
  Washington, 813 Santa Barbara St., Pasadena, CA 91101, USA}
\altaffiltext{8}{Infrared Processing and Analysis Center, Caltech/Jet
  Propulsion Laboratory, Pasadena, CA 91125, USA}
\altaffiltext{9}{Department of Astrophysical Sciences, Princeton University, 4 Ivy Ln., Princeton, NJ 08544}
\altaffiltext{10}{George P. and Cynthia Woods Mitchell Institute for
  Fundamental Physics and Astronomy, Department of Physics and
  Astronomy, Texas A\&M University, College Station, TX 77843, USA}
\altaffiltext{11}{Departamento de Ciencias Fisicas, Universidad Andres
  Bello, Avda. Republica 252, Santiago, Santiago RM, Chile} 
\altaffiltext{12}{Departamento de F\'isica y Astronom\'ia, Universidad de Valpara\'iso, Av. Gran Breta\~na 1111, Valpara\'iso, Chile}
\altaffiltext{13}{Argelander Institut f\"ur Astronomie, Universit\"at
  Bonn, Auf dem H\"ugel 71, D-53111 Bonn, Germany} 
\altaffiltext{14}{Leiden Observatory, Leiden University, PO Box 9513,
  NL-2300 RA Leiden, The Netherlands}
\altaffiltext{15}{National Snow and Ice Data Center, University of Colorado, Boulder, CO 80309, USA}
\altaffiltext{16}{Max-Planck-Institute f\"ur Extraterrestrische Physik, Giessenbachstra\ss{}e 1, 85748 Garching, Germany} 
\altaffiltext{17}{Department of Particle Physics and Astrophysics,
Weizmann Institute of Science, Israel} 
\altaffiltext{18}{Department of Physics and Astronomy, University of
  Pennsylvania, Philadelphia, PA 19104, USA} 
\email{gaston.folatelli@ipmu.jp}

\setcounter{footnote}{18}

\begin{abstract}
\noindent This is the first release of optical spectroscopic data
of low-redshift Type Ia supernovae (\sneia) by the {\em Carnegie
Supernova Project} including 604 previously unpublished
spectra of 93 \sneia. The observations cover a range of
phases from 12 days before to over 150 days after the time of $B$-band 
maximum light. With the addition of 228 near-maximum spectra
from the literature we study the diversity among \sneia\ in a
quantitative manner. For that
purpose, spectroscopic parameters are employed such as expansion
velocities from spectral line blueshifts, and pseudo-equivalent widths
(\ew). The values of those parameters at maximum light are
  obtained for 78 objects, thus providing a 
characterization of \sneia\ that may help to improve our understanding
of the properties of the exploding systems and the thermonuclear flame
propagation. Two objects, namely SNe~2005M and 2006is, stand out from the 
sample by showing peculiar \ion{Si}{2} and \ion{S}{2} velocities but
otherwise standard velocities for the rest of the ions. 
We further study the correlations between spectroscopic and
photometric parameters such as light-curve decline rate and color. In
agreement with previous studies, we find 
that the \ew\ of \ion{Si}{2} absorption features are very good 
indicators of light-curve decline rate. Furthermore, we demonstrate
that parameters such as \pwtwo\ and \pwsix\ provide precise
calibrations of the peak $B$-band luminosity with dispersions of
$\approx$$0.15$ mag. In the search for a secondary
parameter in the calibration of peak luminosity for \sneia, we find a
$\approx$2--3-$\sigma$ correlation between $B$-band Hubble residuals
and the velocity at maximum light of \ion{S}{2} and \ion{Si}{2}
lines.
\end{abstract}



\section{INTRODUCTION}
\label{sec:intro}

Type Ia supernovae (\sneia) are extremely important astrophysical
objects due to their connection with stellar evolution in binary
systems, and with the chemical enrichment and energy deposition of
the interstellar medium. They are also very useful in the
determination of extragalactic distances and thereby in the study of
the cosmological expansion rate. Thanks to an empirical calibration of
the luminosity of \sneia\ based on the decline rate of the light curve
\citep{phillips93}, it has been possible to measure precise relative
distances out to very large look-back times. This led to the
surprising discovery of the accelerated expansion of
the universe and the introduction of dark energy as its currently
dominant component \citep{riess98,perlmutter99}. 

Present \sneia\ experiments 
employ a two-parameter luminosity calibration by adding a color term
in the cosmological fits, as first introduced by \citet{tripp98}. Such
an approach has allowed the equation-of-state parameter,
$w$, of dark energy to be constrained to within 10\% by comparing
low- and high-redshift \sneia\ samples \citep[see,
  e.g.][]{riess07,freedman09,hicken09,amanullah10,sullivan11,suzuki12}. This
is thanks to a 
precision of about 0.1 -- 0.15 mag in the peak luminosity after
calibrating with respect to decline rate and color. 
Efforts aimed at further improving the calibration have searched for
yet a third parameter\footnote{Usually referred to as ``secondary
  parameter''.} which would correlate with luminosity independently of
the other two parameters. As detailed at the end of this
Introduction, many such attempts involve optical spectra,
which provide more detailed information than broad-band
photometry. Moreover, spectra are a key to understanding the
physical properties of the explosion and therefore are fundamental for
providing theoretical support for the use of \sneia\ as distance
indicators. 

The existence of \sndia\ subtypes has been known for the last two
decades. Not long after the definition of \sneia\ as a spectroscopic SN class,
examples of diversity in spectral properties started to be found as
more data were being gathered. \citet{branch87} showed the case of
SN~1984A, an object whose spectrum displayed the same lines as the
rest of \sneia\ and seemingly no difference in peak brightness, but
whose expansion velocities were significantly higher. This was followed by the
discovery of SN~1986G \citep{phillips87}, SN~1991T
\citep{filippenko92a,phillips92}, and SN~1991bg 
\citep{filippenko92b,leibundgut93}, which gave evidence of a variety in line 
strengths, this time related with differences in peak luminosity and
color. 

With time it was observed that the spectroscopic diversity was not
due to rare exceptions but that some \sneia\ subclasses could be
identified \citep{li01}. SNe~1991T and 1991bg became prototypes of
their own subgroups comprising significant fractions of the
total \sndia\ population \citep[e.g.][find fractions of $>$9\% and
  18\%, respectively, for these subclasses]{li10}. While 1991bg-like
SNe are clearly identifiable through their photometric
properties, the distinction of 1991T-like objects from normal
\sneia\ relies on pre-maximum spectroscopy, which makes it more
difficult to establish. 

\citet{nugent95} first quantified the spectral diversity using flux and
line strength ratios, and provided an explanation by means of a temperature
sequence due, in turn, to different amounts of synthesized
$^{56}$Ni. Spectral line ratios such as \rsi, defined
between the absorption depths of \ion{Si}{2} $\lambda$5972 and
$\lambda$6355 lines, were found to be correlated with the decline rate
parameter, \dm \citep{phillips93}, and thus with luminosity. The
diversity in expansion velocity did not fit into this picture, however,
and it was pointed out that \ion{Si}{2} velocities do not correlate with
\rsi\ or \dm\ \citep{hatano00}. A one-parameter scenario regulated
solely by ejecta temperature is further complicated by opacity and
line-blanketing effects \citep{hoeflich96,pinto00,kasen07a}.

Based on line ratios, expansion velocities
and light-curve decline rates, \citet{benetti05} provided a
quantitative classification of \sneia\ into three subtypes. Their high- and
low-velocity-gradient classes (HVG and LVG, respectively) are
distinguished by the post-maximum rate of \ion{Si}{2} $\lambda$6355 
velocity decline, named $\dot{v}$. The third subtype, FAINT, includes
objects with large \dm\ which also show large $\dot{v}$
values. An alternative classification scheme was introduced by
\citet{branch06} based on absorption equivalent widths of the
\ion{Si}{2} $\lambda$5972 and $\lambda$6355 lines at maximum
light. They distinguished between ``Core Normal'' (CN), ``Cool'' (CL),
``Broad Line'' (BL), and ``Shallow Silicon'' (SS) \sneia. The
prototypes of the BL and CL classes are SN~1984A and SN~1991bg,
respectively, while the SS class includes SN~1991T- and SN~2002cx-like
objects. There is a rough correspondence between the BL and 
HVG groups, and also between the CL and FAINT groups. LVG SNe roughly
include CN and SS classes. More recently, \citet{wang09} divided their
sample into ``Normal'', ``1991bg-like'', ``1991T-like'', and ``HV''
(High-Velocity) SNe. HV SNe are differentiated from Normal events
using the \ion{Si}{2} $\lambda$6355 velocity within one week from
maximum light, with HV objects 
being somewhat arbitrarily defined as those SNe whose velocity lies
3~$\sigma$ above the average velocities of 10 fiducial objects in the
Normal group. Using this definition, the HV and HVG subclasses roughly coincide.

In spite of their spectroscopic differences, both LVG and HVG objects
(or normal and HV objects) are usually employed for determining
distances, as they seem to obey the same luminosity-decline rate
calibration. However, \citet{wang09} showed that host-galaxy 
extinction of HV \sneia\ may obey a different wavelength dependence than
that of normal \sneia. Such a difference would affect the color
correction and the derived distances. More recently, \citet{foley11a}
re-analyzed the same dataset and suggested a difference in intrinsic
color between HV and Normal \sneia, possibly as a consequence of
increased line blanketing in HV SNe. They show that the scatter in the
Hubble diagram is reduced when using only Normal
\sneia. Interestingly, \citet{maeda10b} proposed a unification
scenario based on asymmetric explosion and 
line-of-sight effects to explain the observed velocity differences
among LVG and HVG SNe. We note that signatures of unburned material in
the ejecta of \sneia---namely \ion{C}{2} lines in pre-maximum light
spectra---have been found with larger incidence in LVG SNe than in HVG
objects \citep{parrent11,thomas11b,folatelli12,silverman12d}. This may
indicate a different explosion mechanism for both groups.

Several attempts have been made to quantify the spectroscopic properties of
\sneia\ subtypes and to look for an improved calibration of the peak
luminosity adopting different
approaches. Pseudo-equivalent widths (\ew) of absorptions were
introduced by \citet{folatelli04} as a way to quantify the differences
in line strengths. \citet{hachinger06} employed this type of
measurements and expansion velocities to study correlations with
\dm\ on a sample of nearby \sneia. Being less sensitive to noise and
flux-calibration issues than flux ratios and line depths,
\ew\ measurements have been 
used to compare general properties of low- and high-redshift SNe
\citep{garavini07,arsenijevic08,bronder08,walker11}. A similar
analysis based on wavelet decomposition was presented by
\citet{wagers10}. Flux ratios at 
selected optical wavelengths have been studied as luminosity
indicators by \citet{bongard06}. More thorough analyses of flux ratios
have been presented recently \citep{bailey09,blondin11a} in
search of a luminosity calibrator which would reduce the scatter
in the \sneia\ Hubble diagram. \citet{foley11b} used a large sample of
\sneia\ to measure expansion velocities and \ew\ and find correlations
with intrinsic color. They find a larger scatter in intrinsic color
for high-velocity SNe than for low-velocity ones and suggest that the
latter are more precise distance indicators.

The issue of \sneia\ diversity has been quantitatively addressed
  by \citet{blondin12} in the analysis of a large sample of spectra
  obtained by the Centre for Astrophysics (CfA) Supernova
  Program. These authors presented a detailed examination of the spectroscopic
  properties and relations between spectroscopic and photometric
  parameters for different \sneia\ subtypes defined in the terms of Branch
  et~al.~and Wang et~al classification schemes. The Berkeley Supernova Ia Program (BSNIP) has
  also released a significant amount of spectra and provided their
  careful analysis in a series of papers
  \citep{silverman12a,silverman12b,silverman12c,silverman12d}. The
  second and third papers of this series deal with the quantitative
  characterization of \sneia\ spectra near maximum light, and with the
possible improvement of distance determinations based on spectroscopic
parameters.

In this paper we present optical spectra of low-redshift
\sneia\ gathered by the {\em Carnegie Supernova Project}
\citep[CSP;][]{hamuy06} between 2004 and 2009. Expansion velocities and
\ew\ measurements are used to revisit the definition of
\sneia\ subtypes and to search for correlations with photometric
properties. 

In Section~\ref{sec:data} we present the sample of SNe and the
spectroscopic observations and data reduction performed by the CSP. In
Section~\ref{sec:meas} we describe the methods utilized to obtain
spectroscopic parameters such as expansion velocities and
pseudo-equivalent widths. Section~\ref{sec:spdiv} is devoted to a quantitative
analysis of the spectroscopic diversity among \sneia\ with special
attention to the definition of subtypes and the correlations
between different spectral parameters. Section~\ref{sec:spphot}
focuses on the connection between spectroscopic and photometric
parameters with the aim of improving the precision in the luminosity
calibration of \sneia. Finally, in Section~\ref{sec:concl} we
discuss the results and provide some concluding remarks.

\section{DATA}
\label{sec:data}

We present optical spectroscopic data for a sample of 93
\sneia\ observed by the CSP between 2004 and 2009. The dataset amounts
to 832 optical spectra, 569 of which were obtained by the CSP, 35 were
provided by other observers, and 228 were taken from the
literature. In the following we describe the SNe included in this
work, and the observing and reduction procedures applied to the CSP
data. 

\subsection{Supernova Sample}
\label{sec:sne}

The CSP carried out follow-up campaigns to obtain optical and
near-infrared (NIR) light curves and optical spectroscopy of nearby
($z \lesssim 0.08$) SNe of all types. Between September 2004 and May
2009 over 250 objects were monitored, among which 129 were
\sneia. High-quality optical and NIR light curves of 36 of these were
published by \citet{contreras10}, and their analysis was presented in
the work of \citet{folatelli10}. Light curves of additional 50 \sneia\
were made available by \citet{stritzinger11}. 

Along with well-sampled light curves, the CSP collected spectra for most
of the SNe. However, due to the relative scarcity of spectroscopy time, the
temporal sampling obtained was more sporadic than that of the photometry.
In the present work we have collected the spectra of the \sneia\ with
light curves published by \citet{contreras10} and \citet{stritzinger11}
plus several SNe observed more recently for which clean photometry was
obtained without performing subtraction of the host-galaxy light
---i.e., because they were bright enough and isolated from their 
 hosts. Note that we do not include SN~2005hk --- a 2002cx-like object 
 observed by the CSP \citep{phillips07} --- in the analysis 
 due to the lack of CSP spectra near maximum light.

\begin{figure}[htpb]
\epsscale{1.0}
\plottwo{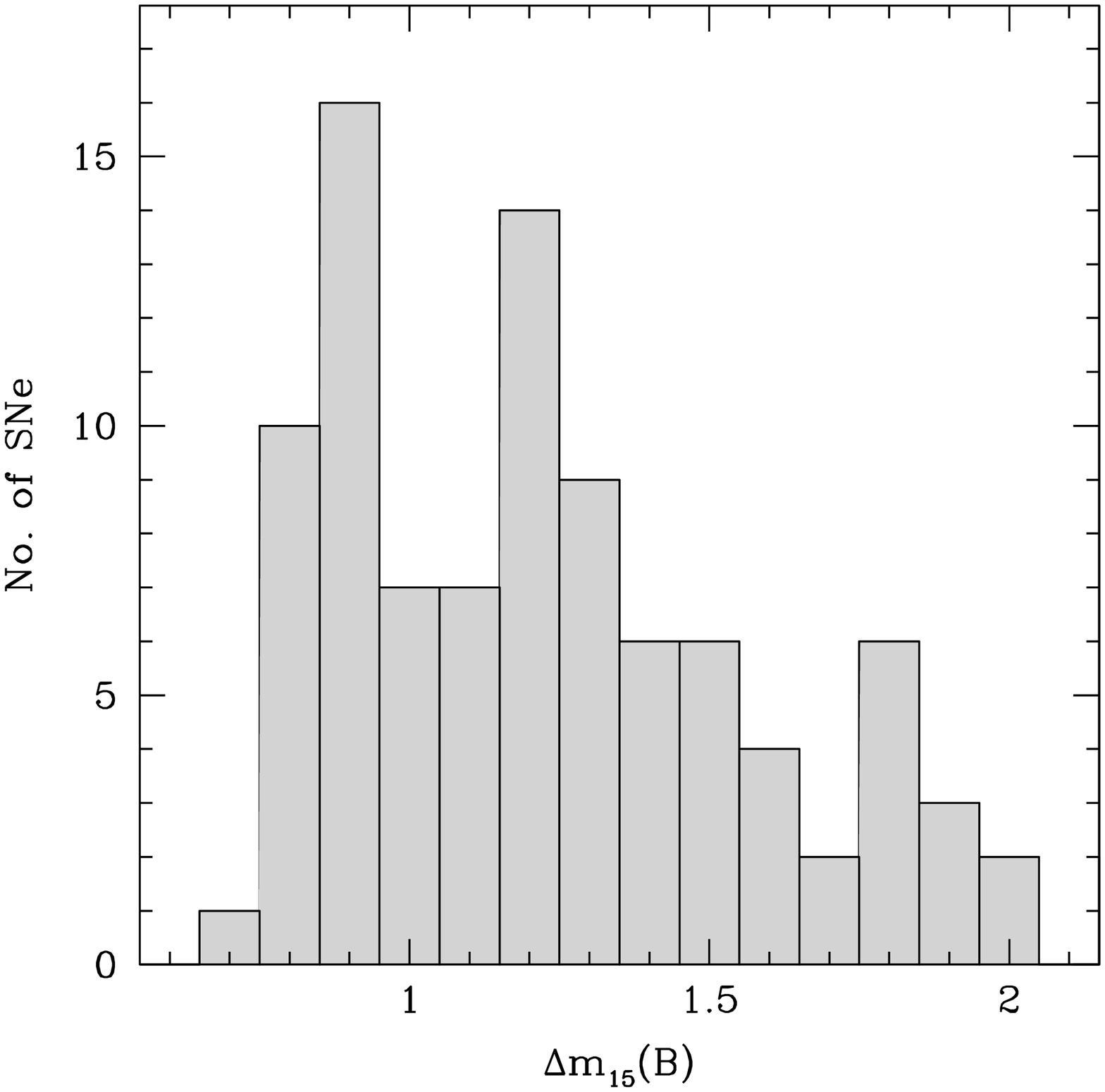}{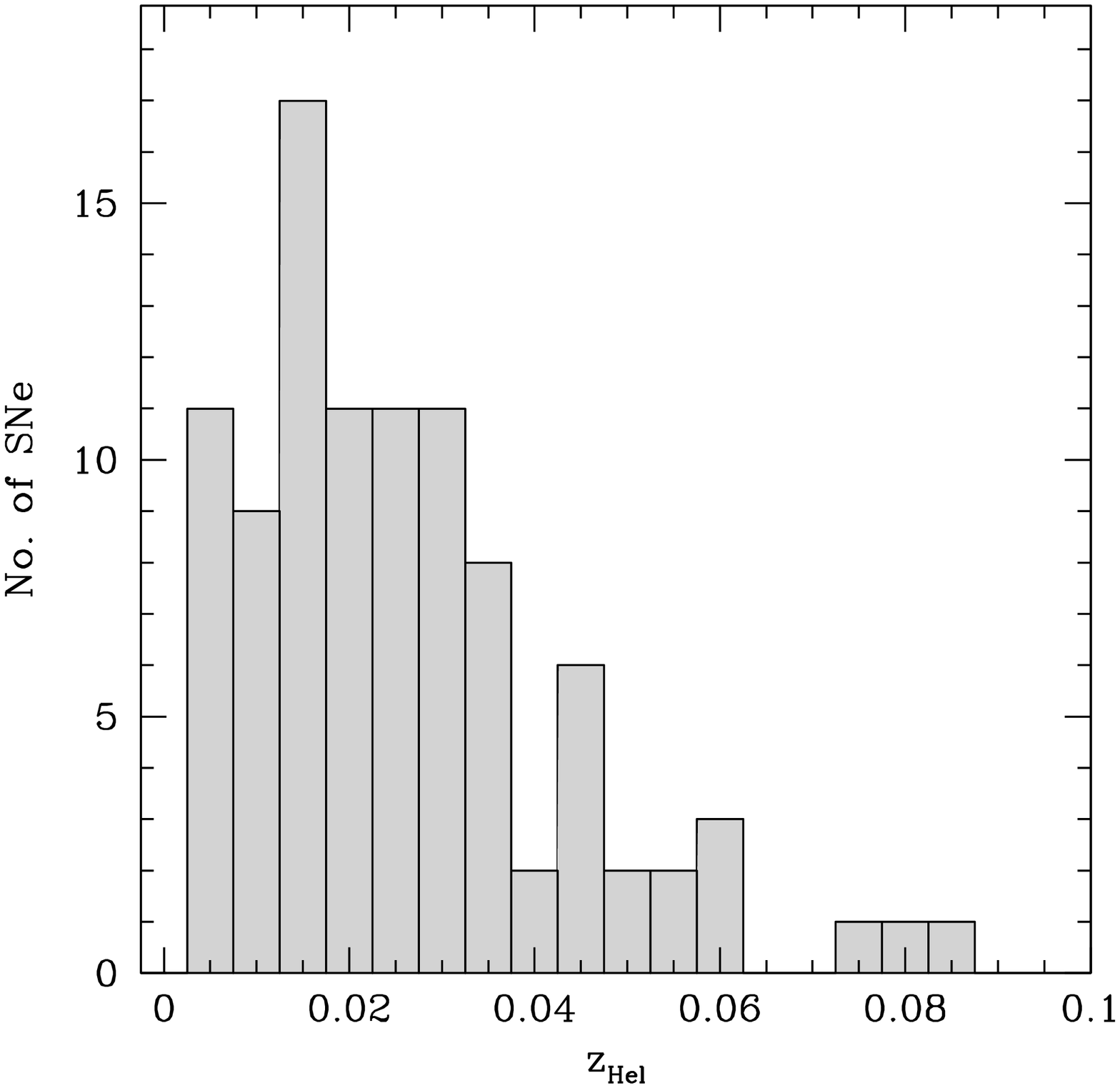}\\
\plottwo{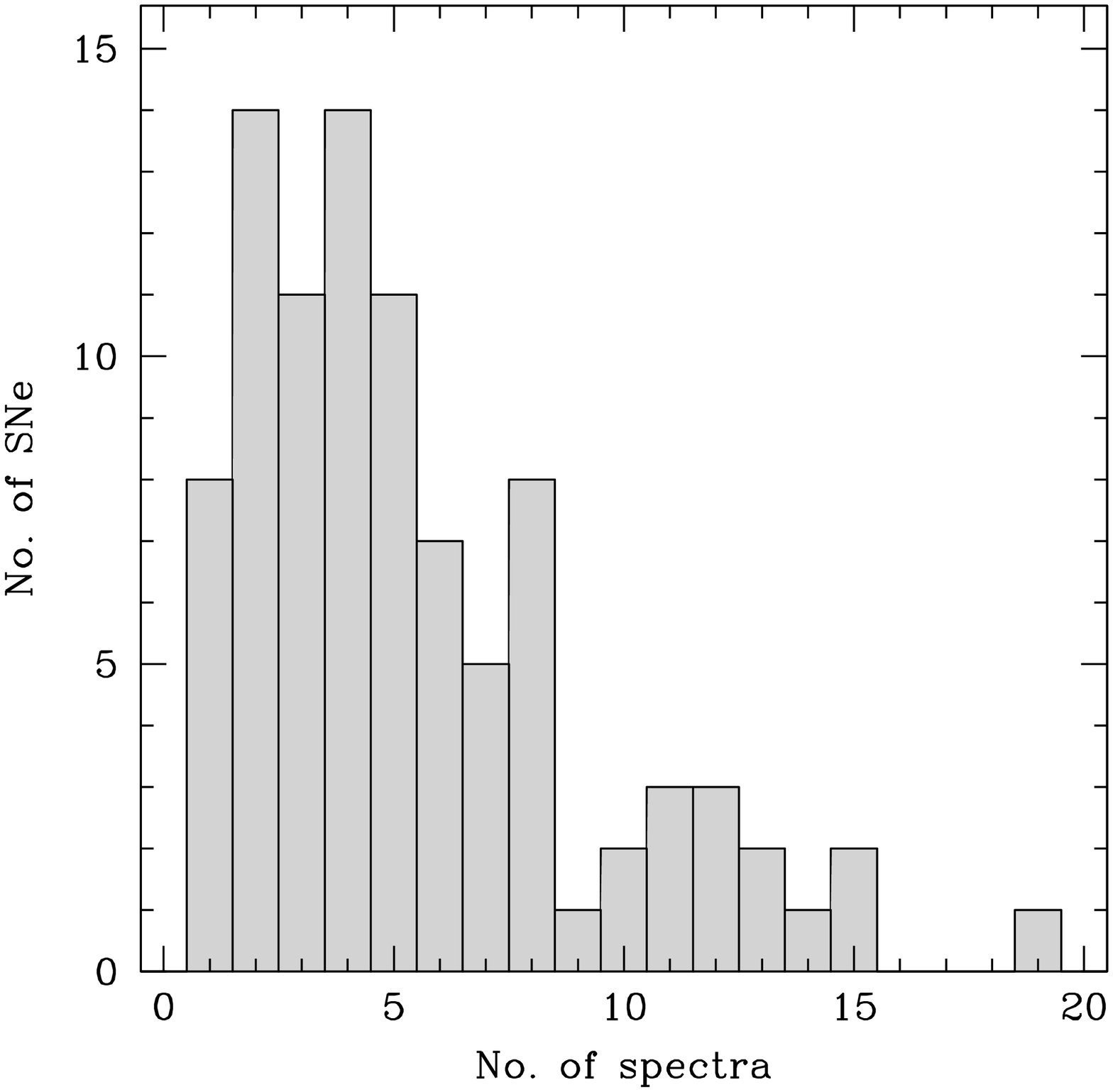}{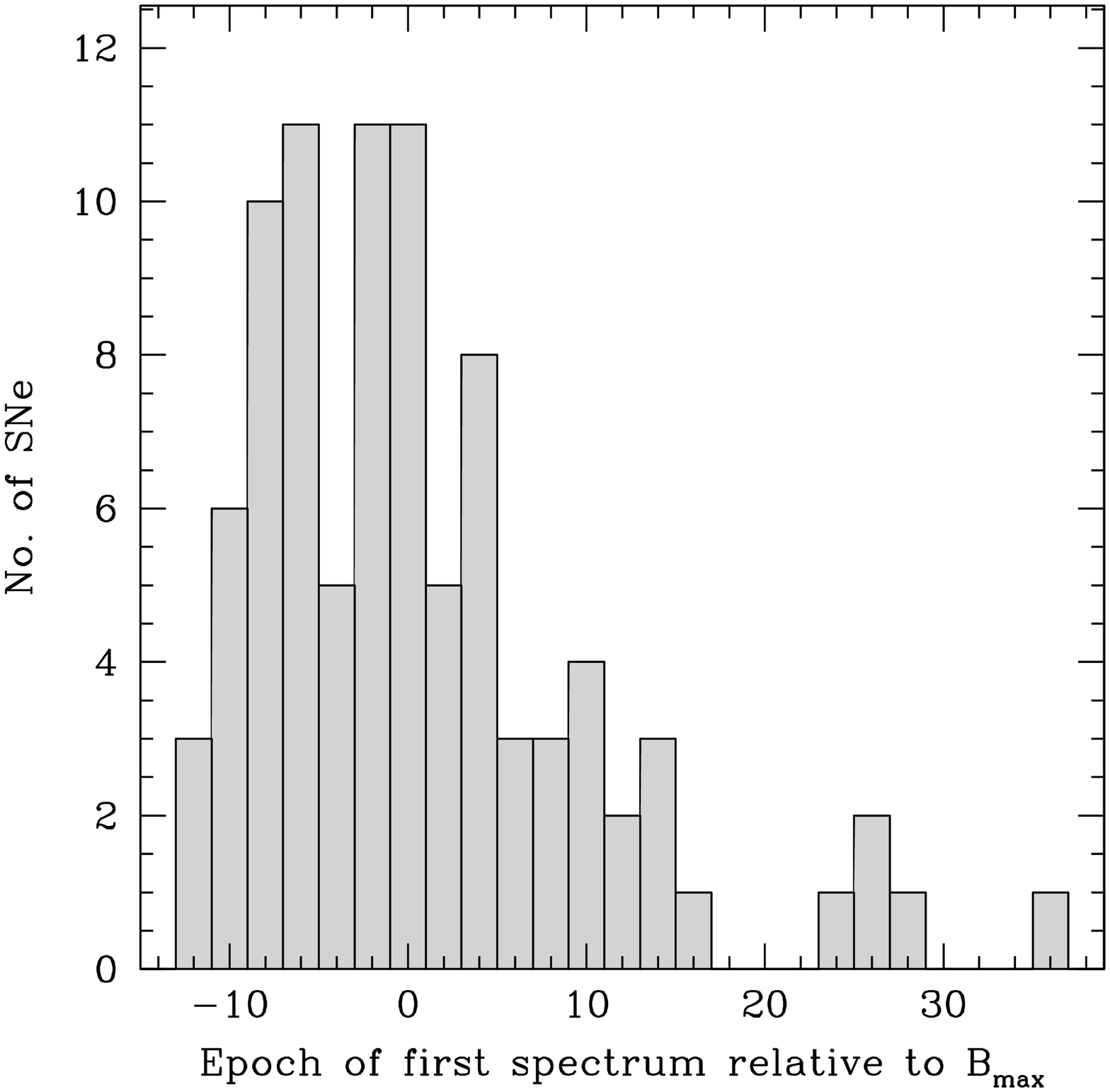}
\caption{({\em Top left\,}) Distribution of decline rates
  parameterized as \dm\ for the present SN sample. A wide range is
  covered between $0.6$ and $2.0$ mag. ({\em Top right\,})
  Distribution of heliocentric redshifts of the present SN sample.
  ({\em Bottom left\,}) Distribution of the number of spectroscopy
  epochs per SN. Most SNe were observed between 2 and 5 times. A few
  were observed on more than 10 epochs. ({\em Bottom right\,}) Distribution 
  of the epoch when the first spectrum was obtained for each SN. About
  77\% of the SNe were first observed spectroscopically
  before five days past maximum light.
  \label{fig:hist}}
\end{figure}

Table~\ref{tab:sne} lists the sample of \sneia\ selected for this
work along with the amount and span of spectroscopy epochs, and
spectral classification as defined in
Section~\ref{sec:spdiv}. Additional information in
  Table~\ref{tab:sne} includes redshift and photometric parameters.
The sample includes a large variety of \sneia, as can be inferred from
the wide distribution of light-curve decline rates parameterized by
\dm\ and shown in the top left panel of Figure~\ref{fig:hist}.  

For most of the objects in the sample, the heliocentric redshifts given in
Table~\ref{tab:sne} were obtained from the NASA/IPAC Extragalactic
Database (NED). Whenever possible, we have used CSP spectra to verify
the quoted values by measuring host-galaxy lines. For 55 spectra
of 34 SNe, we obtained a very good agreement with an average
difference in $z$ of 0.00009 and a dispersion of
0.00055 (165 km s$^{-1}$). 
For SNe~2005ag, 2006is and 2006lu, no information on the host galaxies or
their redshifts is available in NED, so we have adopted our measured
values, as indicated in Table~\ref{tab:sne}. These redshift values
were used to put the spectra in the rest frame. As shown in
Figure~\ref{fig:hist}, the SN sample covers a range of redshifts up to
$z=0.083$, with $\approx$80\% of the objects closer than $z=0.04$.

\subsection{Observations and Reductions}
\label{sec:obs}

The procedures followed by the CSP for spectroscopic observations and 
reductions were described by \citet{hamuy06}. As mentioned there,
most of the data were
obtained with the 2.5~m du Pont Telescope at Las Campanas  
Observatory, using the Wide Field CCD Camera (WFCCD) in 
long-slit spectroscopy mode. Other instruments used to
improve our spectroscopic time coverage were the Las Campanas
Modular Spectrograph at the du Pont, the Low Dispersion Survey Spectrograph
\citep[LDSS2;][]{allington94} on the Magellan Clay 6.5~m
telescope, and the Ritchey-Chr\'etien spectrograph at 
the 1.5~m CTIO telescope, operated by the SMARTS consortium. In
addition to these, more recently we have also employed,
at Las Campanas: LDSS3 on the Magellan Clay telescope (an upgrade of
LDSS2, with new grisms and different 
long slits); the Inamori Magellan Areal Camera and Spectrograph 
\citep[IMACS;][]{dressler11} on the Magellan Baade 6.5~m telescope, in
its long (f/4) and short 
(f/2) camera modes, with different combinations of gratings/grisms and
slits; the Boller and Chivens spectrograph at the du Pont telescope,
and finally, a few spectra have been obtained using the Magellan
Echellette \citep[MagE;][]{marshall08} spectrograph on the
Magellan Clay telescope.
We have also obtained single nights with the New Technology Telescope
(NTT) and the 3.6~m Telescope at ESO-La Silla,  
using the ESO Multi-Mode Instrument \citep[EMMI;][]{dekker86}
in medium resolution spectroscopy mode (at the NTT) and the ESO Faint Object
Spectrograph and Camera \citep[EFOSC;][]{buzzoni84} at the 3.6~m and
NTT telescopes. 

About 80\% of our spectra were obtained with the WFCCD, with most of
the remaining 20\% secured with EMMI, LDSS2/3, and IMACS.
In Table~\ref{tab:spec} we provide a complete journal of the spectroscopic
observations considered in the present work, giving for each spectrum the
spectral coverage, FWHM resolution, exposure time and airmass in the middle
of the observation. UT and Julian dates are also provided along with the
estimated rest-frame phase with respect to maximum light in the $B$
band\footnote{Hereafter the term ``maximum light'' refers to the time
  of $B$-band maximum, and all epochs will indicate the amount of
  rest-frame days with respect to that date.}. 

In its spectroscopic monitoring, the CSP obtained at least one
spectrum of the vast majority of the SNe which were selected for
photometric follow-up. Among the 36 \sneia\ included in
\citet{contreras10}, only SN~2005ir at $z=0.076$ was not observed 
spectroscopically by our program. Among the additional 50 objects published by
\citet{stritzinger11}, no spectra were obtained for only five SNe,
namely 2005hj, 2005mc, 2006bt, 2007hx, and 2007mm. As can be seen in the
bottom left panel of Figure~\ref{fig:hist} which shows the distribution
of the number of spectroscopic epochs for all the SNe in 
the present sample, for most of the objects we obtained between 2 and
5 spectroscopic epochs. Moreover, several SNe 
were followed more intensively and for longer intervals, which allowed
us to gather spectra at 10-15 different epochs.

We tried to concentrate our spectroscopic observations around the time
of maximum light. For most SNe we obtained the first spectrum
before or around maximum light. Specifically, for 72 out of
the 93 SNe the first spectrum was obtained earlier than five days
after $B$-band maximum light (see Figure~\ref{fig:hist}). In many
cases the monitoring was extended up to $\sim$50 days after the time
of $B$-band maximum. For the brightest objects we were able to obtain
spectra to approximately $+150$ days.  

Although most of the observations were performed without order-sorting
filters, the effect of second-order contamination is
negligible in most cases because \sneia\ in general do not show
extremely blue colors. The flux calibration was generally performed
using several spectrophotometric standards to reduce the risk of
introducing second-order contamination in the calibration of the red
part of the spectrum. We have evaluated the effect of second-order
contamination by comparing spectra obtained with
and without order-sorting filter using the Boller and Chivens
spectrograph at the du Pont telescope. We have done this with several
SNe observed between 2007 and 2009. In all cases, the results
are in agreement within a few percent between the two
instrumental setups. Unfortunately, the WFCCD spectrograph does not
allow observations to be made with order-sorting filters. However the
results of the Boller and Chivens spectrograph ---which has
significantly greater sensitivity in
the blue--- indicate that the effect is negligible also in the case of
the WFCCD. 

Telluric features were removed from almost all the spectra using 
appropriate standards observed each night with the same slit used for 
the supernova observations. Note that this procedure often left 
residuals from the strongest telluric bands which were not further 
corrected, as explained in detail in \citet{hamuy06}. Only 22 of the 
total of 604 CSP spectra were not telluric-corrected because the
necessary standards were not obtained.

\begin{figure}[htpb]
\epsscale{1.0}
\plotone{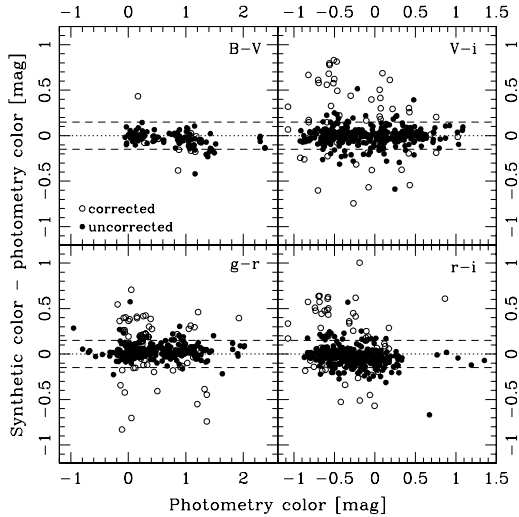}\\
\caption{Comparison of synthetic and photometry colors for the CSP
  spectroscopic sample. Clockwise from the top-left panel, colors
  shown are $(B-V)$, $(V-i)$, $(g-r)$, and $(r-i)$. Open symbols
  indicate spectra that were subsequently corrected to match the
  photometry as described in Section~\ref{sec:obs}. Filled symbols
  correspond to spectra that required no correction or that did not
  cover the minimum of three bandpasses to allow the correction. The
  dotted lines indicate zero deviation, and the dashed lines indicate
  $\pm$$0.15$ mag deviation.
  \label{fig:syncol}}
\end{figure}

To avoid light loss due to differential refraction, we carried out the
vast majority of the observations by aligning the slit at the
parallactic angle, especially when observing at low elevation. Also,
special care was taken to fit and subtract the underlying emission
from the host galaxy in order to obtain a clean SN spectrum. In order
  to assess the quality of the spectrophotometry, we compared
synthetic photometry computed 
through different bandpasses with the corresponding broad-band
photometry. For most of the spectra 
the wavelength range covered allows comparison with several
bandpasses. In general we find an agreement within a few percent
between synthetic and observed fluxes after removal of a constant flux
term. In the last column of Table~\ref{tab:spec} we provide the rms of
the differences in magnitudes between the bandpasses. In some cases we
were not able to perform the comparison, either because the spectrum
had a wavelength coverage that was too restricted, or because the
photometric data did not cover the epoch of the spectrum. When the rms
was larger than $\approx$$0.15$ mag among at least three
  bandpasses we used a low-order polynomial
function to correct the overall shape of the spectrum. As a
further consistency check, we compared synthetic colors obtained from
the spectra with corresponding colors measured from the photometry and
interpolated to the same epochs of the spectra. The comparison is
shown in Figure~\ref{fig:syncol}. The figure shows that the majority
of the spectra show color deviations within $\pm$$0.15$ mag. Most of
the spectra that show larger deviations in any given color are the
ones that were corrected by the procedure described above. A few
spectra with a large deviation in one color index were not corrected
because their wavelength span did not cover other bandpasses.

\begin{figure}[htpb]
\includegraphics[width=0.15\textwidth]{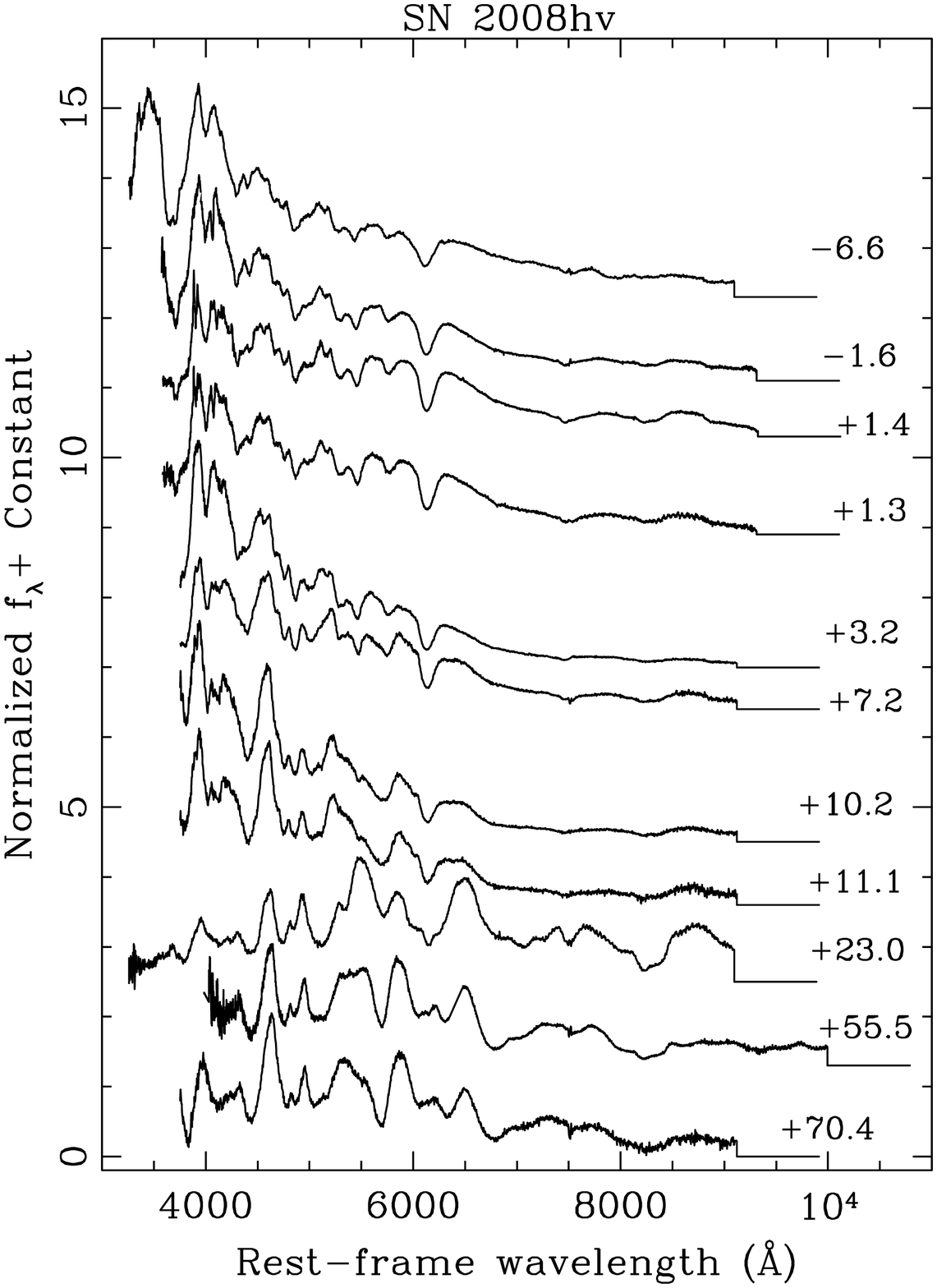}
\includegraphics[width=0.15\textwidth]{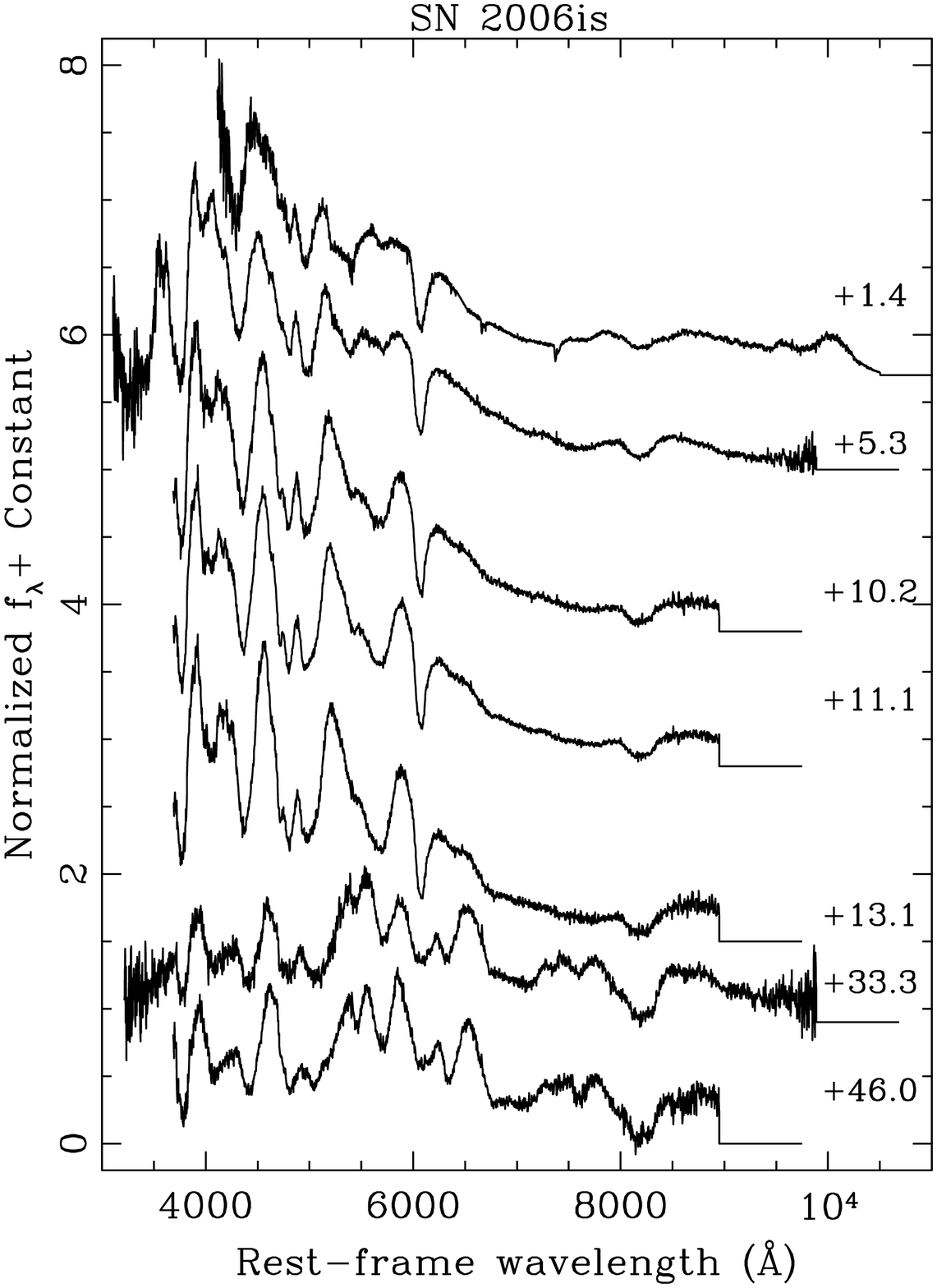}
\includegraphics[width=0.15\textwidth]{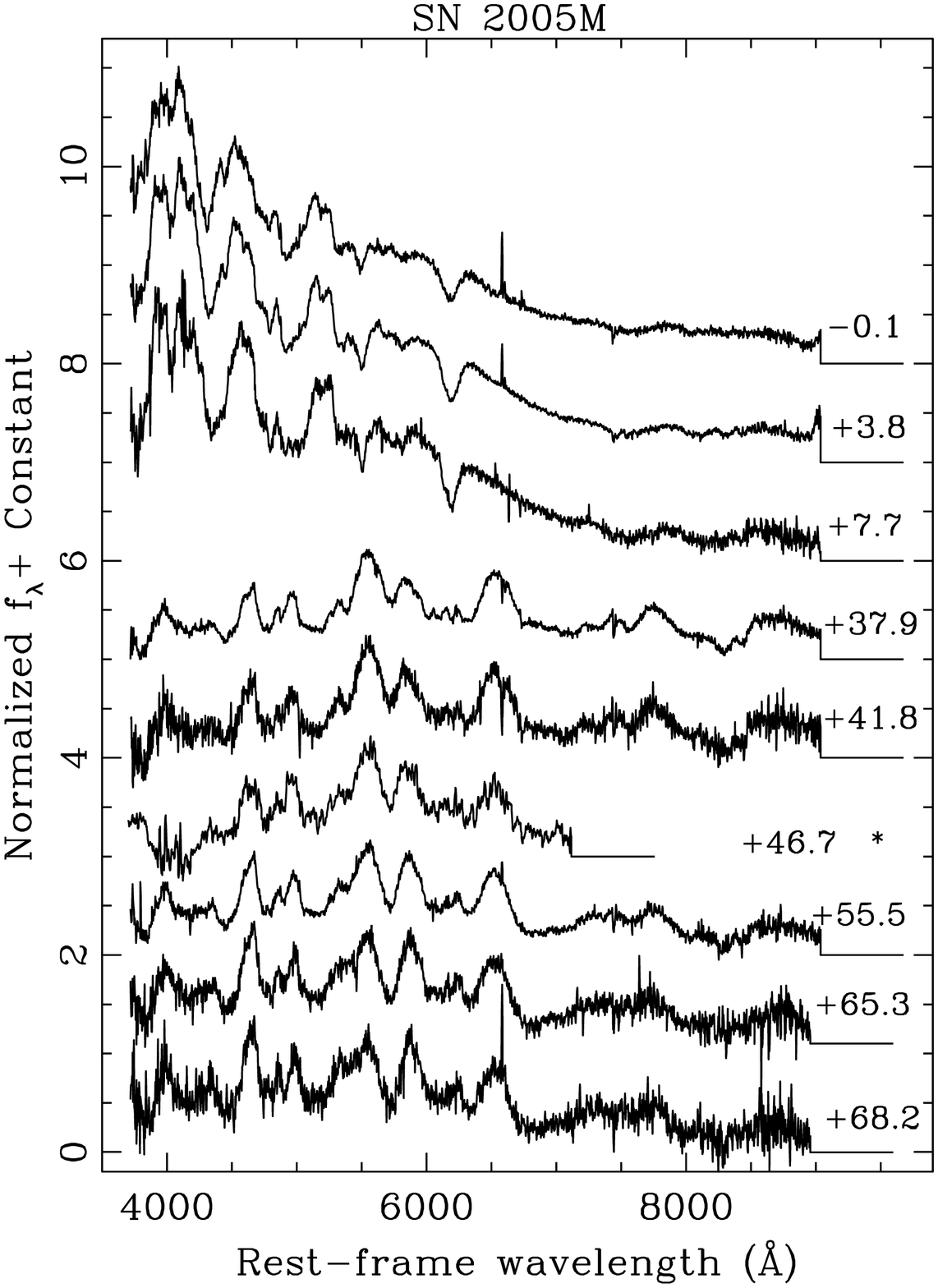}\\
\includegraphics[width=0.15\textwidth]{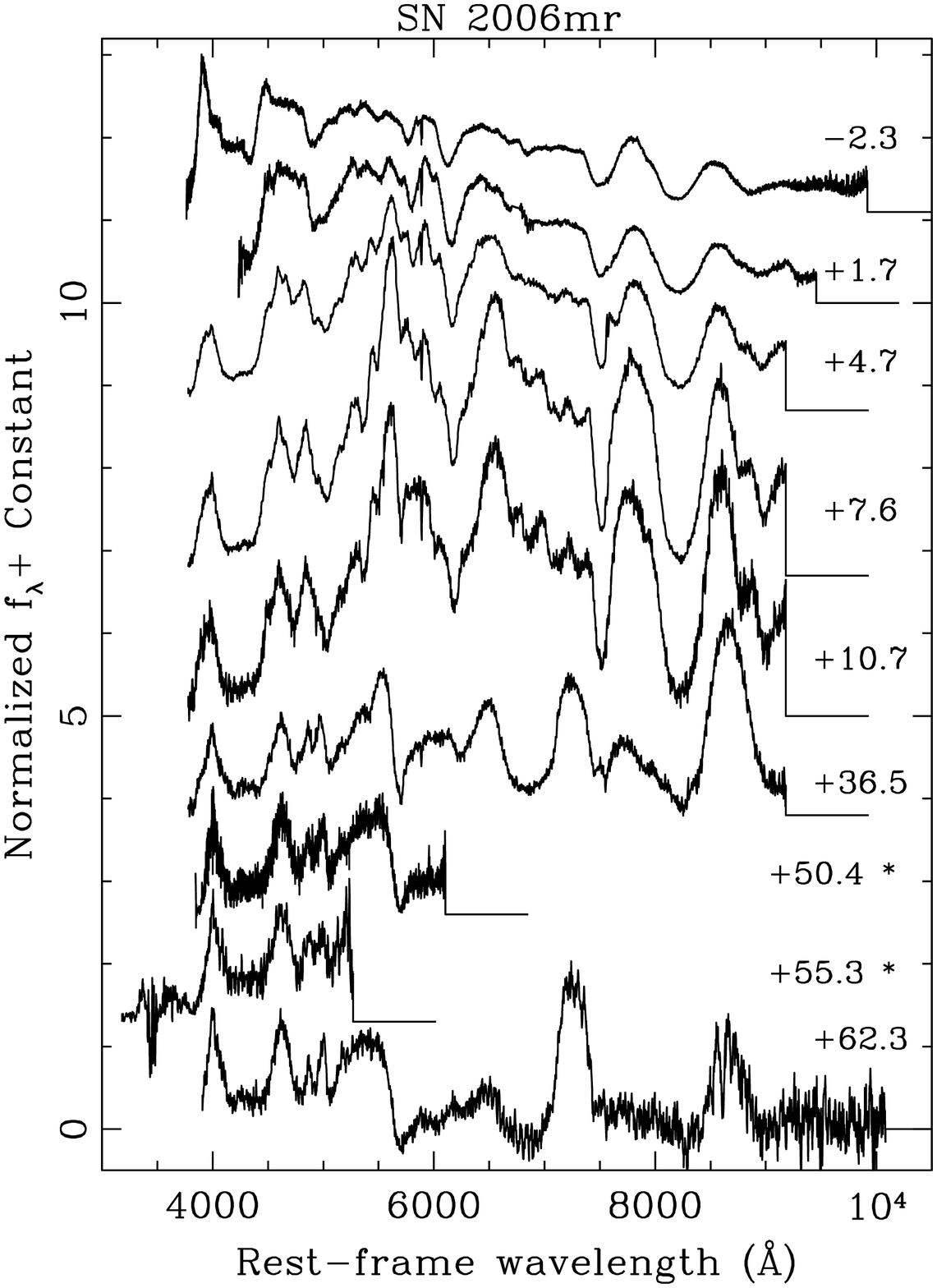}
\includegraphics[width=0.15\textwidth]{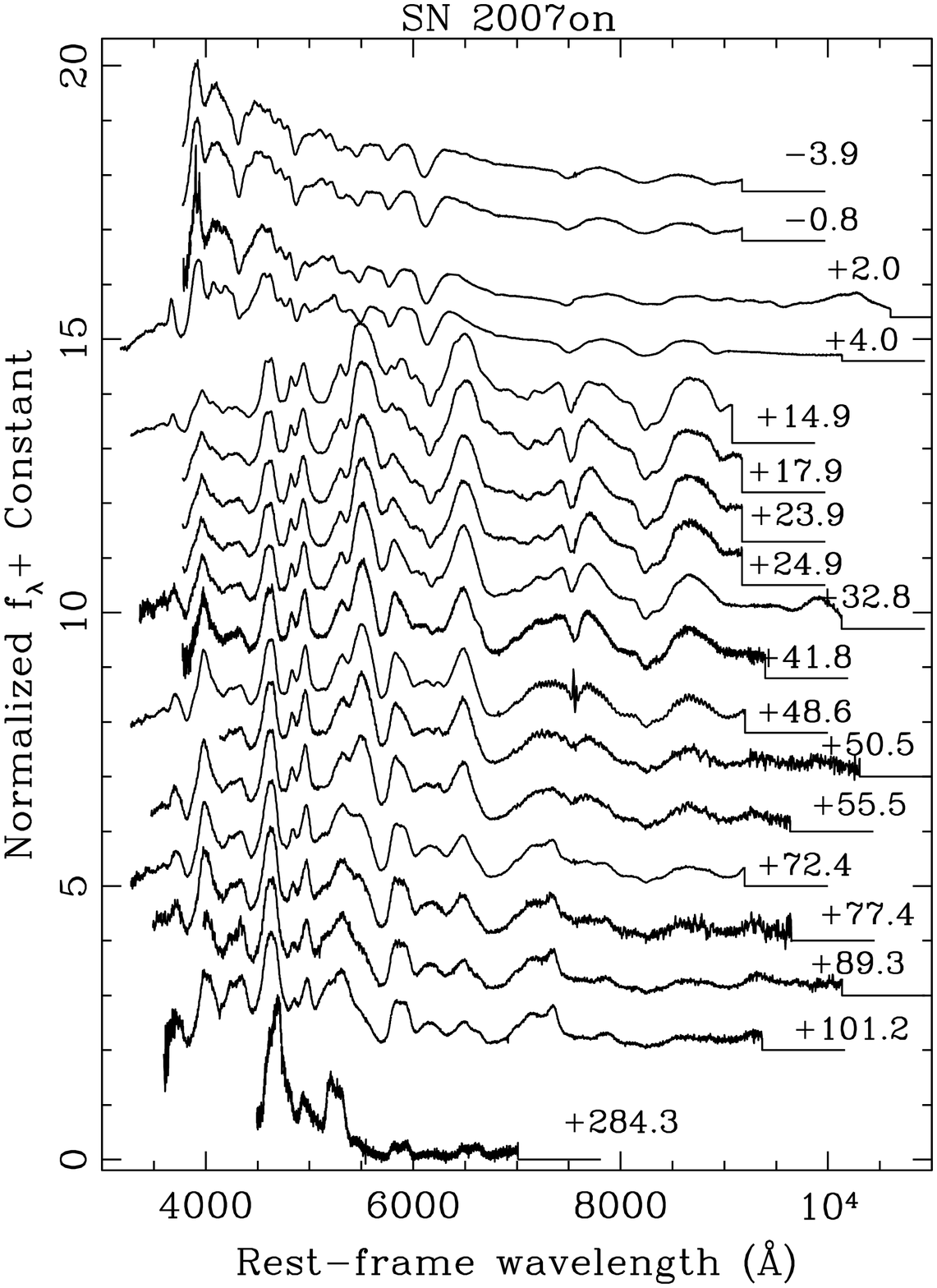}
\includegraphics[width=0.15\textwidth]{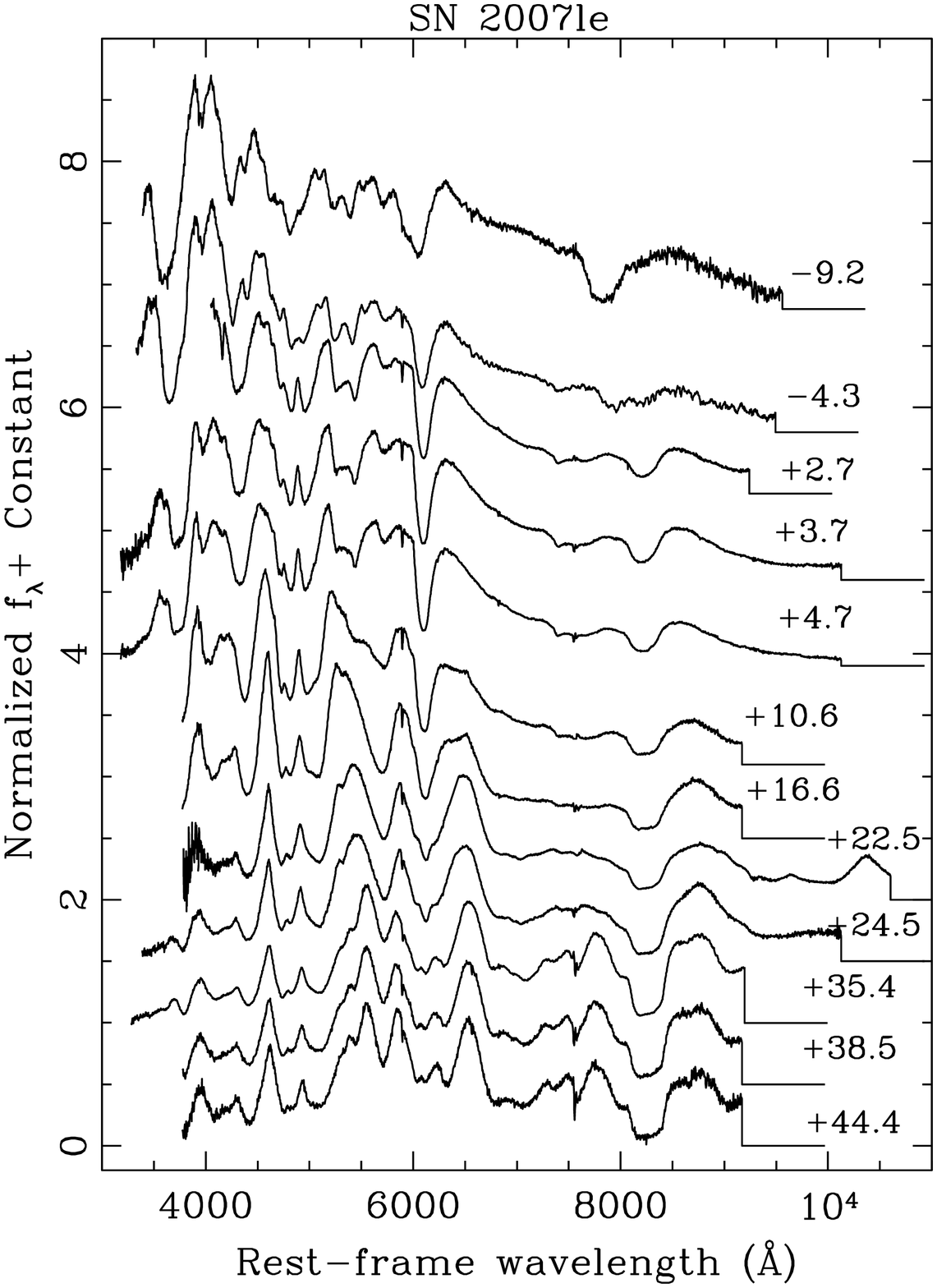}
\caption{Examples of spectral time-series obtained by the CSP. The
  spectra are shown in units of $f_\lambda$ and have been normalized
  by the average of the flux in a common wavelength range. An
  arbitrary additive constant was applied to each spectrum for
  clarity. The zero flux level is shown by a horizontal line to the
  right of each spectrum. The labels on the right-hand side show the
  epoch of each spectrum in days with respect to $B$-band maximum
  light. Asterisks indicate spectra that were smoothed. Examples shown
  in this figure represent the different Branch subtypes (see
  Section~\ref{sec:spdiv}): SNe~2008hv and 2006is are CN; SN~2005M is
  SS; SNe~2006mr and 2007on are CL (the former is eCL); and SN~2007le
  is BL. Similar plots for the complete sample of SNe presented in
  this paper will be made available along with the data on the CSP web
  site. 
\label{fig:spec}}
\end{figure}

Figure~\ref{fig:spec} shows examples of spectral
time-series for some of the SNe in the current sample. The complete
set of plots are provided as online-only material and will be
made available together with the spectroscopic data in the CSP web
site (\url{http://csp.obs.carnegiescience.edu}). 

A number of non-CSP spectra have been included in this paper. 
Some of them are unpublished, including: a total of 30 spectra of
  SNe 2006dd, 2006ef, 2006ej, 2006et, 
2006gj, 2006hb, 2006is, 2006kf, 2006lu, and 2006mr obtained with the
Boller \& Chivens CCD spectrograph at the Hiltner 2.4~m Telescope of
the MDM Observatory; four spectra of SNe 2005hc, 2005ku and 2006fw
obtained during the SDSS-II Supernova Survey \citep{frieman08} with
with the Boller \& Chivens 
CCD spectrograph mounted on the McGraw-Hill 1.3~m Telescope of the MDM
Observatory, and EMMI at the NTT telescope of ESO-La Silla; and one
spectrum of SN~2005A obtained with the 
Double Spectrograph \citep[DBSP;][]{oke82} mounted on the Hale 200~inch
Telescope at Palomar Observatory. In addition to this, 84 spectra of
SN~2004dt \citep{altavilla07}, SN~2004eo \citep{pastorello07}, SN~2005bl
\citep{taubenberger08}, 
SN~2005hj \citep{quimby07}, and SN~2006X \citep{wang08,yamanaka09}
were retrieved from The Online Supernova Spectrum Archive 
(SUSPECT, \url{http://suspect.nhn.ou.edu/$\sim$suspect/}). 
Near-maximum spectra of the SNe included in this work were added
from the CfA Supernova Program and the BSNIP. This comprises
106 spectra from the CfA sample \citep{blondin12}, and 38
from the BSNIP sample \citep{silverman12a}.
The SUSPECT, CfA and BSNIP spectra are not listed in Table~\ref{tab:spec}.

\section{SPECTRAL MEASUREMENTS}
\label{sec:meas}

Around maximum light, optical \sndia\ spectra present a strong
continuum marked by P-Cygni lines characteristic of dense
expanding material located around an emitting body. The most common ions
producing such features are \ion{Fe}{2}, \ion{Ca}{2}, \ion{S}{2},
\ion{Si}{2}, \ion{Na}{1}, \ion{Mg}{2}, and \ion{O}{1}. The
distribution of the lines and the large Doppler broadening causes
significant blending, particularly at blue wavelengths. For this
reason, the actual continuum flux and the individual line profiles are
difficult to determine. 

In order to quantify the spectroscopic properties of \sneia\ we have
performed measurements on different spectral features that are
identifiable during the evolution of the object around maximum
light. These measurements are: (1) line expansion velocities obtained
from Doppler shifts of the absorption minima of several lines, and (2)
pseudo equivalent widths of ``absorptions'' surrounded by local flux
maxima. 

\subsection{Line Expansion Velocities}
\label{sec:vel}

The shift of the absorption minimum of a line with respect to the rest
wavelength of the corresponding transition provides an estimate of the
average expansion velocity of the material producing the
absorption. Wavelength shifts can be converted to velocities via the
Doppler formula. We adopted the relativistic Doppler formula
\citep[see Equation~(6) of][]{blondin06}. Accurately measuring the
absorption minimum  
is not always easy in the case of SNe because the large speed of the
material broadens the lines and often makes them blend together. We
have selected a number of multiplets which are the most easily 
identifiable and isolated in the spectra of \sneia\ around maximum
light. For these, effective rest-frame wavelengths 
were computed based on air wavelengths and oscillator strengths
given by the NIST Atomic Spectra Database (version 3.1.5; available
from \url{http://physics.nist.gov/asd3}).
These ions and resulting effective wavelengths are: \ion{Ca}{2} H$\&$K
$\lambda$3945.02 and the IR triplet $\lambda$8578.79; 
\ion{Si}{2} $\lambda$4129.78 (4130) $\lambda$5971.89 (5972), and 
$\lambda$6356.08 (commonly referred to as 6355);
\ion{S}{2} $\lambda$5449.20 (5449) and $\lambda$5622.46 (5622),
where numbers between parentheses will be used as line identifications
in what follows. The \ion{O}{1} $\lambda$$\lambda$7772,\,7775
  doublet is commonly found in \sneia\ spectra. We have however not
  included it in this analysis because its absorption can be affected
  by residuals of the telluric A-band.

The observed line positions were derived via Gaussian fitting of the 
absorption minimum performed with the IRAF routine
{\tt splot}\footnote{IRAF is distributed by the National Optical Astronomy
  Observatory, which is operated by the Association of Universities
  for Research in Astronomy, Inc., under cooperative agreement with
  the National Science Foundation.} after removing the redshift
introduced by the host-galaxy recession velocity. Since the
  Gaussian function is generally not a good approximation of the
  complete absorption profile the fits were restricted to the core of
  the lines. This way we were able to obtain the location of the
  absorption minimum in a reproducible way. In cases when the minimum
presented a flat shape the Gaussian provided an approximation of the
central position. Whenever there was a double profile, we found the
location of both minima using two local Gaussians. In the following
analysis we consider the velocity of the redder component (with lower
velocity) whenever the line had a double minimum.

Measurement uncertainties, which are very much dependent on the width of
the line, its signal-to-noise ratio and spectral resolution, 
were estimated by performing repeated Gaussian fits
around the originally measured value. The limits of the fitting
regions on each side of the measured minimum were allowed to vary
between 40 and 70 \AA\ for the weakest lines (\ion{Si}{2}
$\lambda$4130, \ion{S}{2} $\lambda$5449 and 
$\lambda$5622), and between 40 and 100 \AA\ for the rest of the
lines. The limiting points of the fitting range were varied by three
pixels between repetitions, which implies that the number of Gaussian
fits was determined by the spectral sampling. The median absolute
deviation (MAD) of all Gaussian minima was adopted as an estimate of the
measurement uncertainty. In a conservative approach, we also compared
the median central wavelength 
of all the Gaussian fits with the originally measured position derived
from splot. In cases where these values differed by more than the
computed MAD, we adopted the absolute value of the difference as the
uncertainty. 

In the following sections we will use values of expansion velocities
at the time of $B$-band maximum light, as listed in
Table~\ref{tab:vels}. Since the observations very rarely coincided with
that exact phase, we have derived such values using data obtained near
maximum. In cases when there were several observations within one week
before and after maximum, we performed low-order (first or second)
polynomial fits. When only two observations encompassing maximum light
were available, we interpolated to the time of maximum. If the
  observations did not encompass the time of maximum light, we allowed for
  an extrapolation if there was at least one measurement obtained in
  the range of $[-1,1]$ day. Otherwise, we extrapolated any
  measurement obtained within four days from maximum using average slopes
for the velocity evolution obtained for well-sampled SNe. These slopes
are summarized in Table~\ref{tab:velslp}. If more than one such
extrapolation was done for a given SN, then the results were combined
using a weighted average.

We tested the robustness of the fits and interpolations to
maximum light by repeating the calculations described above after
removing data points from the best-observed SNe. Such tests
confirmed that the derived velocities at maximum light are stable
and that median deviations from the originally fit values are
smaller than $\approx$300 km s$^{-1}$ for any number of data points
down to two. In the end, we were able to
obtain expansion velocities at maximum light for 78 of the
SNe in our sample.

\subsubsection{Velocity decline rate}
\label{sec:dv}

Based on the work of \citet{benetti05}, we have parameterized the rate of
change of the \ion{Si}{2} $\lambda$6355 velocity after maximum
light. Because of the relatively small number of velocity measurements
obtained, to quantify this property we simply define \deltav\ as
the difference between the velocity at maximum light and that 20 days
after \citep[see a discussion on the difficulty of measuring velocity
  gradients in][]{blondin12}. In order to obtain the velocity at the latter epoch for a large
sample of SNe, all available measurements obtained after 7 days
post-maximum were employed. Similarly to the method applied at maximum
light, low-order polynomial (mostly linear) fits were used to derive
the value at +20 days. When the data were too sparse to perform the
fits but one observation was available between +17 and +23 days, an
average slope of $-76 \pm 16$ km s$^{-1}$ d$^{-1}$ 
obtained from 21 SNe with good coverage was employed to correct the
measured velocity to the epoch of +20 days. \deltav\ values are given
in the last column of Table~\ref{tab:vels}.

\subsection{Pseudo-Equivalent Widths}
\label{sec:ew}

\begin{figure}[htpb]
\epsscale{0.5}
\includegraphics[angle=270,width=0.48\textwidth]{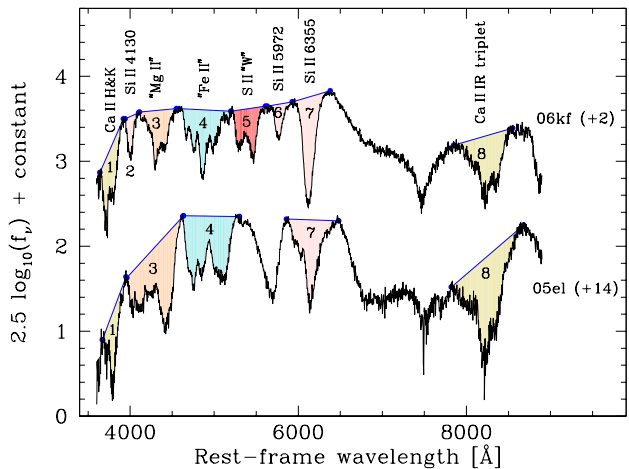}
\caption{Examples of the definition of pseudo-equivalent features used
  in this paper \citep[see also][]{garavini07}. Spectra of
  SN~2006kf at $+2$ days and SN~2005el at $+14$ days are shown with
  black lines. The data are shown
  as $\log(f_\nu)$ for clarity, and pseudo-continuum
  fits are shown as straight lines in the graph, although
  they are defined as such in $f_\lambda$ space. The labels indicate
  feature numbers and adopted names (see Table~\ref{tab:featw}). Eight
  features are defined until about ten few days after maximum light
  (see top spectrum). At later times, fewer features are considered
  (see bottom spectrum). Note that feature 3 by definition
  can include the wavelength region of feature 2.\label{fig:featdef}}   
\end{figure}

A convenient way of quantifying spectral properties of \sneia\ is to
use {\em pseudo-equivalent widths} (\ew) of apparent absorption features. The
prefix ``pseudo'' is used to indicate that the reference ``continuum''
level adopted is generally not the actual continuum emission. We have
measured \ew\ for eight spectral features defined by
\citet{folatelli04} following the prescriptions 
given by \citet{garavini07}. Examples of the \ew\ features are shown in
Figure~\ref{fig:featdef}. Feature definitions are summarized in
Table~\ref{tab:featw}. The first column gives the names that will be
used throughout the text. The second column provides the main lines that
are associated with each absorption feature near the time of maximum
light. Although these line identifications are generally not unique and they
vary with SN phase, we adopt them for the feature names to help the
reader follow our analysis. Nevertheless, the association of features
1 and 8 with \ion{Ca}{2} lines is persistent with phase, although
  near maximum light feature 1 may include a significant contribution
  of \ion{Si}{2} $\lambda$3858 \citep[see][]{blondin12}. Features 3 and 4 are
blends of lines due to Fe and intermediate-mass elements; for convenience, we
call these ``\pwthree'' and ``\pwfour'' in
reference to the ions that produce the strongest absorptions at
maximum light. Feature 5 corresponds to the W-shaped absorption due to
\ion{S}{2} that is observed until about ten days after maximum
light. Finally, features 2, 6, and 7 are associated with \ion{Si}{2}
$\lambda$4130, $\lambda$5972, and $\lambda$6355, respectively, although
other ions also contribute. The former two features are weak and can
only be identified until about ten days after maximum. Feature 7
is measured until approximately two months post maximum, although its
identification with \ion{Si}{2} $\lambda$6355 is only valid until day
$\approx$+10. We have left aside the absorption around 7500 \AA\ that
is mostly due to the \ion{O}{1} $\lambda$$\lambda$7772,\,7775
  doublet because it lies near the telluric A-band absorption. Even
  though most of the spectra were corrected for telluric absorption,
  the residuals of such correction can be large enough to affect the
  measurement of the \ion{O}{1} doublet. 

Each \ew\ measurement is obtained by defining a straight
``continuum'' level between two flux peaks and computing the integral
of the spectrum flux relative to the continuum (with positive sign for
simplicity). The flux peaks which define each feature are
selected within a fixed spectral range so that the maximum allowed
spectral range is spanned. Columns 3 and 4 of Table~\ref{tab:featw} provide the
wavelength ranges allowed for the location of the flux peaks on each
side of the eight features \citep[see][]{garavini07}. Note that the
definition of \pwthree\ permits it to cover the wavelength
range of \pwtwo. This generally happens at times after maximum
light or for the subtype of cool (CL) \sneia\ (see
section~\ref{sec:spdiv}), and in those cases feature 2 is not measured.

Because \ew\ measurements involve an integration over a relatively
wide wavelength region of $\sim$100 \AA,
statistical uncertainties, in relative terms, are usually several
times to one order of magnitude smaller than the uncertainties in the
flux in a wavelength resolution bin. A significant contribution to the
measurement uncertainty arises from systematic errors in the
definition of the continuum points, variations in the flux peak levels
due to an imprecise flux calibration, to contamination by host-galaxy
light or even to noise. Most of our spectra have high signal-to-noise
ratios ($S/N \gtrsim 30$ per 10\AA\ bins for over 90\% of the
spectra), so we have focused on other possible sources of systematic
error. In our measurement
procedure we have included a contribution to the uncertainty estimated
by randomly varying the regions used for the continuum fit and for 
the integration \citep[see][]{garavini07}.

The effect of reddening is to smoothly modify the shape of the 
continuum and thus its slope around the absorption
features. \citet{nordin11a}
find that this effect reduces the \ew\ of some features,
with a magnitude of $<$5\%, for $E(B-V)<0.3$ mag, and assuming the
reddening law of \citet{cardelli89} with $R_V=1.7$. \citet{garavini07}
find similar results adopting a law with $R_V=3.1$. To avoid this
possible systematic error, we corrected the spectra of SNe with
$E(B-V)>0.3$ mag using the same reddening law and $R_V=1.7$ \citep[this
choice of $R_V$ is supported by the results of Section~\ref{sec:Lum} and
those of, e.g.,][]{folatelli10,foley11a,mandel11}.  

Contamination by the host galaxy adds flux to the continuum and
therefore produces systematically lower \ew. Host-galaxy
contamination is not important in our sample because most of the SNe
are nearby and isolated from the bright regions of their hosts. As
explained in Section~\ref{sec:obs}, special care was taken to subtract
the underlying emission from the host galaxies, and synthetic
photometry from the resulting spectra was compared with
broad-band photometry in order to check the quality of the
subtraction. For most of the spectra, the rms of the 
differences between synthetic and observed photometry was $<0.1$
mag (see Table~\ref{tab:spec}), which indicates low degrees of
contamination. \citet{garavini07} estimated roughly 10\% decrease in
\ew\ for every 10\% of contamination from host-galaxy light in the
observed flux. Based on this, the effect of contamination is
negligible for most of the spectra in the present sample. 

In a similar way as was done for the line expansion velocities
(Section~\ref{sec:vel}), we derived \ew\ values at $B$-band maximum
light. Table~\ref{tab:ews} lists these values for 78 SNe. When several
measurements were available within one week before 
and after the time of maximum, a smooth polynomial fit was used. If
only two data points were obtained encompassing maximum light within
$-4$ and $+4$ days, then an interpolation was performed. In a few
cases, only data before or after maximum were available within
$[-7,+7]$ days. In those cases, an extrapolation was allowed if the
closest point to maximum light was not further than one day. For SNe
which only had one measurement in the range $[-4,+4]$ days or when
measurements in that range did not encompass maximum light, average
slopes determined from the best observed SNe were used for correcting
the \ew\ value to the time of maximum light. Table~\ref{tab:ewslp}
provides the average \ew\ slopes used for this purpose (see
Section~\ref{sec:spdiv} for a detailed definition of
\sneia\ subtypes). Similarly to the velocity fits, we performed tests
by removing data points from the best-observed SNe and found that the
resulting values at maximum light are robust. Median deviations were
estimated to be within 5 \AA\ for all \ew\ parameters.

\subsection{Comparison with CfA and BSNIP Data}
\begin{figure}[htpb]
\epsscale{1.0}
\plottwo{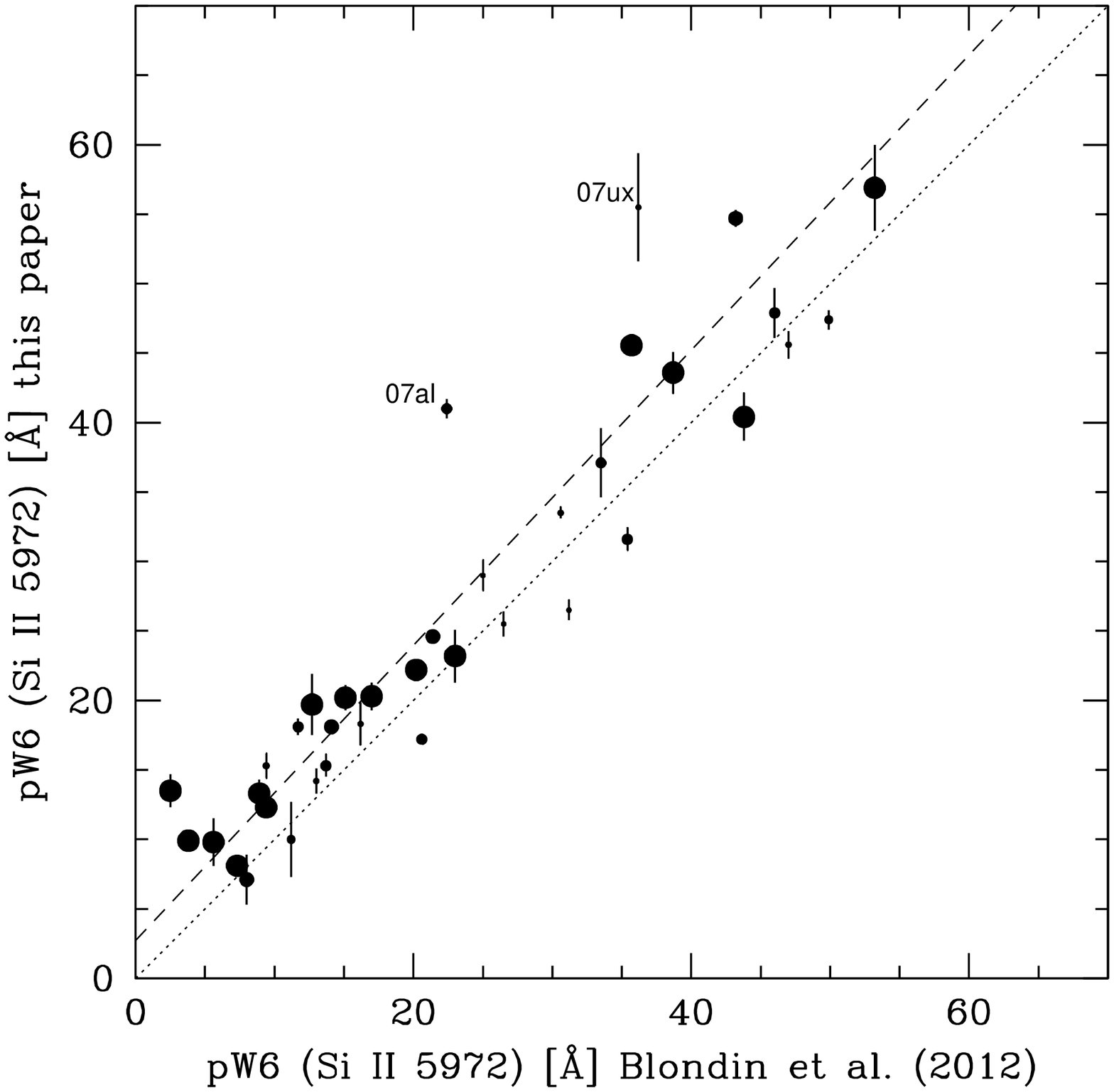}{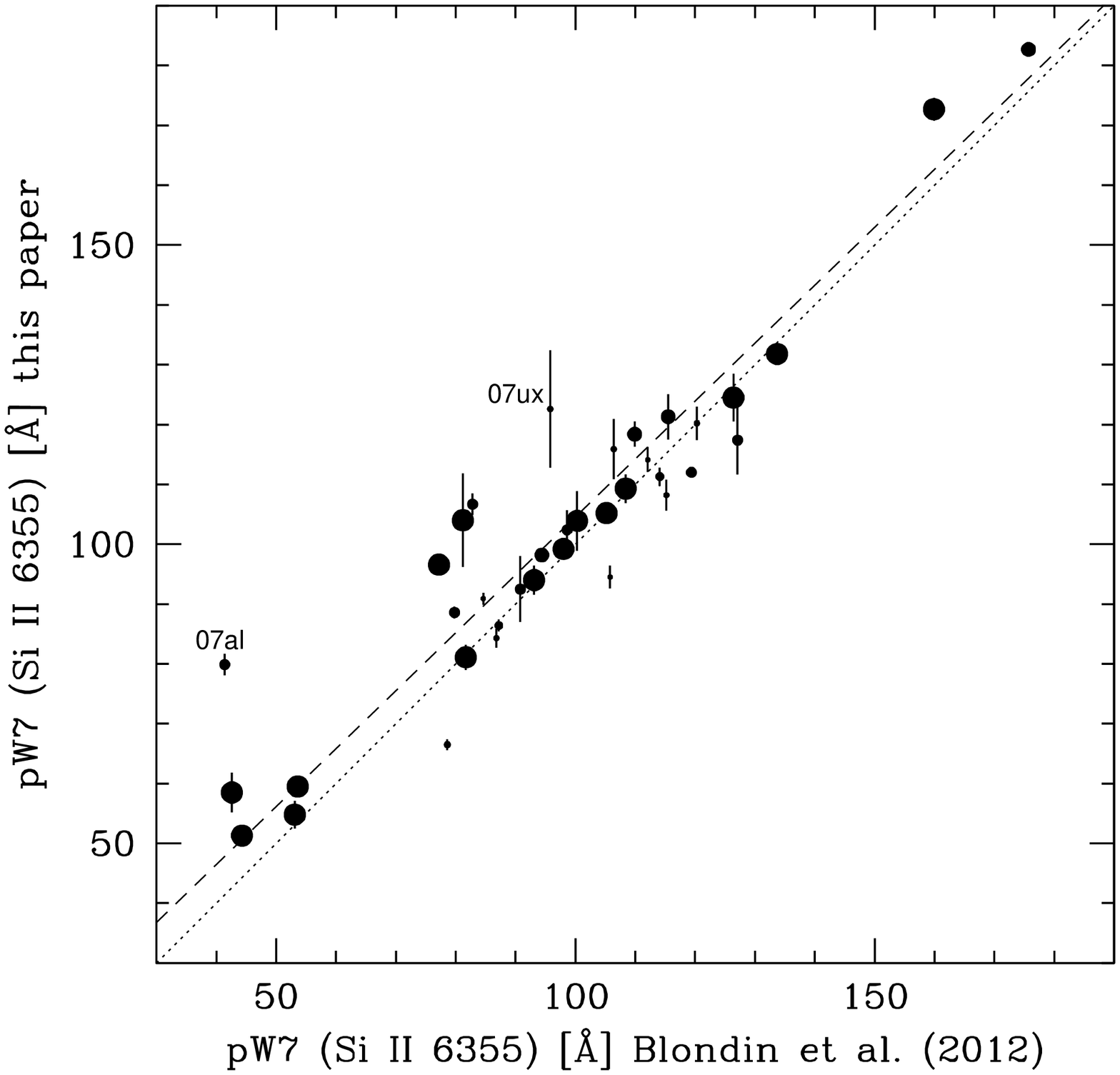}\\
\epsscale{0.5}
\plotone{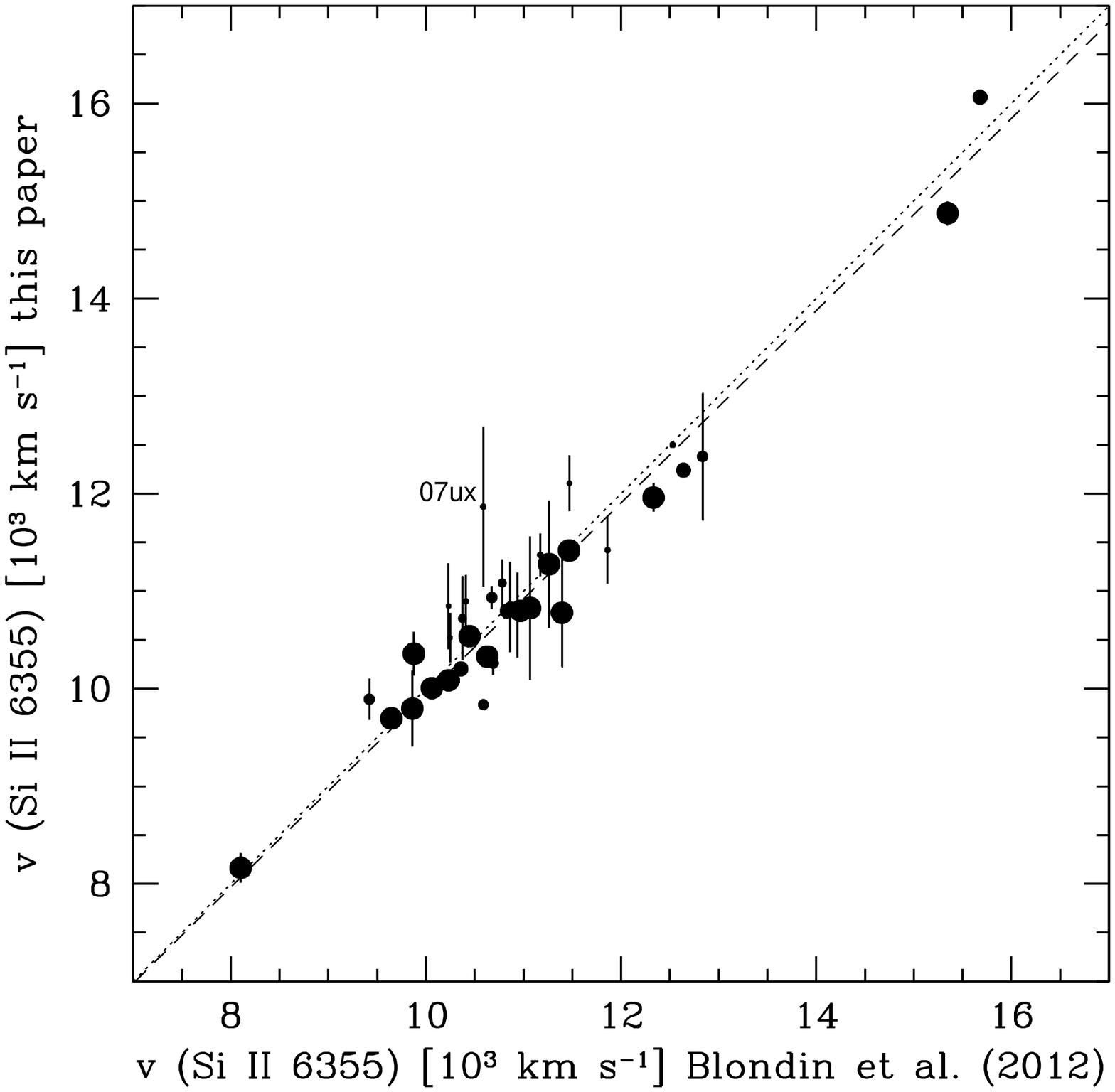}
\caption{Comparison of \pwsix\ ({\em top left}), \pwseven\ ({\em top
    right}) and \vsix\ ({\em bottom}) for the same SNe
  observed by the CSP and the CfA Supernova Program
  \citep{blondin12}. Symbol sizes are inversely proportional to the
  time interval of the CfA spectra relative to $B$-band maximum
  light. The dotted lines indicate the one-to-one relation. The
  dashed lines show fits to the data points.\label{fig:compcfa}} 
\end{figure}

\citet{blondin12} published spectroscopic measurements of line
velocities and pseudo-equivalent widths for an expanded sample of
\sneia\ collected by the Centre for Astrophysics (CfA) Supernova
Program (see their Table~4). We compared the measurements for
thirty-seven objects in common with our
sample. Figure~\ref{fig:compcfa} shows the comparison of \pwsix,
\pwseven, and \vsix, all evaluated at maximum light. We can see that
in general there is a good agreement in the 
$pW$ values. A systematic shift of $\approx 3 \pm 2$ \AA\ in is
observed which indicates that the measurements of \citet{blondin12}
are slightly smaller than our own. Straight-line fits to the
\ew\ measurements yield: $pW$6\,(Si\,II\,5972)$_{\mathrm{CSP}}=2.73
(\pm 0.53) + 1.06 (\pm 0.02) \times$$pW$6\,(Si\,II\,5972)$_{\mathrm{CfA}}$, and
$pW$7\,(Si\,II\,6355)$_{\mathrm{CSP}}=7.8 (\pm 1.4) + 0.97 (\pm 0.01)
\times$$pW$7\,(Si\,II\,6355)$_{\mathrm{CfA}}$.
The differences may be due to a larger incidence
of host-galaxy contamination in the CfA spectra, or by a
systematically different measurement procedure. The largest
discrepancies are found for SNe 2007al and 2007ux. The
CfA spectra of these objects shows systematically redder
  continua as compared with the CSP spectra at
similar epochs. This may be due to host-galaxy contamination which
produces a systematic decrease in the \ew\ of both \ion{Si}{2} lines.

In the case of \ion{Si}{2} velocities, the agreement is very good. A
weighted average of the differences yields $\Delta v \equiv
v_{\mathrm{CfA}}-v_{\mathrm{CSP}} = 78 \pm 63$ km s$^{-1}$. A fit to
the velocity measurements, considering uncertainties in both axes,
yields $v_{\mathrm{CSP}}=90\,(\pm 260) + 0.99\,(\pm 0.02) \times
v_{\mathrm{CfA}}$ km s$^{-1}$. The absolute values of the differences
are $<600$ km s$^{-1}$ for all SNe except for SNe~2005bl and 2007ux,
for which we get velocities at maximum that are $\approx$750 km
s$^{-1}$ lower and $\approx$1300 km s$^{-1}$ higher than those
measured by the CfA team, respectively. These differences can be explained by
the fact that the CfA spectra were obtained at $-2.9$ days and $5.8$
days relative to maximum light for each SN, respectively. In fact, when we 
compare measurements from spectra at similar epochs for these SNe, 
in all cases the velocities agree within the uncertainties with those
of the CfA.

The Berkeley Supernova Ia Program (BSNIP) recently published a series
of papers with their \sndia\ spectroscopic sample. \citet{silverman12b}
presented spectroscopic measurements of 432 near-maximum spectra of
261 \sneia. Among those SNe, 52 objects are in common with
our sample. We used the data in their Tables~B1 to B9 to compare
velocity and \ew\ measurements. We compared each of their measurements
with an average of all available measurements of the same SN from spectra
obtained by the CSP within four days of the BSNIP spectrum. This was done
for all eight of our \ew\ features and for the velocities of \ion{Si}{2},
\ion{Ca}{2}, and \vfour\ (the latter corresponds to their \ion{S}{2}
``W'' velocity.) The agreement for the velocities is good, with
no significant deviations between the two samples. However, the weighted rms
scatter of the differences was $\approx$500 km s$^{-1}$ for the
\ion{Si}{2} and \ion{S}{2} lines, and $\approx$1000 km s$^{-1}$ for
the \ion{Ca}{2} lines. These dispersions indicate that assuming
flat uncertainties of 100 km s$^{-1}$ for individual measurements may
be an underestimation.

For the \ew\ measurements, the agreement is again good within the
quoted uncertainties for \pwone, \pwtwo, \pwfive, \pwsix, and
\pweight. The weighted dispersion of the differences is 2--3
\AA\ for the weak 
\ion{Si}{2} features, and 15 \AA\ and 33 \AA\ for \pwone\ and
\pweight, respectively. These values provide an indication of the
actual measurement uncertainties. The relative dispersion of
\pwfive\ values is large (about 15 \AA) and may be due to differences
in the continuum-fitting regions that define the \ew\ measurements
\citep[compare Table~\ref{tab:featw} in this paper with Table~1
  of][]{silverman12b}. Non-negligible systematic differences of
3--4~$\sigma$ appear in the cases of \pwthree, \pwfour, and
\pwseven. Our \ew\ are larger on average than those of
\citet{silverman12b} by $18.5\pm4.7$ \AA\ for \pwthree, $15.5\pm4.3$
\AA\ for \pwfour, and $7.0\pm2.4$ \AA\ for \pwseven, based on 
28, 37 and 52 measurements, respectively. In the first two cases, we
suspect that the discrepancy arises from differences in the regions of
continuum fitting (our regions are wider and may thus lead to larger
\ew\ values). Although the discrepancy in \pwseven\ is smaller, we
could not find a clear explanation for it. Since this parameter
measured at maximum light is useful for characterizing \sneia\ as will
be shown in the following sections, we performed the comparison with
measurements obtained only within one week of maximum and found a
still-significant, although smaller, discrepancy of $5.6\pm2.1$
\AA\ among 30 measurements.

\section{SPECTROSCOPIC DIVERSITY}
\label{sec:spdiv}

\begin{figure}[htpb]
\epsscale{1.0}
\plotone{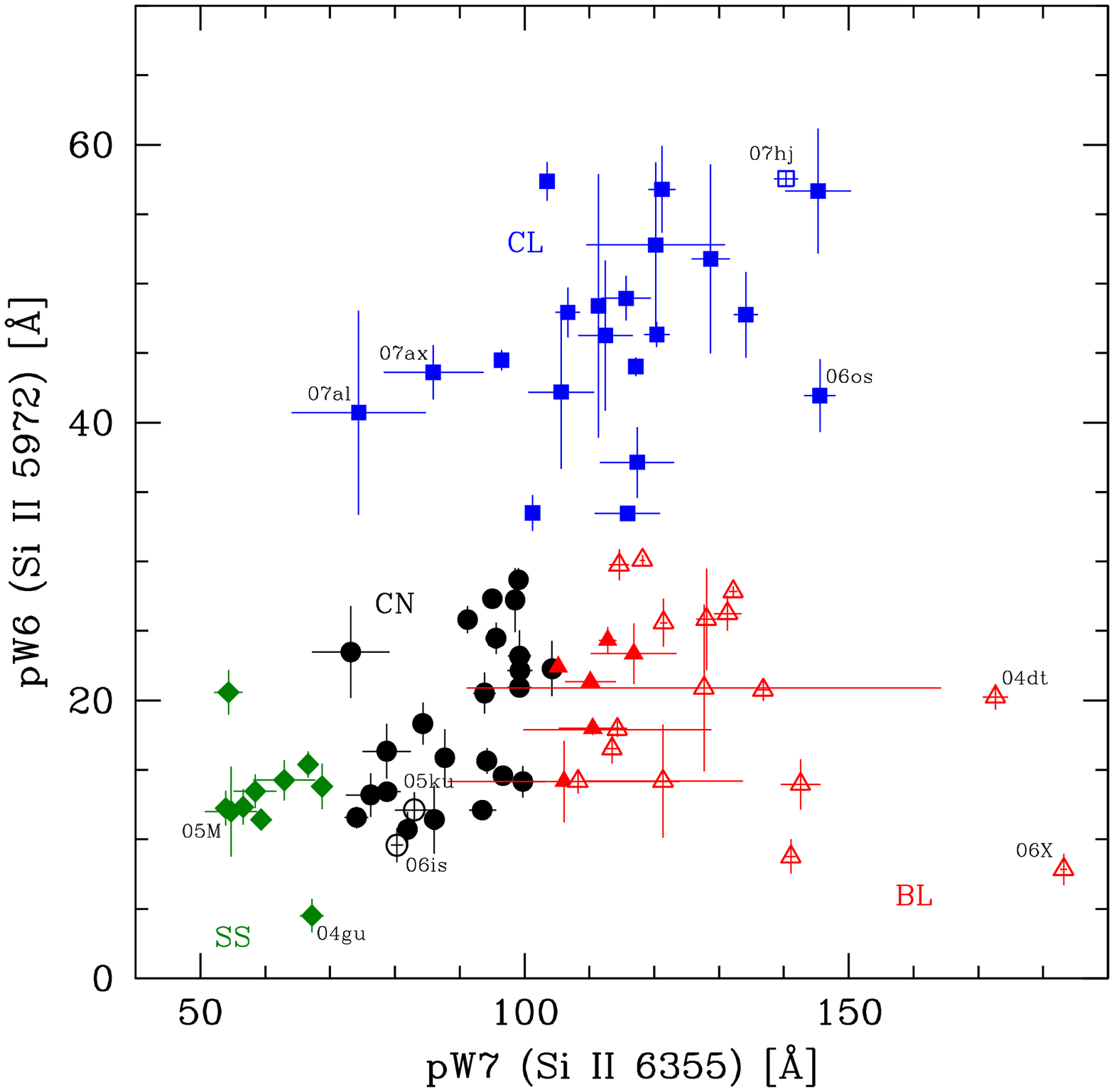}
\caption{Subtypes of \sneia\ as defined by \citet{branch06} (see
  also Section~\ref{sec:spdiv}) can be
  identified in the diagram of \pwsix\ versus \pwseven. Core Normal
  (CN; {\em black circles}), Shallow Silicon (SS; {\em green diamonds}), Broad
  Line (BL; {\em red triangles}), and Cool (CL; {\em blue squares}) subtypes
  are distinguished based on a larger sample presented by
  \citet{blondin12}. Open symbols indicate SNe with \vsix\,$>12000$
  km s$^{-1}$ at maximum light, that is HV objects \citep{wang09}. 
\label{fig:ew6ew7}} 
\end{figure}

In spite of the relative spectroscopic homogeneity among
\sneia\ as compared with other SN types, some variations exist that
are most notable in spectra obtained before or around the time of
maximum light. These variations are quantified here in terms of the
spectroscopic measurements described in the previous section. This
study serves to distinguish different subtypes of \sneia\ and to
detect peculiar cases. In Figure~\ref{fig:ew6ew7} we present the
\pwsix\,--\,\pwseven\ diagram, which was introduced by \citet{branch06}
to identify the subtypes of Core Normal (CN), Broad Line (BL),
Shallow Silicon (SS) and Cool (CL) \sneia. Recent versions of this
diagram have been presented by \citet{blondin12} for the CfA sample,
and by \citet{silverman12b} for the BSNIP sample. In a way this
scheme is a summary of the spectroscopic diversity discovered thus
far: a decreasing temperature sequence from 1991T-like to normal and
1991bg-like SNe (SS - CN - CL), plus the high-velocity SNe (BL). We
note that the SS group includes 2002cx-like and 1991T-like SNe which
have been suggested to be physically distinct types of objects, with
the spectroscopic differences being most evident in the expansion
velocities \citep{li03,branch04,phillips07}.

\citet{blondin12} presented the classification of several hundred SNe
observed by the CfA Supernova Program in the scheme of
\citet{branch06}. To facilitate comparison with this work, we adopt
the following criteria to define the different subclasses, which are
very similar to those used by \citet{blondin12}:

\begin{itemize}
\item
  CL SNe: \pwsix\,$>30$ \AA,
\item
  BL SNe: \pwseven\,$>105$ \AA\ and \pwsix\,$<30$ \AA,
\item
  SS SNe: \pwseven\,$<70$ \AA,
\item
  CN SNe: $70\leq$\,\pwseven\,$\leq 105$ \AA\ and \pwsix\,$\leq 30$ \AA.
\end{itemize}

\noindent As this scheme is based on \pwsix and \pwseven\ at maximum
light, we were able to classify 78 of the \sneia\ in our sample in
the following groups: 25 CN, 21 CL, 22 BL, and 10 SS objects. The
classification of each SN is given, when available, in
Table~\ref{tab:sne}.

The divisions given above are somewhat arbitrary and, as noted by
\citet{branch09}, there is no evident discontinuity in the overall
spectroscopic properties as we move along the
\pwsix\,--\,\pwseven\ diagram. If we 
compare the fraction of objects within each subtype with the 
sample of \citet{blondin12} (see their Table~4 which includes the CfA
sample and ten previously published \sneia), we obtain similar
distributions. There are 32\% (38\%) CN, 28\% (30\%) BL, 27\% (18\%)
CL, and 13\% (14\%) SS SNe in the CSP (CfA) sample. All these
fractions are within the statistical uncertainties of the limited
samples. If anything, the CSP sample contains relatively fewer
CN and BL SNe and more CL objects. Comparing our classification
  with that of \citet{silverman12b} for twelve SNe in common, we find
agreement except for SN~2006D, which they classify as BL and we
  include in the CN group. This is explained by a slightly different
boundary between the CN and BL subclasses in \pwseven\ as adopted by
  \citet{silverman12b}.

\begin{figure}[htpb]
\epsscale{1.0}
\plotone{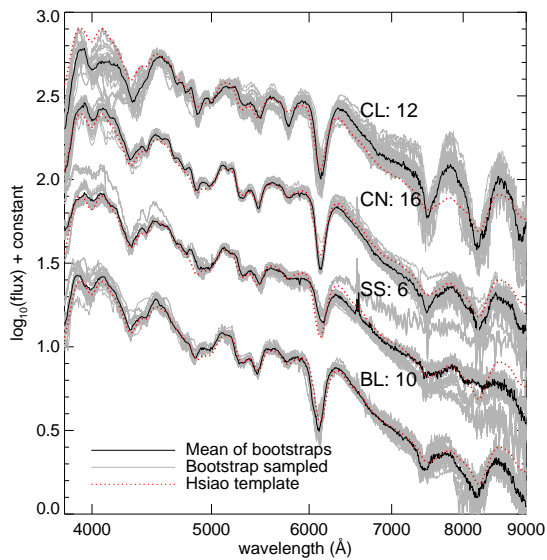}
\caption{Composite spectra at maximum light for different Branch
  subtypes. The labels above each spectrum indicate the subtype and
  the number of SNe involved in the calculation for that
  subtype. Each bootstrap combination is shown with a gray line. The 
  black lines show the mean of all bootstrap combinations. The red
  dotted lines show the template spectrum of normal \sneia\ at maximum light
  as computed by \citet{hsiao07}.\label{fig:spave}} 
\end{figure}

Figure~\ref{fig:spave} shows composite spectra at maximum light for
each Branch subtype \citep[cf. Figure~11 of][]{blondin12}. The
  composite spectra were constructed using CSP observations obtained
in the range of $[-4,+4]$ days with respect to maximum
light. We ensured that a SN only contributed one spectrum in a given
bin, such that the composite spectra are not dominated by a few SNe
with many time-series observations. The calculations involved 16 CN,
12 CL, 6 SS, and 10 BL SNe. The spectra were all converted to
the rest frame and corrected by extinction in the Milky Way and the
host galaxy. Extinction corrections were done using the law of
\citet{cardelli89} and assuming $R_V=3.1$ for the Milky Way component
and the host galaxy component 
when $E(B-V)_{\mathrm{Host}}<0.3$ mag, and $R_V=1.7$ for the host
galaxy component when $E(B-V)_{\mathrm{Host}}>0.3$ mag 
  \citep[see][]{folatelli10,foley11a,mandel11}.  
The composite spectra were computed using a bootstrapping
algorithm. For each Branch subtype we constructed repeated realizations
of the average spectrum by randomly selecting spectra from the available
sample. In each realization the resampling was performed with
replacement, so a single spectrum could enter more than once in the average. 
The bootstrap procedure was repeated 100 times for each Branch
subtype. Each bootstrap mean is plotted in the Figure as a
gray line. The mean of all individual realizations is plotted as a
black line. As opposed to \citet{blondin12}, we have not normalized
the spectra by a pseudo-continuum function, and thus we can see the
{\em actual} differences in color. 

As pointed out by \citet{blondin12}, the CN group shows the highest
degree of homogeneity. In general 
the largest dispersion is seen blueward of 4500 \AA. This is most
noticeable in the BL, SS and CL SNe. Some of this scatter is
probably due to sampling issues or inaccurate corrections for host-galaxy
extinction that, as we show in Section~\ref{sec:spphot}, affects more the
BL and SS groups. The dispersion in the SS subtype can in part be
due to the small number of spectra used in the combination. However,
part of the differences in the blue range of the
spectrum are likely due to intrinsic differences in line strengths.
This is most likely the case for the CL SNe whose spectra at these
wavelengths are sensitive to small differences in effective
temperature \citep{nugent95}. As we explain in
Section~\ref{sec:bv} most of the CL SNe can be considered to have
suffered little reddening by dust. This suggests that the differences
in the blue part of the spectrum among CL SNe are mostly of intrinsic
origin. 

It is interesting to note that the composite spectra of CN, SS, and BL
subtypes show a double-component \ion{Ca}{2} IR triplet
absorption. For CL SNe instead, the strong absorption appears to arise from a
single component. In the cases of CN and BL SNe, the profile is
dominated by a component at 11000--12000 km s$^{-1}$. A shoulder on
the blue wing of the absorption profile is seen at 18,000--20,000 km
s$^{-1}$ that can be associated with a high-velocity feature (HVF).
The same velocity components are seen in SS SNe, although the
absorption is weaker than in the previous cases. Such HVF of
\ion{Ca}{2} are commonly detected in pre-maximum spectra of \sneia.
There have been claims that HVF are ubiquitous to this type of SNe
\citep{mazzali05}. The composite spectra presented here agree with
such a picture. Unfortunately, the wavelength coverage in the blue part
of the spectrum is not long enough to allow a study of the \ion{Ca}{2}
H\&K absorption profiles. The \ion{Ca}{2} IR triplet is well covered
by a large fraction of our spectra and, as opposed to \ion{Ca}{2}
H\&K, it is conveniently isolated from other strong lines\footnote{Although
\ion{O}{1} $\lambda$8446 may appear in that part of the
spectrum.}. However, on this part of the spectrum the continuum flux
is low and the spectrum is affected by sky emission and instrumental
fringing, which complicates a detailed analysis of the line profile.   
In Section~\ref{sec:ewdm} we will further study the variation of strength
and shape of the \ion{Ca}{2} IR triplet feature. 

It can also be noted from Figure~\ref{fig:spave} that SS objects most
clearly show the \ion{Na}{1}~D absorption at the redshift of the host
galaxies. This absorption is also evident for CN SNe, but much less so for BL
objects and almost negligible in the case of CL SNe. This again would
indicate that SS objects suffered from larger extinction by dust
associated with the gas that produced the \ion{Na}{1}~D absorptions.
It is also worth pointing out the strong incidence of narrow emission
lines of H$\alpha$ and [\ion{N}{2}] among SS SNe. This suggests a
closer relation of this subtype of \sneia\ with star forming,
and therefore dusty, regions of their hosts. As seen in
Section~\ref{sec:spphot}, SS SNe show the slowest decline rates and
are more luminous than the average of \sneia. Therefore the
association of SS SNe with star-forming galaxies agrees with the
findings of \citet{hamuy96}, \citet{howell01}, \citet{gallagher08},
\citet{hicken09}, \citet{sullivan10} and \citet{brandt11}, among others.

\begin{figure}[htpb]
\epsscale{1.0}
\plottwo{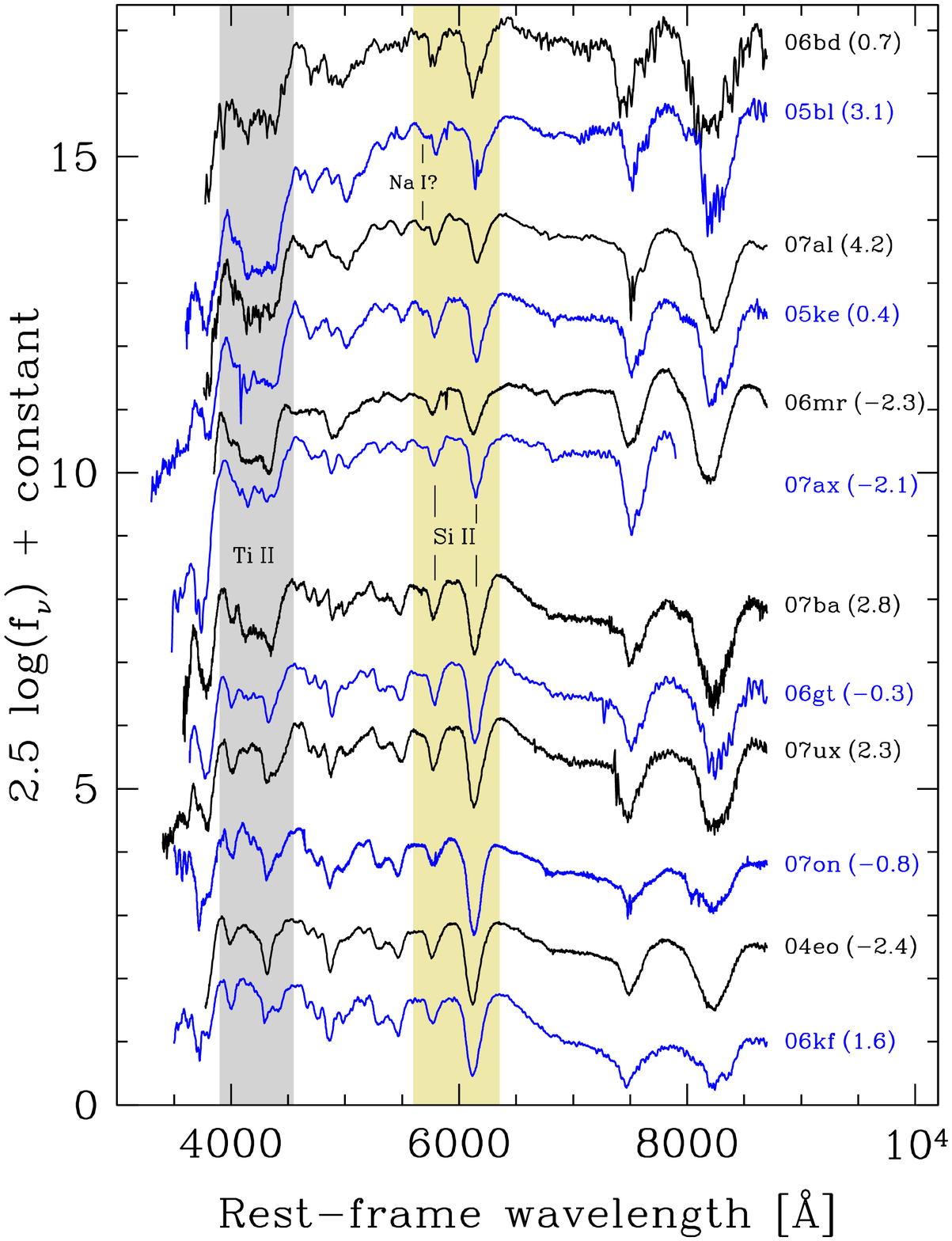}{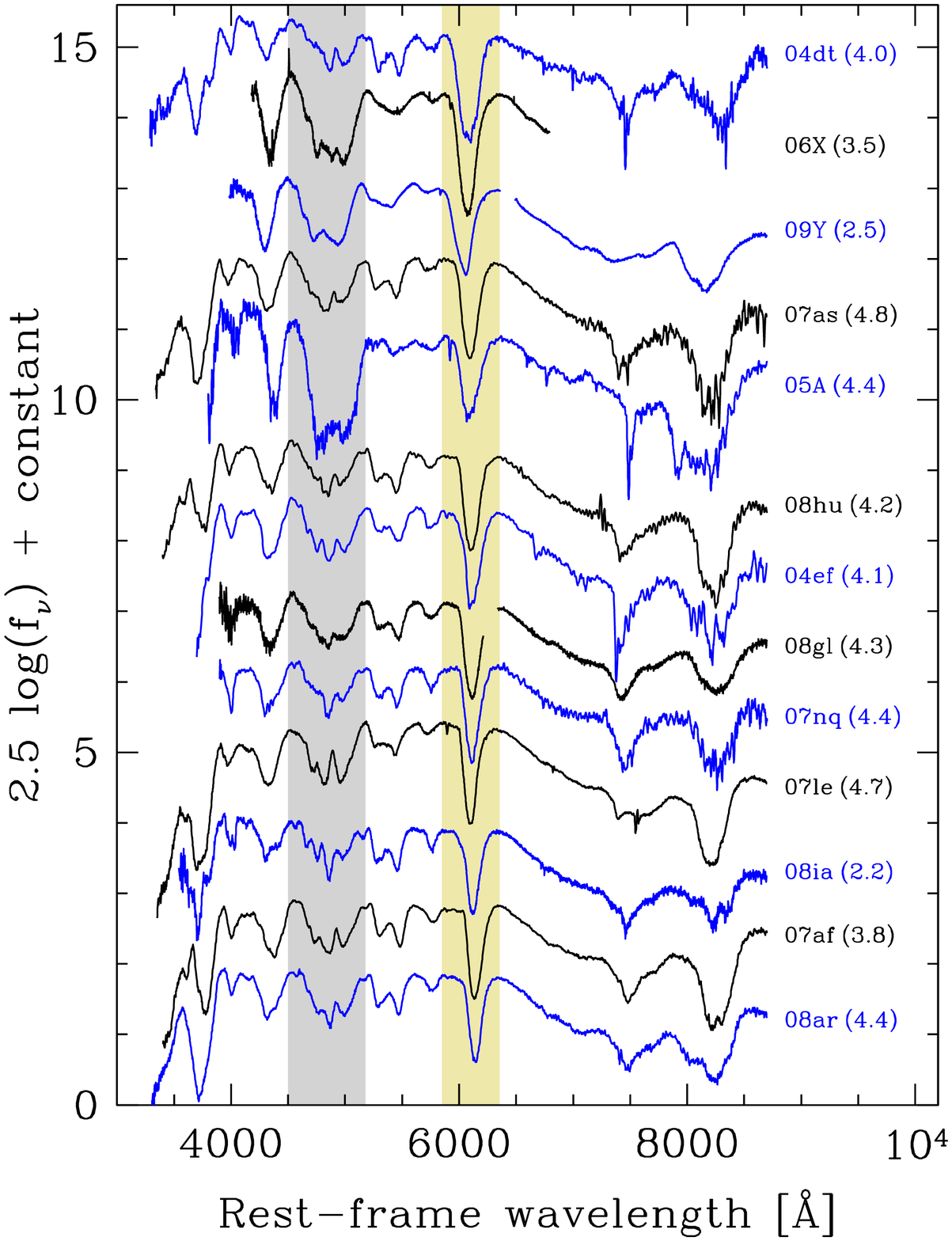}
\caption{({\em Left panel\,}) Spectra of CL and CN
  \sneia\ obtained near maximum light. The spectra are sorted in
  decreasing order of \pwthree. The labels on the right-hand side of
  each spectrum indicate the SN name and the epoch of the observation
  in rest-frame days relative to maximum light. The shaded areas 
  mark the regions of \pwthree\ (to the blue), and of \pwsix\ and \pwseven\
  (near the center). The subgroup of extreme CL (eCL) \sneia\ (top six
  spectra) are distinguished by a strong trough around 4300 \AA, 
  which produces large values of \pwthree\,$>220$ \AA. 
  ({\em Right panel\,}) Spectra of BL \sneia\ and one CN object
  (SN~2008ar) obtained a few days after maximum light. The
  spectra are sorted in decreasing order of \pwseven. The shaded areas
  mark the regions of features 4 (to the blue) and 7 (near the
  center).\label{fig:spclbl}} 
\end{figure}

The left panel of Figure~\ref{fig:spclbl} shows near-maximum-light
spectra of objects in the CL group sorted by decreasing \pwthree. At the
top of the figure some spectra show the strongest absorptions around
4300 \AA\ mostly due to the presence of strong \ion{Ti}{2} lines.
We call these ``extreme Cool'' (eCL) \sneia. They are similar to
the prototypically faint SN~1991bg. It is the eCL spectra that cause
the large dispersion in the composite spectrum of the CL subtype as
shown in Figure~\ref{fig:spave}. We define eCL SNe by having
\pwthree\,$>220$ \AA, which in our sample includes SNe~2005bl, 2005ke,
2006bd, 2006mr, 2007ax, and 2009F. The absorption around 8300
\AA, due to the \ion{Ca}{2} IR triplet, and the one near 7500 \AA, due to the
\ion{O}{1} $\lambda\lambda$7772,\,7775 doublet, are also stronger in CL
objects than in CN ones. These \ion{Ca}{2} IR and \ion{O}{1}
absorptions roughly follow the decrease in strength between eCL and CL
groups as observed for the $\approx$4300 \AA\ absorption. The weighted
average \pweight\ decreases from $325\pm39$ \AA\ for the eCL SNe 
to $238\pm26$ \AA\ for the rest of the CL objects. Interestingly, this
behavior is not seen in the strength of the \ion{Si}{2}\,$\lambda$5972
line; the weighted average \ew\ of this line is $48.6\pm2.3$
\AA\ for eCL SNe and $45.9\pm2.4$ \AA\ for the rest of the
subclass. We note a difference between the eCL objects and the rest
of the CL class in the profile of the absorption near 5800 \AA\ associated
in general with \ion{Si}{2}\,$\lambda$5972. For eCL objects this
absorption has a double structure with a weak component on the blue
side of the main absorption. This component tends to disappear as we
move to CL SNe with smaller \pwthree. The weaker absorption
can be attributed to the \ion{Na}{1}\,D doublet that grows
stronger at lower temperature.

The right panel of Figure~\ref{fig:spclbl} shows spectra of BL
\sneia\ obtained a few days after maximum light, sorted by decreasing
\pwseven. Some BL SNe show large absorption troughs centered at  
$\approx$4800 \AA\ that are quantified by \pwfour, and a blended 
absorption due to \ion{S}{2} at around 5300 \AA\ instead of the usual
W-shaped feature. Based on these characteristics, SNe~2005A, 2006X and
2009Y seem to form a separate group, even though SN~2009Y shows a
shallower \ion{Si}{2} velocity evolution as compared with SN~2006X
[\deltav\,$=1710\pm260$ km s$^{-1}$ for SN~2009Y versus
\deltav\,$=3660\pm110$ km s$^{-1}$ for SN~2006X]. We note that none of
these three SNe are included in the composite BL spectrum shown in 
Figure~\ref{fig:spave}. SN~2005A appears
to be an extreme case, with very large absorption troughs. It
  shows by far the largest \pwfour\ in our sample ($306\pm11$ 
\AA\ versus $241\pm1$ \AA\ for the next highest, SN~2006X), and the
third largest \pweight\,$=401\pm19$ \AA.
Both SN~2005A and SN~2006X share photometric
properties that, as presented by \citet{folatelli10}, indicate the
presence of a peculiar reddening law.
At lower \pwseven\ values, BL SNe 
continuously resemble CN objects, represented by the spectrum of
SN~2008ar in the bottom of the figure. BL SNe with the largest
\ion{Si}{2} $\lambda$6355 absorption strengths tend to show low
\pwsix\ (see Figure~\ref{fig:ew6ew7}). 

In terms of the classification introduced by \citet{benetti05}
\citep[see also][]{hachinger06}, we see
that their LVG group roughly includes the CN and SS subtypes,
while their HVG and FAINT groups approximately correspond to the BL
and CL types, respectively. \citet{wang09} set the division between HV
and normal 
\sneia\ based on the velocity of \ion{Si}{2} $\lambda$6355 between $-7$ and
$+7$ days relative to maximum light and comparing with the average of
ten well observed ``normal'' \sneia. They considered any SN with
silicon velocity above 
3~$\sigma$ from the average in that epoch range to be HV. At the time of
maximum light, we obtain an average of $\langle$\vsix$\rangle=10800
\pm 400$ km s$^{-1}$, which closely matches the value of $10700 \pm
400$ km s$^{-1}$ obtained by \citet{wang09}. Wang et~al.~defined
SNe with \vsix\,$\gtrsim 11800$ km s$^{-1}$
to be of type HV. With a larger sample, \citet{blondin12} place the
division at $\approx$12200 km s$^{-1}$. Our average would set the
limit at $\approx$12000 km s$^{-1}$. In the following we will consider
HV SNe as those that have \vsix\,$>12000$ km s$^{-1}$ at maximum
light. Table~\ref{tab:sne} provides, when
available, the classification of each SN according to the scheme of
\citet{wang09}. Nineteen of the SNe in our sample belong to the HV
subclass. This number would change to 21 and 16 if
we adopted the limits of \citet{wang09} and \citet{blondin12},
respectively. Most of the HV SNe in our sample (16 out
of 19) belong to the BL subtype. Two of the other three (SNe~2005ku
and 2006is) fall in the CN subtype, and one belongs to the CL group
(SN~2007hj). Several objects appear near the velocity boundary and
have measurement uncertainties that make the classification
marginal. The objects that are within 1~$\sigma$ of the division are
SNe~2006eq, 2006os, and 2008gl in the Normal group, and SNe~2004ef,
2005ku, 2006ef, 2006ej, 2007nq, and 2008hu in the HV group.
Because the distribution of velocities between
normal and HV groups is continuous, small differences in the
definition of the limit and small measurement uncertainties can lead to
differences in the classification of some objects.

The fraction of 24\% HV SNe in our sample is 
similar to that of 21\% found by \citet{blondin12}. In order to
compare with the fraction of HV SNe in the sample of \citet{wang09},
we consider only ``normal'' \sneia, i.e. we exclude from our sample 11
1991bg-like objects in the CL group, and six 1991T-like objects
in the SS group (see Table~\ref{tab:sne}). We thus obtain a fraction
of 31\% HV SNe among 63 ``normal'' \sneia, which
is similar to that of 35\% (55 out of 158 SNe) found by
\citet{wang09}. If we adopt the limit of 11800 km s$^{-1}$ to define
the HV class, our fraction is 33\% and the agreement is
improved. The sample of \citet{wang09} has 33 SNe in common with
ours. Among these, we have velocity measurements within one week from
maximum light for 30 objects. Our classification agrees
for all but two SNe, namely SNe~2005am and 2006os. The only
significantly discrepant classification is that of
SN~2005am, for which we measure \vsix\,$=12160\pm70$ km s$^{-1}$.

Comparing 28 SNe in common with the sample of
\citet{silverman12b}, our classifications in the scheme of Wang
  et~al.~agree for 23 objects. Two of the discrepant
objects are so in terms of their silicon velocities at maximum:
SNe~2005am and 2007bd, which are classified as HV in the present work
and as Normal by Silverman et~al. Other three objects, namely
SNe~2005M, 2005hj, and 2007S, are classified as
Normal by Silverman et~al., while we place them in the 1991T-like
category. For the latter two objects the difference may 
arise from the availability, in our case, of earlier spectra. The
spectroscopic distinction between 1991T-like and normal \sneia\ tends
to disappear after maximum light.

In Table~\ref{tab:sne} we also provide the classification that is
derived from the SuperNova IDentification code \citep[{\sc
    snid};][]{blondin07}. We used the spectra closest to maximum light
in order to fit the best-match 
subtype. We employed version 5.0 of {\sc snid} that does not distinguish
between ``1991T-like'' and ``1999aa-like''. 

Based on the classifications above, in the following subsections we
further study the spectral properties of the different subtypes of
\sneia\ using the measurements described in
Section~\ref{sec:meas}. 

\subsection{Velocity and pseudo-equivalent width evolution}
\label{sec:ewvelt}
 
\begin{figure}[htpb]
\epsscale{1.0}
\plotone{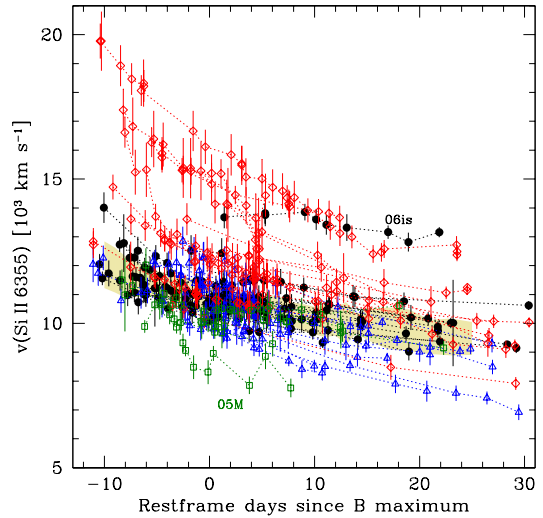}
\caption{Expansion velocity as measured from the \ion{Si}{2} $\lambda$6355
  absorption as a function of epoch with respect to $B$-band maximum
  light. The shaded band shows the average and 1-$\sigma$ 
  dispersion of normal \sneia\ (see text). Data of individual SNe are
  connected by dotted lines. Colors and symbols indicate different subtypes (see
  Section~\ref{sec:spdiv}): CN ({\em black circles}), BL ({\em red triangles}), SS ({\em
    green diamonds}), and CL ({\em blue squares}).\label{fig:vel6}}
\end{figure}

Figure~\ref{fig:vel6} shows the \ion{Si}{2} $\lambda$6355 line
expansion velocity as a function of phase relative to $B$-band maximum
light. Similar plots can be found in Figure~1 of \citet{foley11b},
Figure~15 of \citet{blondin12}, and Figure~5 of
\citet{silverman12b}. We have shaded the 1-$\sigma$ region about the
average of Normal \sneia\ in the classification system of Wang et~al.~and
over-plotted the data of individual objects of different Branch subtypes. For
Normal \sneia, an almost constant decrease is seen between 
$-10$ and $+30$ days, going from about 12000 to 10000 km
s$^{-1}$. While for the non-HV CN SNe the dispersion at any given
epoch is about 1000 km s$^{-1}$, for the complete sample this increases to 
over 5000 km s$^{-1}$. BL SNe show the largest velocities, and span a
range of 5000 to 7000 km s$^{-1}$ depending on the epoch. They also
show the largest velocity decline rates. CL SNe show similar
velocities and dispersions than CN objects, and somewhat larger
velocity gradients after maximum light. The evolution of SS SN 
velocities is rather flat. Among these, SN~2005M shows the lowest
velocity in the whole sample, with \vsix\ $\approx 8000$ km s$^{-1}$
at maximum light, which is about 1500 km s$^{-1}$ below the minimum
value of the rest of the sample. The CN SN~2006is stands out by showing
substantially higher \ion{Si}{2} velocity at maximum light than the
rest of the subtype, and shallower evolution than that of BL SNe with
similar \vsix. SNe~2005M and 2006is will be analyzed separately in
Section~\ref{sec:05M06is}.

\begin{figure}[htpb]
\epsscale{1.0}
\plotone{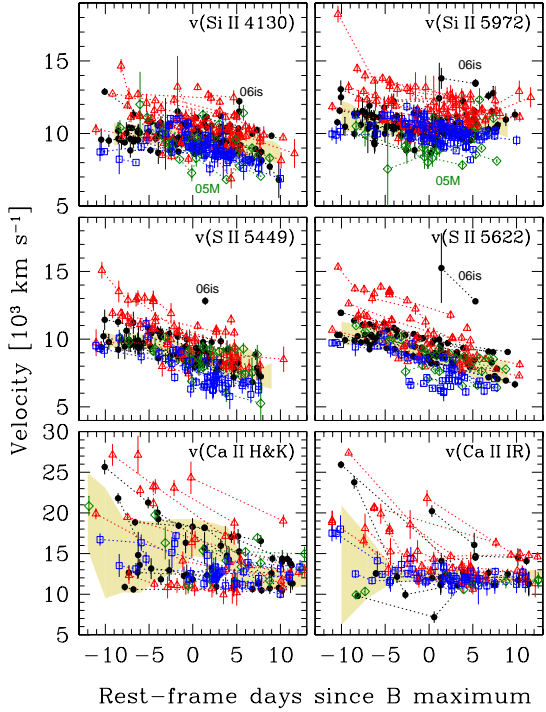}
\caption{Expansion velocity evolution as measured from lines of
  \ion{Si}{2} ({\em top panels}), \ion{S}{2} ({\em middle panels}),
  and \ion{Ca}{2} ({\em bottom panels}). The shaded bands show the
  average and 1-$\sigma$ dispersion of normal \sneia. Data of individual
  SNe are connected by dotted lines. The symbols are the same as in
  Figure~\ref{fig:vel6}.\label{fig:vel251347}} 
\end{figure}

Figure~\ref{fig:vel251347} shows the time evolution of the expansion
velocities for the other two \ion{Si}{2} lines, plus the \ion{S}{2}
and \ion{Ca}{2} lines near maximum light \citep[cf. Figure~4
  of][]{silverman12b}. For the weak \ion{Si}{2} and 
\ion{S}{2} lines, the range of epochs shown is the interval when these
features are clearly distinguishable in the spectra. A continuously
declining behavior is observed for all \sneia, with the exception of \vfive\ after maximum
light when the absorption may become contaminated by \ion{Na}{1}~D lines. The
smallest velocities correspond to the weakest of these lines, namely
\ion{Si}{2} $\lambda$4130, and the two \ion{S}{2} absorptions. The relative
behavior of different subtypes is similar to that of \vsix\ shown in
Figure~\ref{fig:vel6}.  \ion{Ca}{2} velocities show the largest
dispersions. Part of this may be due to the difficulty of measuring a
velocity based on the absorption minimum of these very broad lines that, as
pointed out by \citet{foley11b} and \citet{blondin12}, commonly
present composite profiles. As will be shown in Section~\ref{sec:qsp},
the dispersion of \ion{Ca}{2} velocities is not correlated with
variations in the velocity of other species.

\begin{figure}[htpb]
\epsscale{1.0}
\plotone{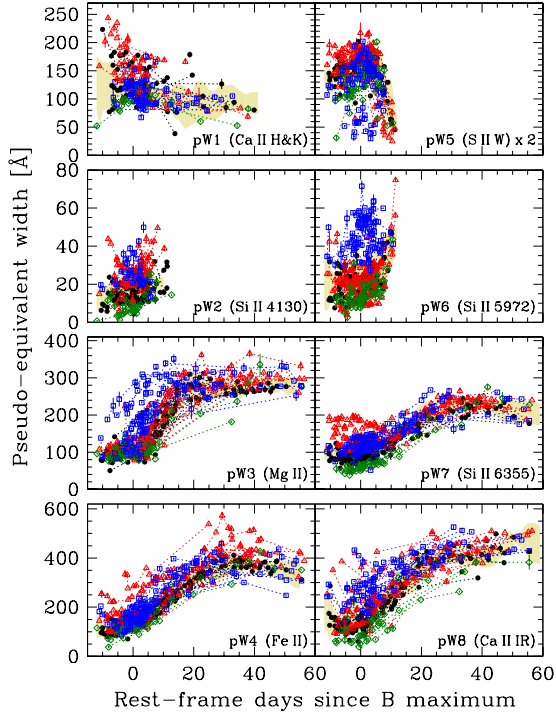}
\caption{Pseudo-equivalent width evolution of the eight
  absorption features defined in this work (see Section~\ref{sec:ew}
  and Table~\ref{tab:featw}). 
 The shaded band shows the average and 1-$\sigma$
  dispersion of normal \sneia. Data of individual SNe are connected by
  dotted lines. The symbols are the same as in
  Figure~\ref{fig:vel6}. \label{fig:ewall}}
\end{figure}

Figure~\ref{fig:ewall} shows the evolution of the eight
\ew\ parameters in the range of epochs where they can be
measured \citep[cf. Figures~7 and 8 of][]{silverman12b}. This
extends to about $+10$ days for the weakest features, 
namely \pwtwo, \pwfive, and \pwsix. The rest of the features are defined
until about two months after maximum light, although their association
with specific species is no longer valid---with the possible exception
of the \ion{Ca}{2} absorptions. The evolution of \pwthree, \pwfour,
\pwseven\ and \pweight\ is roughly similar, i.e., a phase of 
nearly constant or slightly decreasing \ew\ before maximum light,
followed by an increase, and a subsequent flattening. The
increase is the fastest for \pwthree, lasting for about 15
days, while it lasts for about 25 days for \pwfour,
\pwseven\ and \pweight. \pwone, as opposed to all the other features,
decreases almost continuously with time. For \pwfive\ there is a steep
decrease that occurs after about day $+5$. 

CL SNe are clearly distinguished by their higher than normal
\pwsix\ values. They also show large \pwtwo, \pwthree\ and \pweight,
and small \pwfive\ values. For eCL SNe the
increase of \pwthree\ occurs earlier than for normal \sneia, even as
early as $-10$ days. The subgroup of SS SNe shows relatively low values
of \ew\ for all features. The rise of \pwthree\ for SS objects occurs
at later times as compared with the rest of the subgroups. 

\subsection{Quantified spectroscopic properties}
\label{sec:qsp}

\begin{figure}[htpb]
\epsscale{1.0}
\plotone{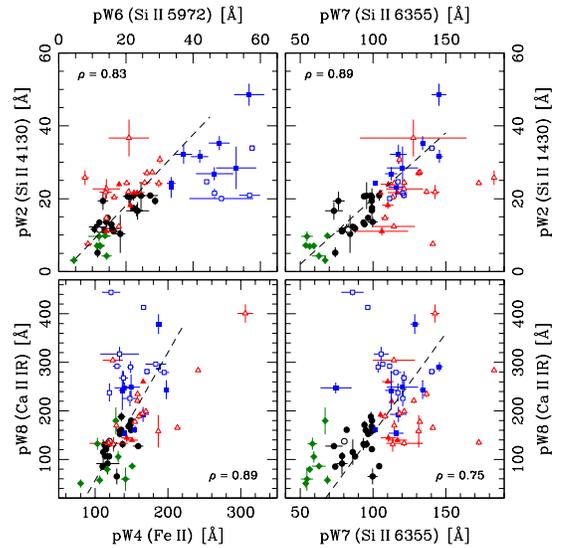}
\caption{Examples of strong correlations between
  \ew\ parameters. Different Branch subtypes are
  represented by black circles (CN), blue squares (CL), red triangles
  (BL), and green diamonds (SS). Filled symbols indicate SNe with low expansion
  velocities [\vsix\,$<12000$ km s$^{-1}$] and \dm\,$<1.7$ mag, while
  the rest of the objects are shown with open symbols. The dashed
  lines show straight-line fits to the filled data points. Pearson
  correlation coefficients $\rho$ are indicated in each
  panel.\label{fig:ewhicorr}}  
\end{figure}

\begin{figure}[htpb]
\epsscale{1.0}
\plottwo{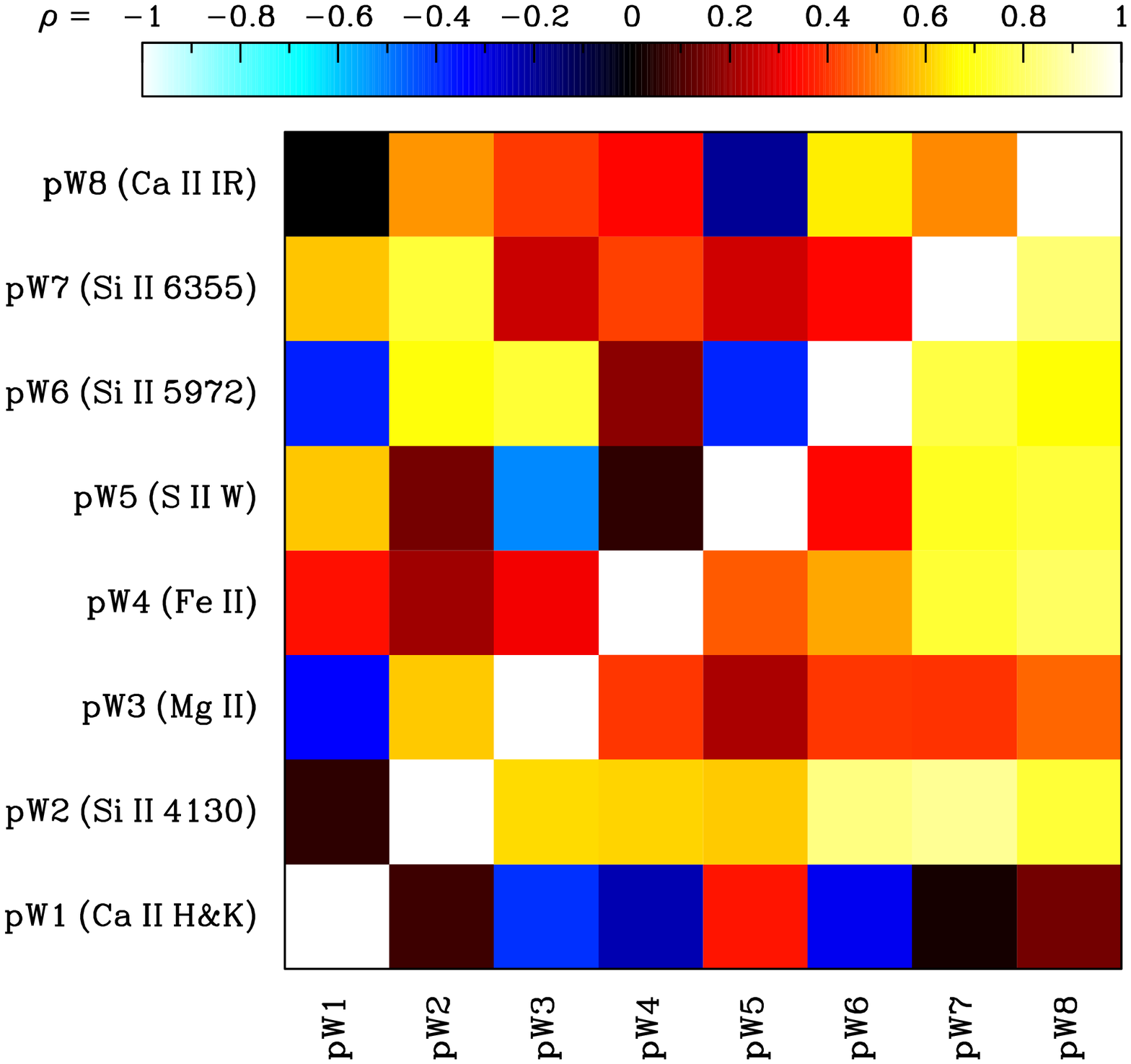}{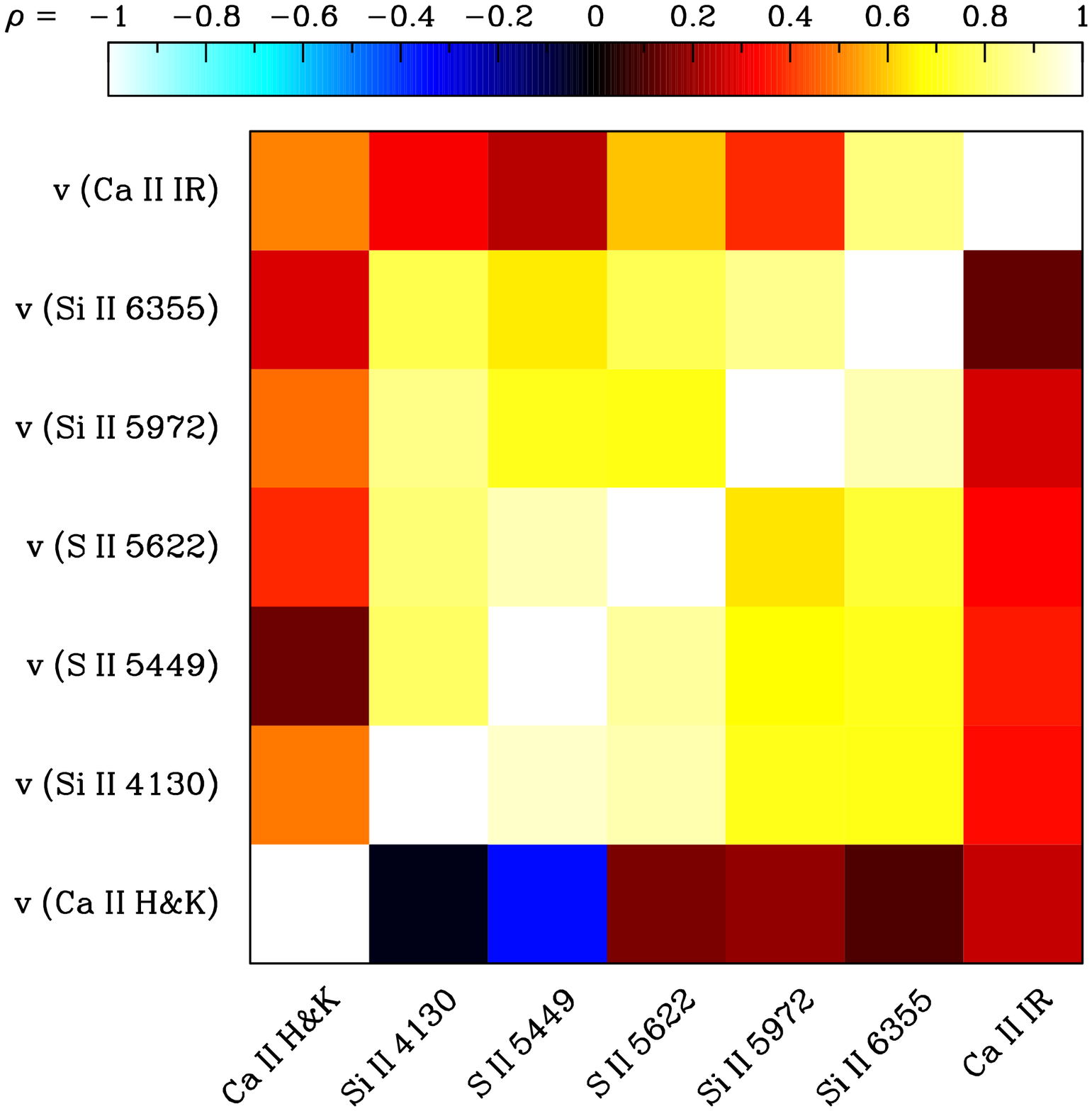}
\caption{Correlation between pairs of \ew\ parameters ({\em left}) and
  of expansion velocities ({\em right}) at maximum
  light. Colors indicate the Pearson correlation coefficient
  $\rho$. Lighter colors indicate larger degrees of correlation (or
  anti-correlation). In the 
  upper-left off-diagonal triangle all SNe are included. In the bottom-right
  off-diagonal triangle, only objects with low expansion velocities
  [\vsix\,$<12000$ km s$^{-1}$] and light-curve  
  decline rates [\dm\,$< 1.7$ mag] are considered.\label{fig:ewvelcorr}} 
\end{figure}

We now analyze the general spectroscopic properties of \sneia\ of
different subtypes in a quantitative manner, concentrating on the
interrelationships between velocity and
\ew\ measurements. Figure~\ref{fig:ewvelcorr} shows the Pearson correlation
coefficients, $\rho$, between all pairs of \ew\ (left panel) and velocity (right
panel) measurements at maximum light. In each panel, the upper-left
triangle corresponds to the complete sample of SNe with available
measurements. The lower-right triangles show the coefficients for
objects with low silicon velocities [\vsix\,$<12000$ km s$^{-1}$]
and light-curve decline rates [\dm\,$< 1.7$ mag]. When all \sneia\ are
considered, no strong correlations between 
\ew\ parameters are found ($|\rho|<0.75$). Some strong correlations
stand out if we exclude HV and fast-declining
SNe. Figure~\ref{fig:ewhicorr} shows the strongest of
such correlations involving \ion{Si}{2} features, and \pweight\ with
\pwfour\ and \pwseven. In
general the correlation coefficients between \ew\ parameters for this
restricted sample are positive. That is, the change in spectral line 
strengths from object to object tends to be in the same direction for
all features. The exception to this is \pwone, which shows low and
sometimes negative correlation coefficients with other \ew\ parameters.

The right panel of Figure~\ref{fig:ewvelcorr} shows that expansion
velocities in general correlate positively with each other. The exceptions again
involve \ion{Ca}{2} lines whose velocities do not correlate with those
of other features. This is likely due in part to difficulty of
measuring a velocity from the absorption minimum of \ion{Ca}{2} lines,
as pointed out before and also stressed by \citet{foley11b} and
\citet{blondin12}. \ion{Si}{2} and \ion{S}{2} line velocities
correlate strongly, with coefficients near $\rho=0.9$. 

\subsection{SNe 2005M and 2006is}
\label{sec:05M06is}

In Section~\ref{sec:ewvelt} we pointed out the existence of two objects
whose expansion velocities, as measured with the \ion{Si}{2}
$\lambda$6355 line, were atypical. These objects are SNe~2005M and
2006is. Their spectral time-series are shown in
Figure~\ref{fig:spec}. As can be seen in Figure~\ref{fig:vel6}, the
former SN shows a lower  \ion{Si}{2} $\lambda$6355
velocity than any other object in the sample,
while SN~2006is shows a large velocity, compared with the rest of the CN
sample, but shallower velocity evolution than typical HV SNe. The
behavior of SN~2006is is similar to the one observed for SN~2009ig
after maximum light \citep{foley12}. 

The velocity shifts in the spectra of SNe~2005M and 2006is of the other 
\ion{Si}{2} lines as well as \ion{S}{2} $\lambda\lambda$5449,5622
with respect to the measurements for the rest of the sample are noticeable, 
though less pronounced (see Figure~\ref{fig:vel251347}). However, the
shape of the spectrum outside 
the \ion{Si}{2} $\lambda$6355 line is similar to that of other SNe of
the corresponding subtype (SS for SN~2005M and CN for SN~2006is). We thus
studied the distribution of different species in the ejecta in order
to confirm the peculiar behavior of \ion{Si}{2} and \ion{S}{2}. 

\begin{figure}[htpb]
\epsscale{1.0}
\plotone{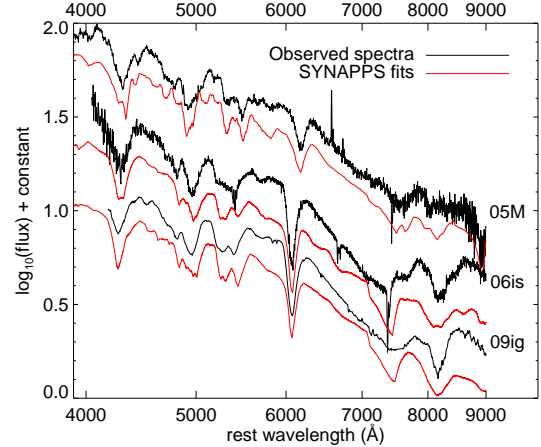}
\caption{Spectra of SNe~2005M, 2006is, and 2009ig near maximum light
  ({\em black lines\,}) and corresponding SYNAPPS fits ({\em red lines\,}). See
  details in Section~\ref{sec:05M06is}.\label{fig:05M06is}} 
\end{figure}

For this purpose, we fit the maximum-light spectra of SNe~2005M, 2006is
and 2009ig with the automated spectrum synthesis code SYNAPPS
\citep{thomas11a}, derived from SYNOW \citep{fisher00}. SYNAPPS
uses a highly parameterized, but fast spectrum synthesis technique,
useful for identifying the ions that form the observed features. The
best-fit synthetic spectra are compared with the observed spectra in
Figure~\ref{fig:05M06is}. The best-fit \ion{Si}{2} and \ion{S}{2} velocities for
SN~2005M are 8100 km s$^{-1}$ and 9400 km s$^{-1}$, respectively. These are
substantially lower than the best-fit median velocity of 10900 km s$^{-1}$
for the rest of the ion species, notably \ion{Mg}{2}, \ion{Ca}{2} and
\ion{Fe}{3}. 

The best-fit \ion{Si}{2} velocity for SN~2006is is 14400
km s$^{-1}$. \ion{O}{1} and \ion{S}{2} also show high velocities at 13300
km s$^{-1}$ and 12300 km s$^{-1}$, respectively. The remainder of the
ion species has a median velocity of 11300 km s$^{-1}$. SN~2009ig has
very similar 
spectral features as SN~2006is, and the best-fit velocities are also
very similar. The best-fit \ion{Si}{2} and \ion{S}{2} velocities are
14800 km s$^{-1}$ and 12300 km s$^{-1}$, respectively, while the
remainder of the ions has a median velocity of 11500 km
s$^{-1}$. Note that for SN~2006is and SN~2009ig, constraining the
velocities of all the ions to have the same value produces poor fits,
especially for the \ion{Si}{2} $\lambda$6355 feature. Although
for SN~2006is we adopted the redshift measured from our own deep
spectrum of the host galaxy (see Section~\ref{sec:sne}), the
behavior of line velocities cannot be attributed to an error in the
redshift as it would equally affect all lines. Moreover, the close
similarity with SN~2009ig provides further confidence that the
measured redshift is accurate.

The light curves of SNe~2006is and 2009ig were both more slowly
declining than average, with \dm\,$=0.80$ mag \citep{stritzinger11}
and $0.89$ mag \citep{foley12}, respectively. Interestingly, SN~2005M was 
also a slow decliner with \dm\,$=0.82$ mag \citep{contreras10}. 
We performed SNooPy \citep{burns11} fits to
the $uBgVriYJH$\footnote{These are the bandpasses in the {\em natural}
photometric system of the CSP \citep[see][]{contreras10}.} light
curves of SN~2006is and found that its 
$V-$NIR colors at maximum light are bluer by about $0.2$--$0.3$ mag
than those of unreddened \sneia\ of similar decline rate. On the other
hand, optical colors of 
this SN are normal. The near-infrared light curves of SN~2006is show the 
double-peaked shape that is typical of \sneia, but their brightness
relative to the optical bands appears to be low as compared with template
\sndia\ of similar \dm. SNooPy fits to the $uBgVriYJH$ light curves
of SN~2005M do not reveal any obvious photometric peculiarities.

The existence of rare events like SNe~2005M and 2006is whose
departure from the norm is mostly indicated by the peculiar
velocities of certain intermediate-mass elements (IME) 
seems to indicate a new form of spectroscopic diversity, related
with peculiar distribution of elements --- notably
\ion{Si}{2} and \ion{S}{2} which are the products of explosive oxygen
burning --- in the ejecta. An asymmetric explosion may have to be
invoked in the case of SN~2005M to explain 
the presence of IME at lower velocity than that of \ion{Fe}{2}. The
cases of SNe~2006is and 2009ig may indicate the existence of an outer
silicon-rich shell. Since these objects depart from the 
\ion{Si}{2} $\lambda$6355
velocity vs. velocity gradient relation at maximum light
\citep{foley11b}, they do not seem to comply with the same
geometrical picture of \citet{maeda10b} involving asymmetric
explosions.

\section{SPECTROSCOPIC AND PHOTOMETRIC PROPERTIES}
\label{sec:spphot}

Peak luminosities of \sneia,
after correcting for host-galaxy extinction, can be 
calibrated based on a single parameter which measures the initial
decline or ``width'' of the light curve
\citep{phillips93,phillips99}. Such calibration has been successfully
employed to determine extragalactic distances and thereby cosmological
parameters. From the viewpoint of spectroscopy, it has been observed
that the strength of \ion{Si}{2} features near maximum light follows
this luminosity-decline rate correlation
\citep{nugent95,hachinger06,silverman12c}. More 
recently, \citet{bailey09} and \citet{blondin11a} showed that some flux ratios can
reduce the scatter in the luminosity calibration. However,
other spectroscopic properties show a diversity which does not comply
with this one-parameter description. Expansion velocities, for
instance, show large variations which do not correlate with
light-curve decline rate \citep{hatano00,benetti05}.  

In this section, we present a quantitative comparison of spectroscopic
and photometric properties of \sneia. Based on light-curve fits
performed with the SNooPy code of \citet{burns11}, we obtained $K$- and
Galactic-extinction corrected peak magnitudes in $B$ and $V$ bands
($B_{\mathrm{max}}$ and $V_{\mathrm{max}}$, respectively), as well as
decline rates, \dm\footnote{When the $B$-band light curve did not
  span a suitable range of time to directly measure \dm\ from the data
  points, we adopted the $\Delta m_{15}$ fitting parameter of SNooPy
\citep{burns11}. According to Burns et~al., the actual \dm\ may be
slightly different from $\Delta m_{15}$ (see their Figure~6.)}. From
the peak magnitudes we computed $B-V$ {\em 
  Pseudo}-colors at maximum light, $(B_{\mathrm{max}}-V_{\mathrm{max}})$.
Adopting the calibrations of intrinsic color given in Equation~(3) of
\citet{folatelli10}, we derived host-galaxy color 
excesses, $E(B-V)_{\mathrm{Host}}$, as

{\small
\begin{equation}
\label{eq:ebvh}
E(B-V)_{\mathrm{Host}}=(B_{\mathrm{max}}-V_{\mathrm{max}})+0.016-0.12\left[\Delta m_{15}(B)-1.1\right].
\end{equation}
}

\noindent Note that the formula above is valid for SNe with
\dm\,$<1.7$ mag. That is why we provide $E(B-V)_{\mathrm{Host}}$ for
only a small fraction of the CL objects.

Based on the color excesses, reddening-free $B$-band absolute peak magnitudes,
$M_B^0$, were obtained as 

\begin{equation}
\label{eq:mb0}
M_B^0\;=\;B_{\mathrm{max}}\;-\;\mu\;-\;R_B\,\cdot\,E(B-V)_{\mathrm{Host}},
\end{equation}

\noindent where $\mu$ is the distance modulus, and $R_B=3.98$ is converted
from a standard value of $R_V=3.1$, as explained in Appendix~B of
\citet{folatelli10}. This conversion  
assumes the dust extinction law introduced by 
\citet{cardelli89} and modified by \citet{odonnell94} (hereafter, the
CCM+O law), and a template \sndia\ spectrum at maximum light. To
avoid errors in $M_B^0$ larger than $\approx$$0.5$ mag due to
uncertainties in 
the value of $R_B$ (which typical has been found in the literature
to have values in the range of 3--4),
we did not compute $M_B^0$ for SNe with $E(B-V)_{\mathrm{Host}}>0.5$ mag.
Values of $\mu$ were derived from the redshift of
the host galaxies using the second-order Hubble law given in
Equation~(5) of \citet{folatelli10}, with $H_0=72$ km s$^{-1}$
Mpc$^{-1}$, $\omm=0.28$, and $\oml=0.72$ \citep{spergel07}. To avoid
large uncertainties in $\mu$ due to peculiar velocities, this was done
only for SNe with $z_{\mathrm{CMB}}> 0.01$.

\begin{figure}[htpb]
\epsscale{1.0}
\plotone{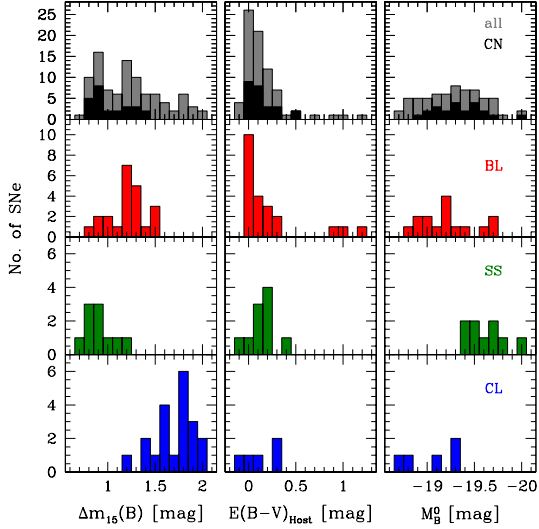}
\caption{Distribution of light curve decline rate \dm\ ({\em left
    column\,}), host-galaxy color excess $E(B-V)_{\mathrm{Host}}$
  ({\em middle column\,}), and reddening-free $B$-band absolute peak
  magnitudes $M_B^0$ for the complete sample ({\em gray histograms\,})
  and for different \sneia\ subtypes ({\em black, red, green and blue
    histograms\,}).\label{fig:dmebvhist}}  
\end{figure}

We now study the properties of the photometric parameters
involved in equations~(\ref{eq:ebvh}) and (\ref{eq:mb0}) for the 
different spectroscopic subtypes. Figure~\ref{fig:dmebvhist} shows
the distribution of \dm\ (left 
column), $E(B-V)_{\mathrm{Host}}$ (middle column), and $M_B^0$ (right
column) for the complete sample and the different subtypes of 
\sneia. Weighted averages of these quantities for each subtype
are listed in Table~\ref{tab:phave}. The \dm\ distribution of the
complete sample has an average of $\langle$\dm$\rangle=1.16$ mag. SS
and CL SNe clearly occupy the extremes of the \dm\ distribution, the
former being slow decliners
[$\langle$\dm$\rangle_{\mathrm{SS}}=0.89\pm 0.11$ mag] and the latter 
showing the fastest decline rates
[$\langle$\dm$\rangle_{\mathrm{CL}}=1.51 \pm 0.12$ mag]. CN and BL
objects show similar averages, with a distribution 
that is slightly more skewed toward larger \dm\ in the case of BL SNe. 
From the middle column of the figure it is seen that SS SNe tend
to have larger color excesses [$\langle
E(B-V)_{\mathrm{Host}}\rangle_{\mathrm{SS}}=0.17 \pm 0.05$ mag] than
CN objects [$\langle E(B-V)_{\mathrm{Host}}\rangle_{\mathrm{CN}}=0.13
  \pm 0.03$ mag] and the bulk of BL SNe [$\langle
  E(B-V)_{\mathrm{Host}}\rangle_{\mathrm{BL}}=0.15 \pm 0.06$ mag, but 
this value is reduced to $0.08 \pm 0.02$ mag when excluding three
extremely reddened BL events, SNe~2005A, 2006X, 2006br]. If confirmed
with larger samples, this result points in the same direction as the
evidence presented in Figure~\ref{fig:spave} of Section~\ref{sec:spdiv} 
for a closer association of SS SNe with gas-rich, dusty, star-forming
regions.

In terms of reddening-free $B$-band absolute peak magnitudes, again
the extremes of the distribution are occupied by SS
and CL SNe (considering the few CL objects for which $M_B^0$ can be
obtained). The same trend is seen as the one found by
\citet{blondin12} of increasing $M_B^0$ (decreasing luminosity) as we
move from SS to CN, BL, and finally to CL SNe. This reflects the sequence of
\dm\ distributions. The weighted averages go from 
$\langle M_B^0 \rangle_{\mathrm{SS}}=-19.56 \pm 0.14$ mag to
$\langle M_B^0 \rangle_{\mathrm{CN}}=-19.38 \pm 0.06$ mag, $\langle
M_B^0 \rangle_{\mathrm{BL}}=-19.25 \pm 0.10$ mag, and $\langle M_B^0
\rangle_{\mathrm{CL}}=-19.05 \pm 0.14$ mag (the latter includes only
five SNe).

\subsection{Pseudo-EW and light-curve decline rate}
\label{sec:ewdm}

\begin{figure}[htpb]
\epsscale{1.0}
\plottwo{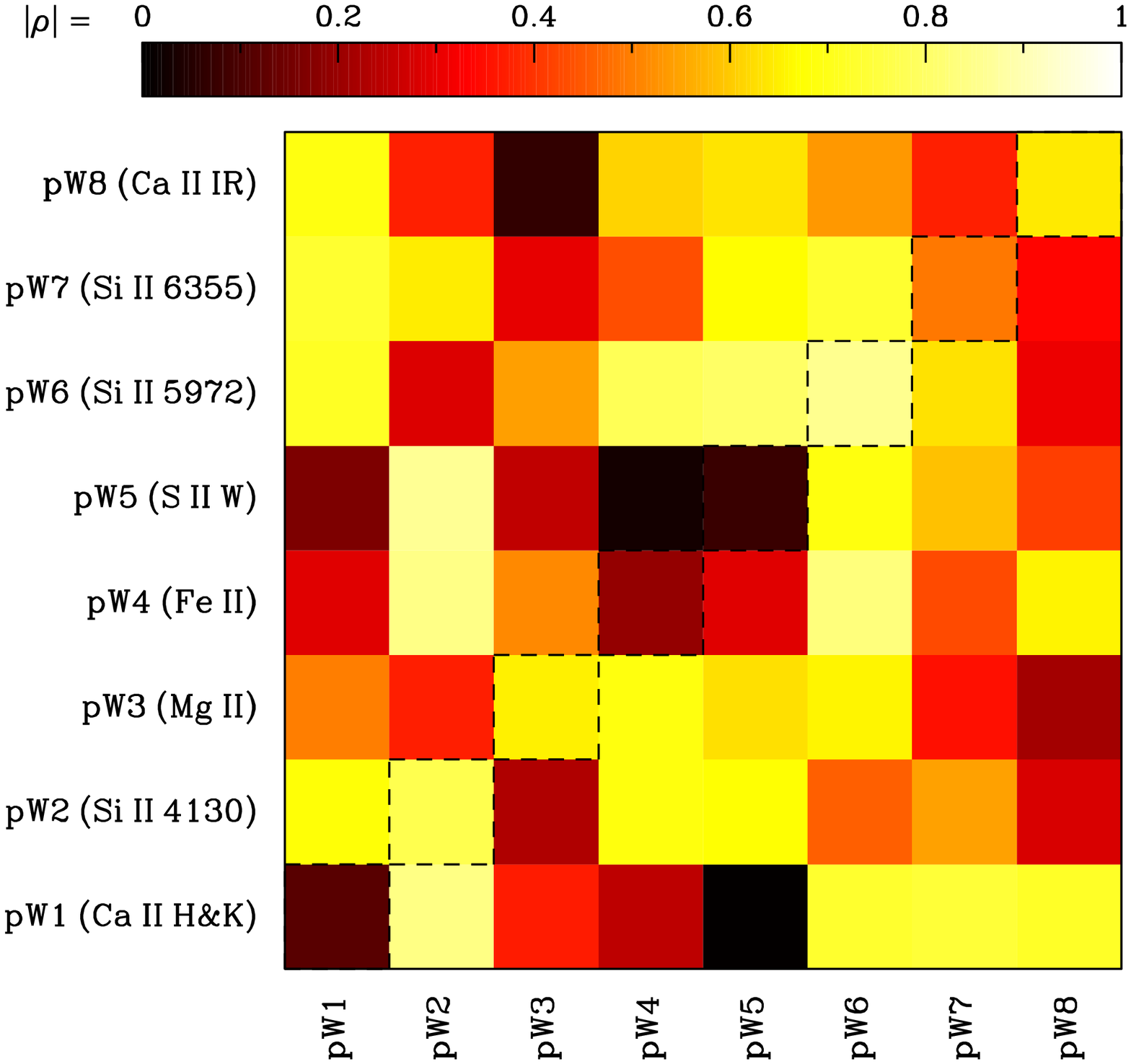}{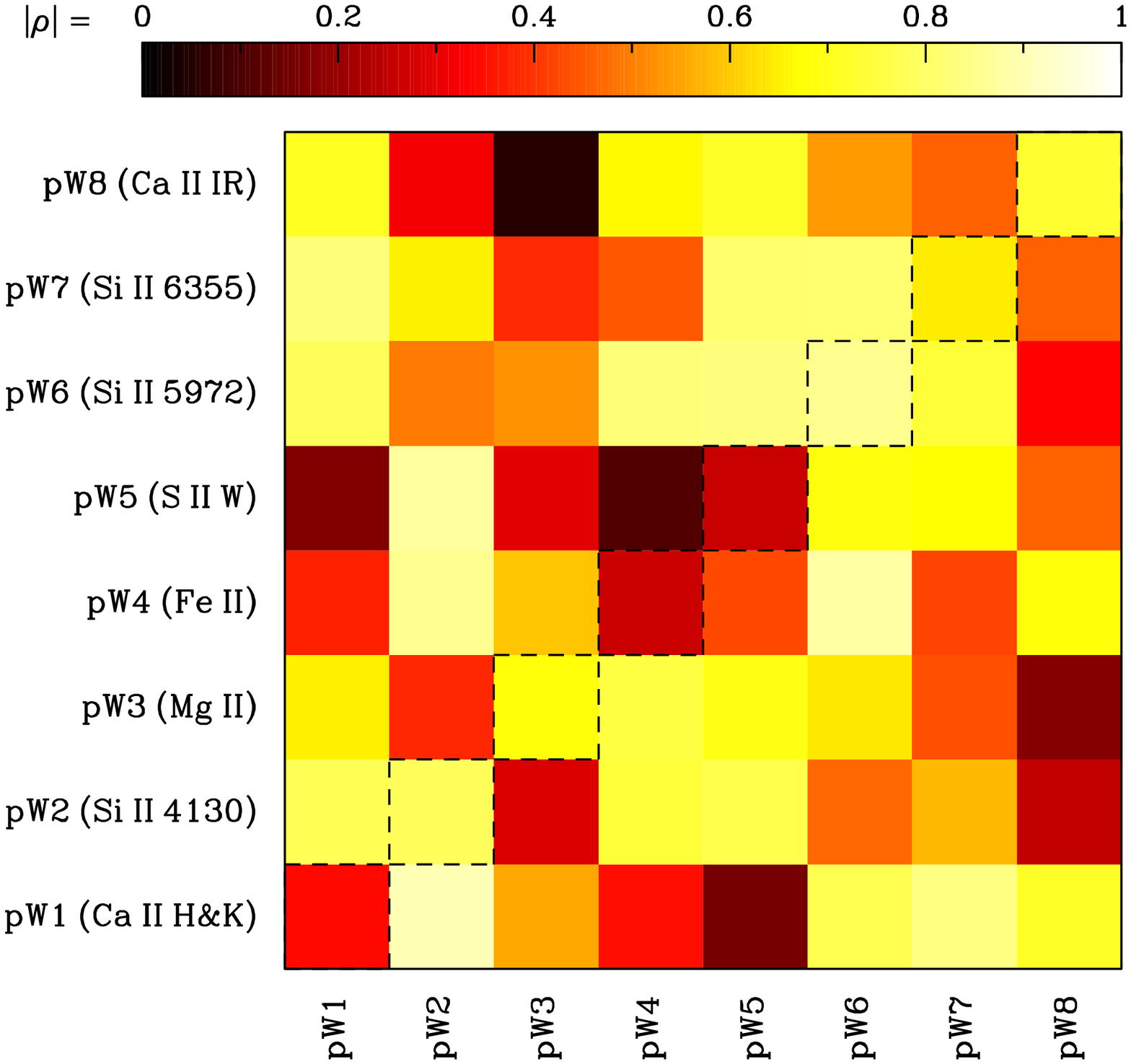}
\caption{Correlation between \ew\ parameters at maximum
  light and light-curve decline rate \dm. Boxes on the diagonal
  show the correlation between \ew\ and \dm. Off-diagonal boxes
  indicate the correlation of \ew\ ratios with \dm, with the ratios
  computed as $pWy/pWx$. Colors indicate the
  absolute Pearson correlation coefficient $\rho$. {\em Left
    panel:\,} all SNe are included. {\em Right panel:\,} only objects
  with low expansion velocities [\vsix\,$<12000$ km s$^{-1}$]
  are considered.\label{fig:ewdmcorr}}  
\end{figure}

\begin{figure}[htpb]
\epsscale{1.0}
\plottwo{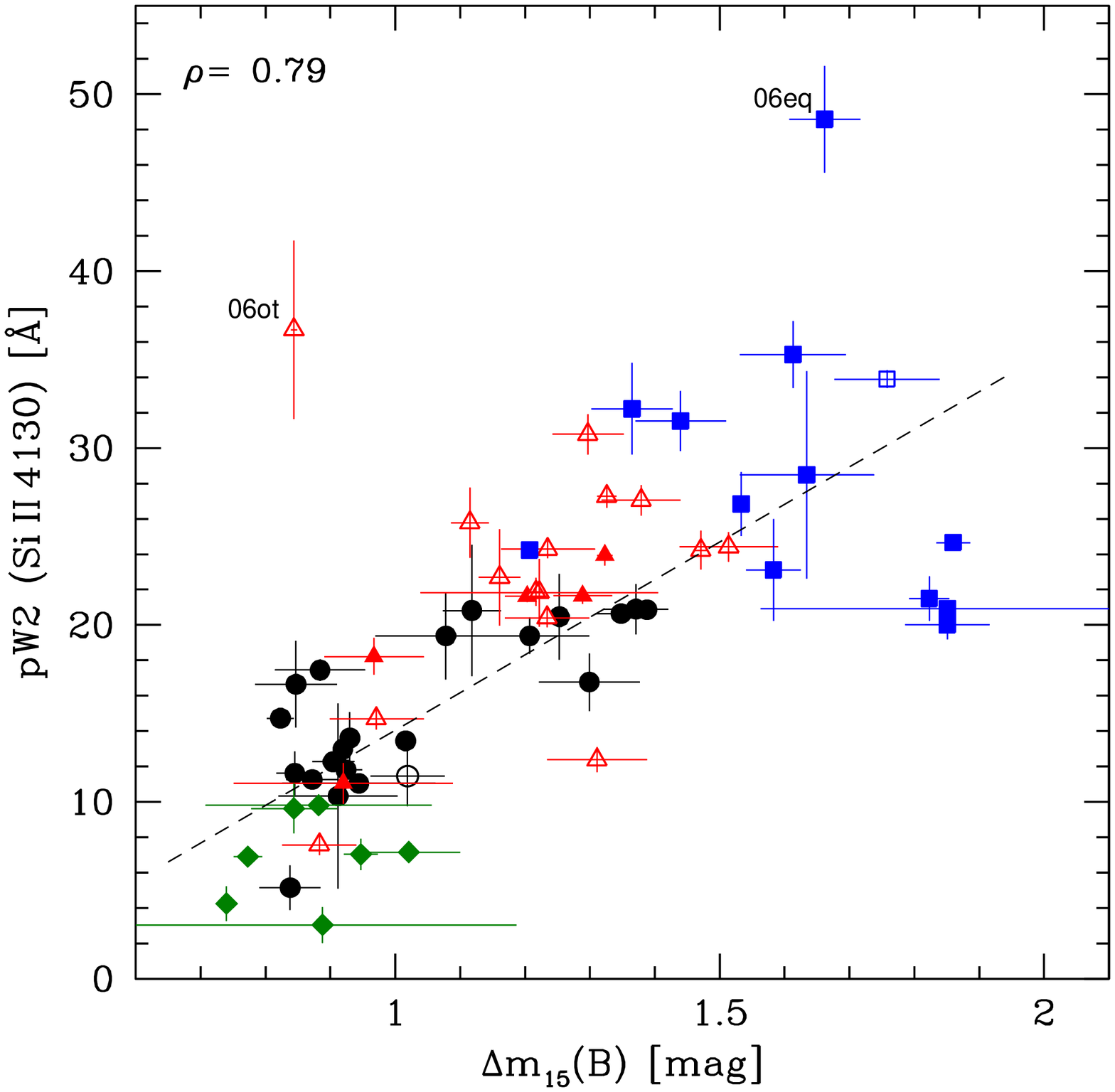}{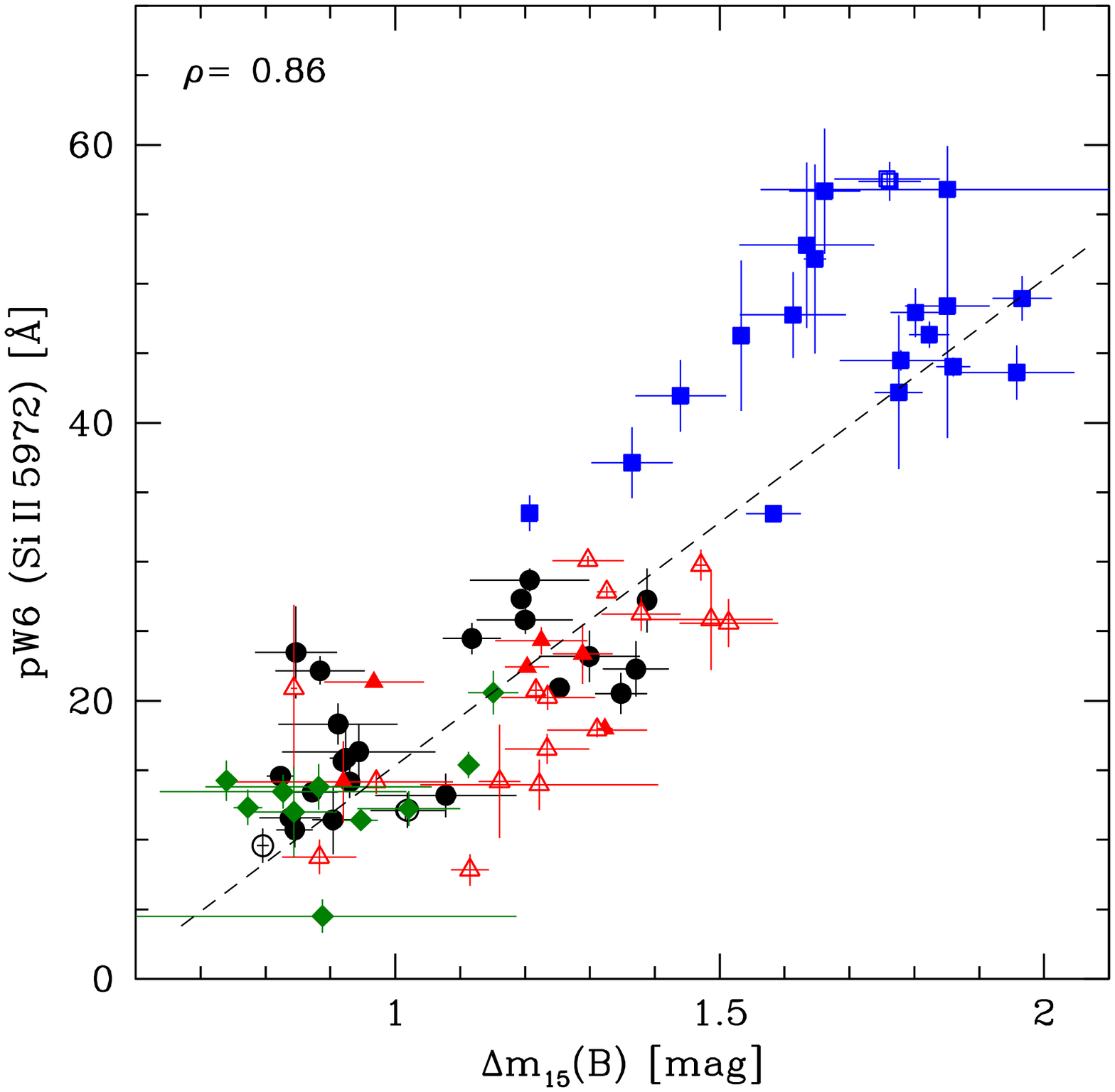}
\caption{Pseudo-equivalent at maximum light width versus decline rate
  for \pwtwo\ ({\em left\,}), and \pwsix\ ({\em right\,}). These features
  show the largest correlation with \dm. Symbols are the same as in
  Figure~\ref{fig:ew6ew7}. The dashed lines indicate straight-line
  fits to the filled data points, i.e., SNe with low expansion
  velocities [\vsix\,$<12000$ km s$^{-1}$]. Pearson correlation
  coefficients $\rho$ are indicated in each panel.\label{fig:ew26dm}}
\end{figure}

Figure~\ref{fig:ewdmcorr} shows the degrees of correlation between
\ew\ parameters and light curve decline rates, \dm, considering both
the \ew\ measurements themselves and their ratios. The
strongest correlation with \dm\ among single \ew\ parameters is that of
\pwsix, with a Pearson coefficient of $\rho = 0.86$ (both for the
  complete sample and when HV SNe are excluded). This relation was previously noted by
\citet{hachinger06}, and more recently by
  \citet{silverman12c}. The resulting dispersion of \dm\ about the  
straight-line fit is of $0.14$ mag. \pwtwo\ also depends strongly on
\dm. Its correlation coefficient of $\rho = 0.77$ for the complete
sample increases to $\rho = 0.84$ if we restrict the
sample to SNe with \dm\,$<1.7$ mag. Both cases of \pwtwo\ and
\pwsix\ are shown in Figure~\ref{fig:ew26dm}. In general, the
correlations become stronger when we discard HV \sneia, as can be seen
in the right panel of Figure~\ref{fig:ewdmcorr}. 

\begin{figure}[htpb]
\epsscale{1.0}
\plottwo{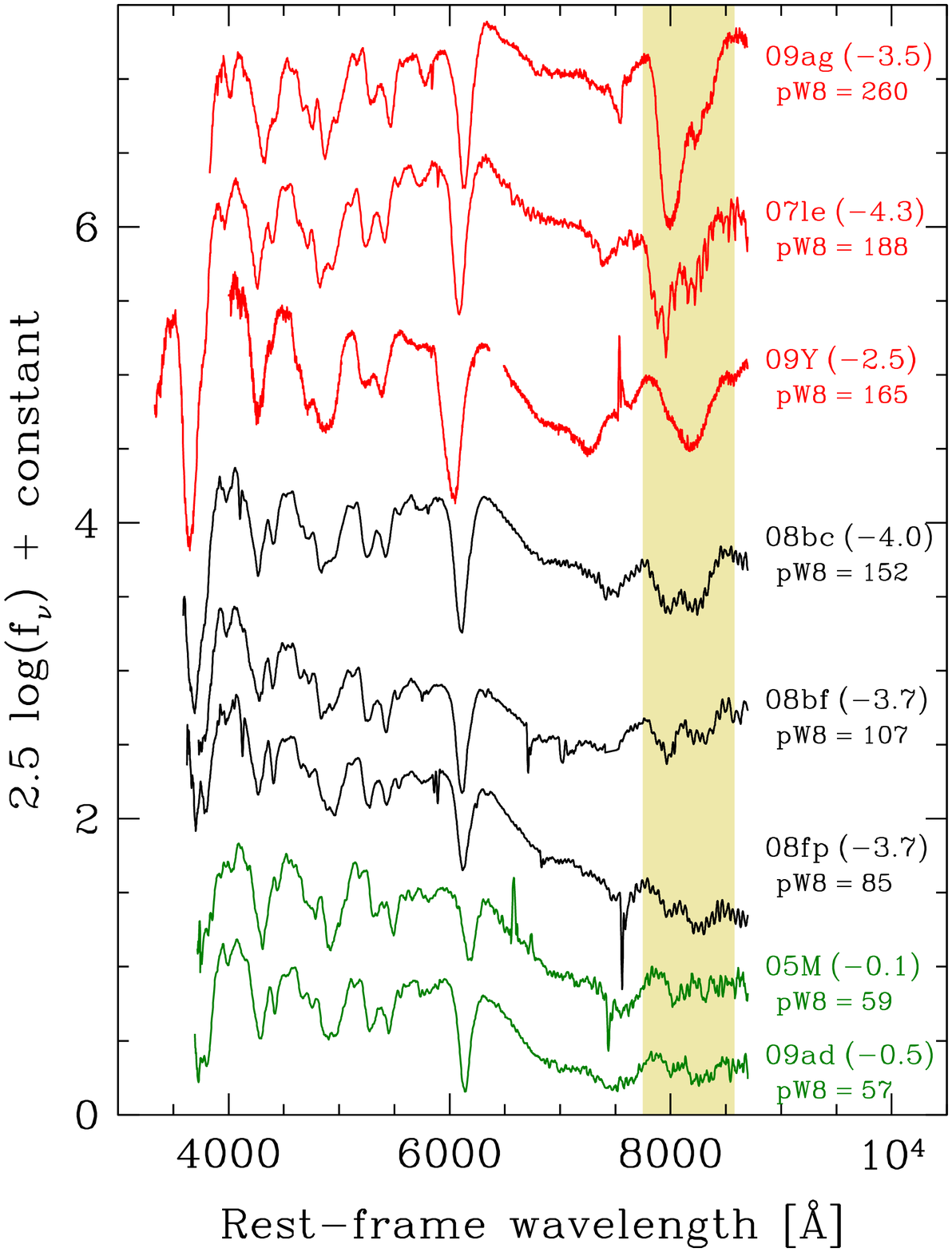}{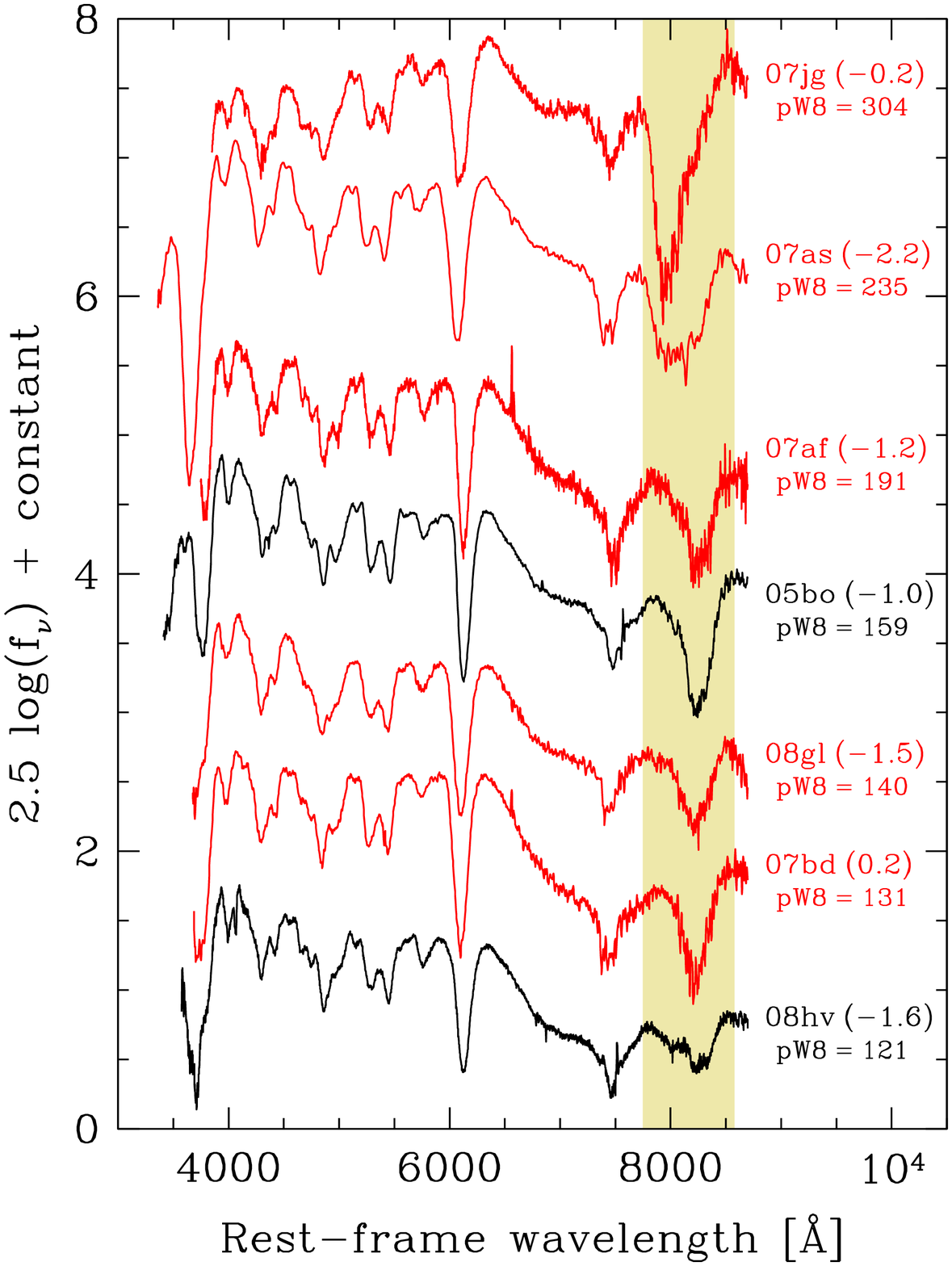}
\caption{Spectral sequences of decreasing \pweight\ for \dm\ $\approx 0.9$
  mag ({\em left}), and \dm\ $\approx 1.3$ mag ({\em right}). The
  Branch subtype of each SN is indicated by the color of the line
  following the same symbol scheme as symbol colors in
  Figure~\ref{fig:ew6ew7}. The labels on the right-hand side indicate
  the SN name, the epoch of the spectrum between parentheses, and the
  value of \pweight\ in \AA\ (derived at maximum light) for each SN. The
  spectra were obtained between four days before and the
  time of maximum light. The shaded region marks the range of
  expansion velocities up to about 30000 km s$^{-1}$. Large variations
  in the \ion{Ca}{2} IR triplet are seen in otherwise homogeneous
  spectra among normal \sneia.\label{fig:spew8}} 
\end{figure}

Interestingly, \pweight\ also shows an increasing trend with
\dm\ ($\rho=0.6$--$0.7$). However, for a fixed \dm\ there is a wide range of
\pweight\ values, with some SNe showing two or three times larger
\ew\ than their counterparts. On the contrary, the other \ion{Ca}{2}
feature, \pwone, shows no correlation with \dm, with $\rho$ in
  the range between $-0.1$ and $-0.3$. Figure~\ref{fig:spew8}
shows example sequences of pre-maximum spectra with different
\pweight\ values at fixed \dm. Among SNe with similar \dm\ the spectra
show quite homogeneous characteristics, except at the location of the
\ion{Ca}{2} IR triplet. The largest \pweight\ values are found when a
strong high-velocity component is present.

\begin{figure}[htpb]
\epsscale{1.0}
\plotone{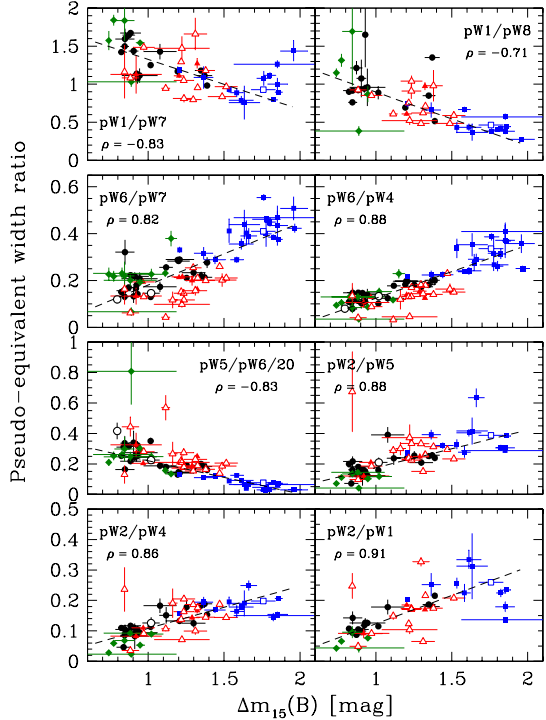}
\caption{\ew\ ratios that correlate with \dm. Symbols are the same as in
  Figure~\ref{fig:ew6ew7}. The dashed lines indicate straight-line
  fits to the data of low-velocity SNe [\vsix\,$<12000$ km s$^{-1}$;
    filled symbols]. Pearson correlation coefficients $\rho$ for those
  data points are indicated in each panel.\label{fig:ewratdm}}
\end{figure}

The strongest correlations between \ew\ ratios and \dm\ are shown in
Figure~\ref{fig:ewratdm} \citep[cf. Figures~13 to 19
  of][]{hachinger06}. Most of these involve \ion{Si}{2} features and
ratios with \pwfour\ and \pwfive\ \citep[see
also][]{silverman12c}. One of the ratios that correlates with
  \dm\ is [\pwsix\,$/$\,\pwseven], which is similar to the line intensity
  ratio introduced by \citet{nugent95}, \rsi. This
  relationship is governed by the stronger 
dependence of \pwsix\ than of \pwseven\ on the decline
rate \citep{hachinger08}. The dispersion is somewhat larger than
that of \pwsix\ versus \dm, mostly due to BL SNe that have relatively
large \pwseven\ and fall below the rest of the sample. 

We note that, although with fewer
data points, \pwone\ is involved in some of these
correlations. Interestingly, as shown in the top right panel, 
the relative strength of \pweight\ with respect to \pwone increases with \dm. 
This is a consequence of the behavior noted above for \pweight\ and
\pwone\ versus \dm. A temperature effect related with
\dm\ may cause the anti-correlation of [\pwone\,$/$\,\pweight] with
\dm. The effect is presumably equivalent to the one described by
\citet{hachinger08} to explain the behavior of 
\rsi\ with \dm. The most abundant ionization state of calcium
in \sneia\ near maximum light is \ion{Ca}{3} \citep{tanaka08}. As
temperature decreases with increasing \dm, the abundance of
\ion{Ca}{2} increases. If both calcium lines are not saturated, this effect
should equally increase their intensities. The observed behavior with
\dm, however, indicates that \ion{Ca}{2}\,H\&K may be saturated and so
it does not react to abundance changes, while the contrary happens with
the \ion{Ca}{2}\,IR triplet.

\subsection{Expansion velocity and light-curve decline rate}
\label{sec:veldm}

\begin{figure}[htpb]
\epsscale{1.0}
\plottwo{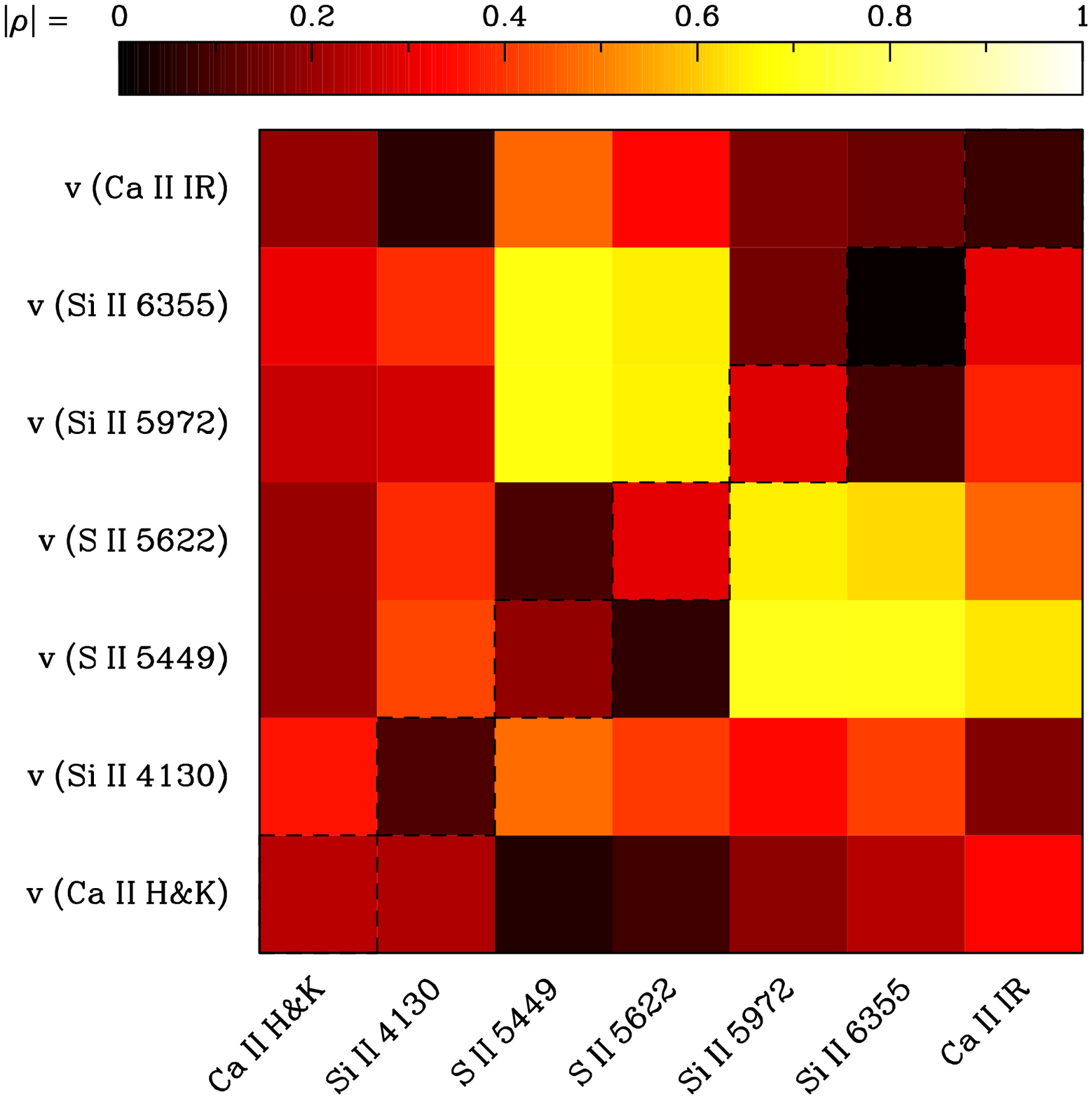}{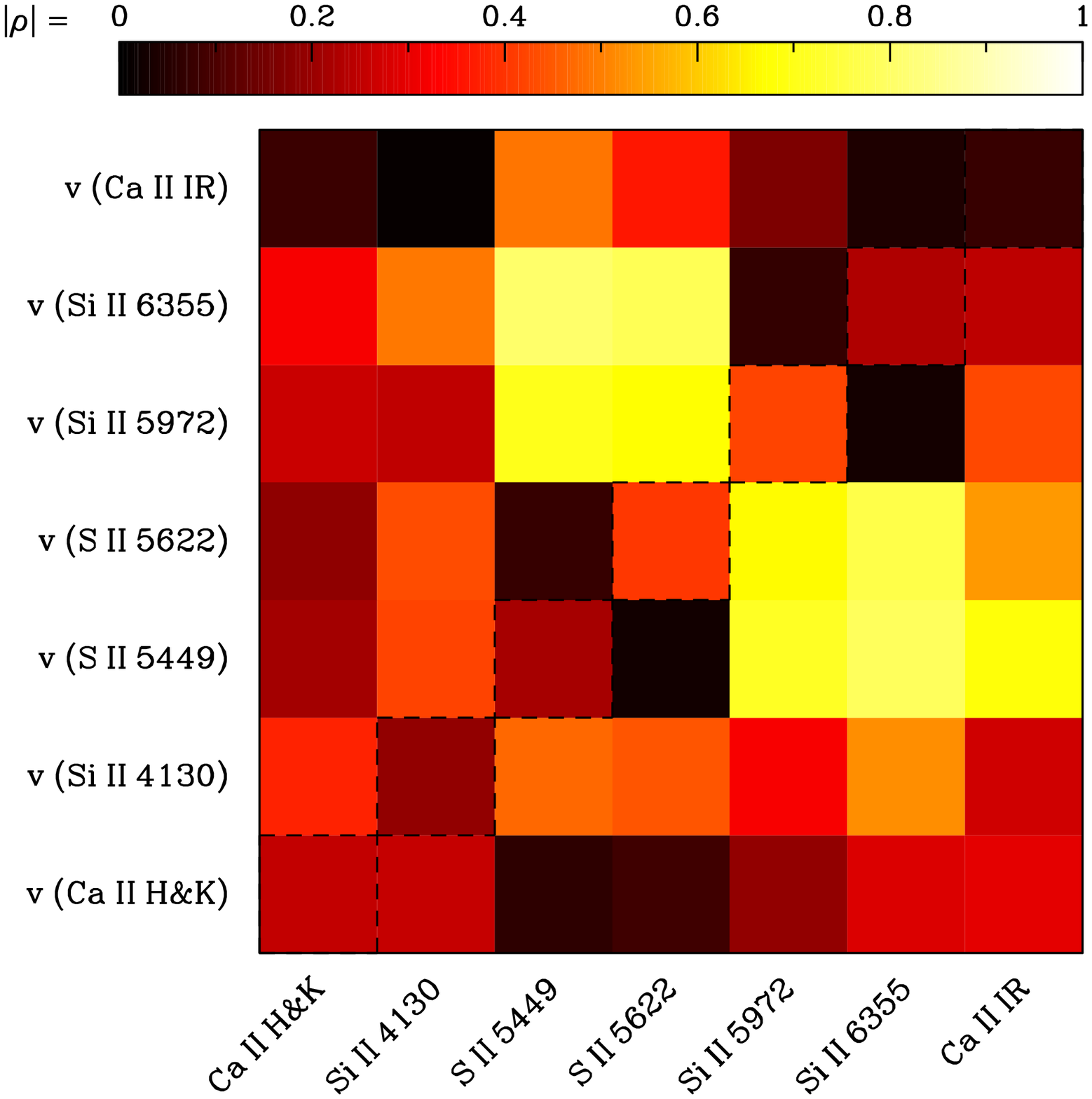}
\caption{Correlation between expansion velocities at maximum
  light (on the diagonal) and their ratios (off-diagonal), and
  light-curve decline rate \dm. Ratios for the off-diagonal boxes
  are computed as $vy/vx$. Colors indicate the absolute Pearson
  correlation coefficient $\rho$. {\em Left panel:\,} all SNe are
  included. {\em Right panel:\,} only objects with low expansion
  velocities [\vsix\,$<12000$ km s$^{-1}$]
  are considered.\label{fig:veldmcorr}}  
\end{figure}

As mentioned in Section~\ref{sec:spdiv}, \sneia\ show a wide
variety of line expansion velocities. Here we study the behavior of
expansion velocities of different lines at maximum light with respect
to light curve decline rate. Figure~\ref{fig:veldmcorr} shows the
degree of correlation between velocity measurements at maximum light
(on the diagonal) or ratios of these (off the diagonal), and \dm. We
can see that none of the line velocities presented here correlate
strongly with decline rate. The ratios of \vthree\,$/$\,\vsix,
\vfour\,$/$\,\vsix, and \vthree\,$/$\,\vtwo\ show the largest degrees of
anti-correlation with \dm\ ($\rho < -0.80$).

The lack of dependence of line velocities on \dm\ is accompanied by a
more or less clearly defined minimum expansion velocity for all
decline rates. This velocity floor is about 10000 km s$^{-1}$ for the
strongest absorptions (\ion{Ca}{2} H\&K and IR triplet, and
\ion{Si}{2} $\lambda$6355), and between 6000 and 8000 km s$^{-1}$ for the
weaker lines. In the case of \vfive, the minimum velocity appears to
be about 1000 km s$^{-1}$ higher for fast-declining \sneia\ [with
  \dm\,$\gtrsim 1.5$ mag] than for the rest. The opposite occurs with
\vthree\ and \vfour\, where SNe with \dm\,$>1.5$ mag present on
average lower velocities by about 1500 km
s$^{-1}$\citep[cf.][]{hachinger06,blondin06}. The SS SN~2005M 
stands out in our sample by having the lowest expansion velocities,
by up to 2000 km s$^{-1}$ for some features (see
Section~\ref{sec:05M06is}). 

\begin{figure}[htpb]
\epsscale{1.0}
\plotone{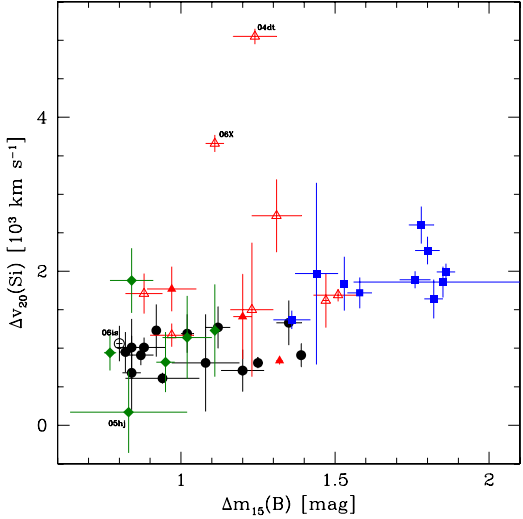}
\caption{\ion{Si}{2} $\lambda$6355 velocity decline rates as a
  function of \dm. Symbol coding is the same as in
  Figure~\ref{fig:ew6ew7}.\label{fig:dvdm}} 
\end{figure}

Figure~\ref{fig:dvdm} shows the velocity decline rates for the
\ion{Si}{2} $\lambda$6355 line, \deltav, versus
\dm. A similar relationship was presented in the left panel of Figure~17 of
\citet{blondin12} by adopting the velocity decline during 10 days after
maximum light. The velocity gradient $\dot{v}$ as defined by
\citet{benetti05} shows a larger dispersion in its relation with
\dm\ (see their Figure~3b) than the one of \deltav\ presented here. The
same happens with the instantaneous velocity decline 
rate defined in the right panel of Figure~17 of \citet{blondin12}. 
In the case of \deltav, there seems to be a continuous link
between CN, SS and CL \sneia. Excluding BL SNe which show the largest
\deltav\ values, a correlation coefficient as large as $\rho=0.86$ is
found between \deltav\ and \dm.

\subsection{Luminosity calibration}
\label{sec:Lum}

\sndia\ peak absolute magnitudes can be calibrated against decline
rate and color \citep{tripp98} with a precision between $0.1$ and
$0.2$ mag \citep{hicken09,folatelli10}. Given the correlation of
\pwtwo\ and \pwsix\ with \dm\ shown in Figure~\ref{fig:ew26dm}
(Section~\ref{sec:ewdm}), we tested the calibration of $B$-band peak
absolute magnitudes versus $(B_{\mathrm{max}}-V_{\mathrm{max}})$ pseudocolor and
these \ew\ parameters \citep[similar analysis can be found
  in][]{blondin11a,silverman12c}. For this purpose, we replaced
\dm\ by \pwtwo\ and \pwsix\ in Equation~(6) of
\citet{folatelli10} for the $B$ band. We obtained

\begin{equation}
\label{eq:modtb}
\mu=B_{\mathrm{max}}-M_B(0)-a_B\,(pW\,-\,pW_0)-\beta_B^{BV}\,(B_{\mathrm{max}}-V_{\mathrm{max}}),
\end{equation}

\noindent where \ew\ corresponds to \pwtwo\ and \pwsix, and $pW_0=15$ \AA\
and $20$ \AA, respectively. The redshift-based distances used for
Equation~(\ref{eq:mb0}) were supplemented with distance to the host
galaxies of SNe~2005ke, 2006X and 2006mr available in the literature.
These objects were excluded from Equation~(\ref{eq:mb0}) because they
either had too large $E(B-V)_{\mathrm{Host}}$ or \dm, but they can be
included in the luminosity calibration. For SNe~2005ke and 2006X we adopted  
  $\mu=31.84 \pm 0.08$ mag and $\mu=30.91 \pm 0.14$ mag, respectively,
  as given in Table~7 of \citet{folatelli10}. For SN~2006mr, we used
  the latest distance modulus to its host, NGC~1316, of $\mu = 31.25
  \pm 0.05$ mag provided by \citet{stritzinger10}.

We fitted for $M_B(0)$, $a_B$, and
$\beta_B^{BV}$ employing the same $\chi^2$ minimization method as
described in the Appendix~A of
\citet{folatelli10}. Table~\ref{tab:lumew} provides the results of the
fits for different subsamples of \sneia. The table lists the fitted
parameters, the estimated intrinsic dispersion $\sigma_{\mathrm{SN}}$ 
\citep[see Appendix~A of][]{folatelli10}, the number of SNe used,
and the sample selected for each fit. The weighted rms about the fit
is also listed, defined as

\begin{equation}
\label{eq:wrms}
\mathrm{WRMS}\;=\;\frac{\sum_i^{N_{\mathrm{SN}}}\left[\mu_i-\bar{\mu}_i\right]^2/(\sigma_i^2+\sigma_{\mathrm{SN}}^2)}{\sum_i^{N_{\mathrm{SN}}}
  1/(\sigma_i^2+\sigma_{\mathrm{SN}}^2)},
\end{equation}

\noindent where $\bar{\mu}_i$ is the distance modulus as derived from
the best fit of Equation~(\ref{eq:modtb}), $\sigma_i$ is the
measurement uncertainty in the distance modulus 
\citep[see Appendix~A of][]{folatelli10}, and $i$ runs along the
sample of $N_{\mathrm{SN}}$ SNe. 

The weighted rms scatter for the complete sample are
$0.17$ mag and $0.19$ mag (fits 1 and 6 for \pwtwo\ and
\pwsix, respectively). Similar results for \pwtwo\ were found by
\citet{blondin11a} and \citet{silverman12c}. The fit and scatter
remain nearly unchanged when HV SNe are removed (fits 2 and 7). When
we consider only CN \sneia, the scatter is reduced to $0.13$
mag for the case of \pwtwo\ (fit 3). If we cut the sample of
\sneia\ to exclude objects redder than
$(B_{\mathrm{max}}-V_{\mathrm{max}})=0.2$ mag---which 
excludes reddened objects but also most CL SNe---the scatter is
reduced to $0.16$ mag (fits 4 and 9).

For comparison, the fits with exactly the same
samples of SNe as in Table~\ref{tab:lumew}, but now with
\dm\ instead of \ew, yield very similar values of WRMS. For
the complete sample and $(B_{\mathrm{max}}-V_{\mathrm{max}})<0.2$ mag
cuts, the resulting WRMS values are $0.20$ mag and $0.18$
mag, respectively \citep[see also Table~8 of][]{folatelli10}. 

The intrinsic dispersions $\sigma_{\mathrm{SN}}$ of the
luminosity--\ew\ fits are $\lesssim$$0.1$ mag in all
cases. Assuming the measurement uncertainties 
are well constrained, this indicates that the actual dispersion not
accounted for in the fits is not large. The slopes, $a_B$, of the
relation between peak absolute magnitudes and \ew\ for all fits are
compatible within uncertainties, except for a slightly larger
slope when using \pwsix\ and SNe with
$(B_{\mathrm{max}}-V_{\mathrm{max}})<0.2$ mag.

The slopes, $\beta_B^{BV}$, of peak luminosity versus color are also
compatible within uncertainties for all fits shown in
Table~\ref{tab:lumew}. Such color dependence of luminosity can be
interpreted as the effect of dust extinction. Nevertheless, the
dependence of luminosity on color for eCL SNe---which are
intrinsically red---seems to agree with that of SNe reddened by
dust. This is indicated by the lack of change in $\beta_B^{BV}$ when
we exclude SNe with \dm\,$>1.7$ mag---that is, basically excluding the
eCL group---, as shown in fits 5 and 10. Assuming the CCM+O reddening
law, the color slope can be converted to the total-to-selective
absorption coefficient $R_V$. The conversion is detailed in Appendix~B of
\citet{folatelli10}. The resulting $R_V$ values are in the range of
$1.6$--$2.1$, depending on the sample employed. This is
slightly larger than the range of $1.0 < R_V < 1.5$ found by
\citet{folatelli10} when fitting versus \dm\ on a subsample of the
\sneia\ presented here. Still, the $R_V$ values found here are smaller
than the Galactic average of $R_V=3.1$., with a significance of at
least 3~$\sigma$.

In the attempt to refine the precision of \sneia\ as distance
indicators, we searched for possible correlations between spectral 
parameters and residuals of the luminosity calibration. 
$B$-band peak absolute magnitude residuals, $\Delta M_B$, were
calculated based on the best fit of luminosity versus \dm\ and
$(B_{\mathrm{max}}-V_{\mathrm{max}})$ pseudocolor at maximum light as
in Equation~(6) of \citet{folatelli10}, 

{\small 
\begin{equation}
\label{eq:modtbor}
\mu=B_{\mathrm{max}}-M_B(0)-b_B\,\left[\Delta m_{15}(B)-1.1\right]-\beta_B^{BV}\,(B_{\mathrm{max}}-V_{\mathrm{max}}).
\end{equation}
}

\noindent The fit included 82 \sneia\ with available
distances and photometric parameters, and yielded
$M_B(0)=-19.16\pm0.01$ mag, $b_B=0.55\pm0.08$,
$\beta_B^{BV}=3.05\pm0.10$ \citep[cf. fit~1 of Table~8
  in][]{folatelli10}. The peak absolute magnitude residuals were
thus computed as

{\small 
\begin{equation}
\label{eq:hres}
\Delta M_B = B_{\mathrm{max}} - \mu + 19.16 - 3.05 \,
(B_{\mathrm{max}}-V_{\mathrm{max}}) - 0.55 \, [\Delta m_{15}(B) - 1.1].
\end{equation}
}

\noindent $\Delta M_B$ values are given in the last column of
Table~\ref{tab:sne}. 

\begin{figure}[htpb]
\epsscale{1.0}
\plotone{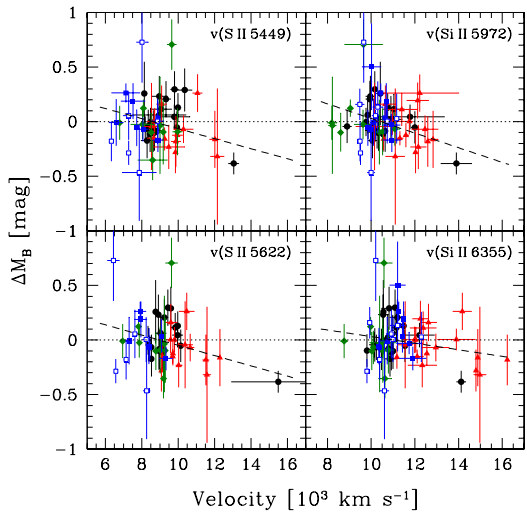}
\caption{Peak absolute $B$-band magnitude residuals versus 
  \ion{S}{2} and \ion{Si}{2} expansion velocities. Different Branch
  subtypes are represented by black circles (CN), blue squares (CL),
  red triangles (BL), and green diamonds (SS). Filled symbols indicate
  SNe with \dm\,$<1.7$ mag while the rest of the objects are shown with open
  symbols. The dashed lines show straight-line fits to the filled data
  points. \label{fig:vel256res}}    
\end{figure}

We investigated possible correlations between $\Delta M_B$ and
spectral parameters. In general agreement with \citet{blondin11a}
  and \citet{silverman12c}, no significant dependence was found with any of
the \ew\ parameters. On the other hand, we found that the residuals
depend slightly on \ion{S}{2} and \ion{Si}{2} velocities, in
particular when considering \sneia\ with \dm\,$<1.7$
mag. Figure~\ref{fig:vel256res} shows the strongest of such correlations.
Table~\ref{tab:velres} summarizes the 
results of the fits. The slopes of $\Delta M_B$ versus
\vfive, \vsix, \vthree, and \vfour\ are different from zero at the
$\approx$2--3-$\sigma$ level. Given the lack of correlations
  found for other \sndia\ samples by \citet{blondin11a} and
  \citet{silverman12c}, the results presented here deserve further
  study using large, homogeneous samples.

A Kolmogorov-Smirnov (K-S) test on the distribution of Hubble
residuals for \sneia\ separated into the HV and Normal classes of
\citet{wang09} gives a $p$-value of 0.07 (for 17 HV
and 31 Normal SNe). There is a non-negligible probability that both
distributions are consistent. However, the difference in average
Hubble residuals is noticeable between HV and Normal \sneia\ (or
between BL and the complete sample), as shown in the last column of
Table~\ref{tab:phave}. In case the differences are confirmed, they may
be related with the finding that the scatter in the Hubble diagram is
reduced when using only Normal \sneia, as shown by \citet{foley11a}.
Part of the difference in Hubble residuals may be
explained by differences in intrinsic colors between Normal and HV SNe. If,
as suggested by \citet{foley11a}, HV objects show redder intrinsic
colors than Normal \sneia, then they could be ``overcorrected'' by
reddening. However, when we compared the distributions of
$(B_{\mathrm{max}}\,-\,V_{\mathrm{max}})$ for both subclasses using a
K-S test, the resulting $p$-value was of $0.85$, indicating that both
color distributions are consistent \citep[cf.][]{foley11b}.

\subsection{Color}
\label{sec:bv}

SN colors play a key role in the determination of extinction and in
the calibration of luminosity. Investigating the relation between
colors and spectroscopic parameters may serve to distinguish reddening
sources of intrinsic and extrinsic nature, and to understand the
origin of color variations. Here we study the relationship between
$(B_{\mathrm{max}}-V_{\mathrm{max}})$ and \ew\ for different spectral
features. In order to search for possible dependence of intrinsic colors
$(B_{\mathrm{max}}-V_{\mathrm{max}})_0$ on \ew, a subset of
``low-reddening'' SNe was identified. This was done following two
requirements: (1) objects with E/S0 host galaxies or 
isolated from nuclei and arms, and (2) no evidence of
\ion{Na}{1} D lines from the host galaxy in the spectra. To this
sample, we added some objects whose late-time $(B-V)$ colors were
compatible with the intrinsic-color law given in Equation~(2) of
\citet{folatelli10}. These SNe are indicated with a ``Y'' in
Column~(8) of Table~\ref{tab:sne}. 

\begin{figure}[htpb]
\epsscale{1.0}
\plotone{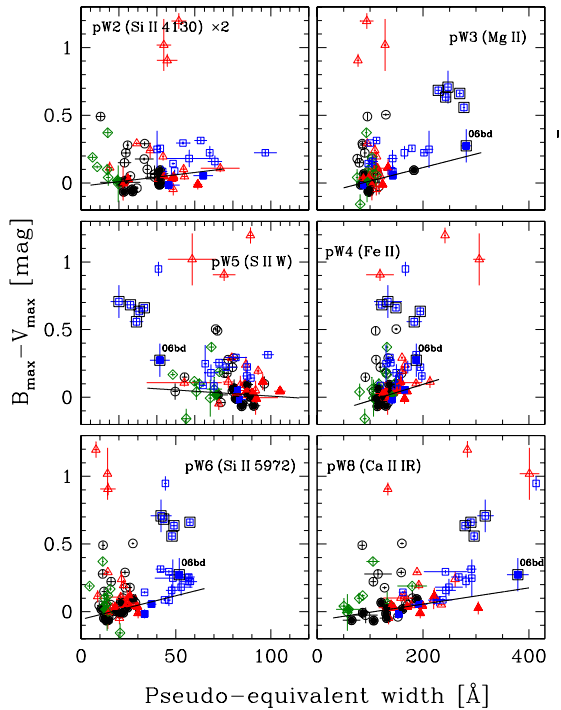}
\caption{Pseudo-colors at maximum light, $(B_{\mathrm{max}}-V_{\mathrm{max}})$,
  corrected by Milky Way reddening, as a
  function of \ew\ for several spectral features. Filled symbols
  correspond to SNe with low host-galaxy reddening (see text) and
  \dm\,$<1.7$ mag. The solid lines show straight-line fits such data
  points, as summarized in Table~\ref{tab:ewbv}. Members of the eCL
  group are indicated with a black square. They present intrinsically
  red colors but most of them are excluded from the fits because they
  have \dm\,$>1.7$ mag.\label{fig:ewbv}}
\end{figure}

It has been shown that intrinsic $B-V$ colors of \sneia\ at maximum
light depend on decline rate
\citep{phillips99,altavilla04,folatelli10}. The dependence is slight
for SNe with \dm\,$<1.7$ mag. For the fastest declining objects,
however, colors are significantly redder than for the rest of \sneia. 
We thus performed straight-line fits to the intrinsic color--\ew\ relations
using low-reddening SNe with \dm\,$<1.7$ mag. This excludes the eCL
\sneia, most of which are in the low-reddening group\footnote{The two eCL
  objects outside the low-reddening sample are SNe~2005bl and
  2006mr. While the former may have suffered significant reddening of
  $E(B-V)_{\mathrm{host}}= 0.17$ mag \citep{taubenberger08}, SN~2006mr
  can be considered to be nearly reddening-free \citep{stritzinger10}.}, with the
exception of SN~2006bd, which has \dm\,$=1.65\pm0.02$ mag and is also
significantly bluer than the rest of the eCL objects. Note that eCL
SNe have no measurements of \pwtwo. Table~\ref{tab:ewbv} summarizes the fit
results for all eight \ew\ parameters. Most features show a
positive dependence of intrinsic color on \ew, except
\pwfive. Figure~\ref{fig:ewbv} shows the relations for that feature
along with those that yielded the strongest dependence. The slopes in these cases
are different from zero at the 2--3-$\sigma$ level. \citet{nordin11b}
showed a similar trend in the 
case of the \ion{Si}{2} $\lambda$4130 line. Based on a large sample of
\sneia\ with measurements of extinction-corrected pseudo-colors at
maximum light, \citet{foley11b} found a correlation between
$(B_{\mathrm{max}}-V_{\mathrm{max}})_0$ and \pwseven. In our case, we
obtain a similar relation, but the fit slope is just $1.7$~$\sigma$
from zero. It should be noted that \citet{foley11b} obtain their
intrinsic colors in a different way than the one adopted here. They
define $(B_{\mathrm{max}}-V_{\mathrm{max}})_0$ as the residuals in the
color axis of the relation between light-curve shape corrected
absolute magnitudes and color. This implies the underlying hypothesis
that the relation between peak luminosity and light-curve shape holds
exactly for all SNe. In our case, we do not resource to such assumption
but define a low-reddening sample, independently of the
luminosity calibration. This procedure has the price of reducing the
working sample of SNe. We confirm the lack of dependence of intrinsic
color on \pwone\ also shown by \citet{foley11b}.

An interesting case is that of \pwfive, which shows a {\em negative}
slope, although with a low significance. It should be noted, however, that
eCL \sneia\ seem to follow the trend found for the rest of the
low-reddening sample, with redder colors and lower \pwfive, although
deviating from the straight-line fit. Something of the same sort can
be seen in the case of \pwthree, with eCL SNe showing the largest
colors and pseudo-equivalent widths. This is expected since the
increase in \pwthree\ by the presence of strong \ion{Ti}{2}
absorption affects the $B$-band magnitude and is related with a
temperature effect. The explanation is not so straight-forward for \pwfive.

We have also searched for correlations between
$(B_{\mathrm{max}}-V_{\mathrm{max}})_0$ and expansion velocities at
maximum light based on the same sample of low-reddening SNe. Contrary
to what \citet{foley11b} presented, we find no significant relation
for any of the line velocities analyzed here. In order to
directly compare with the latter work, we restrict the sample to SNe with
  $1.0\leq$\,\dm\,$\leq 1.5$ mag. For \vsix\ we find a
slope of $0.012 \pm 0.016$ mag $/$ $10^3$ km s$^{-1}$, which is
lower and less significant than the value of $0.033 \pm 0.004$ mag
$/$ $10^3$ km s$^{-1}$ given by \citet{foley11b}. 
Part of the discrepancy may be due to the different definition of
$(B_{\mathrm{max}}-V_{\mathrm{max}})_0$, as was pointed out above. 
\citet{blondin12} find milder relations than \citet{foley11b},
although still with 2--3-$\sigma$ significance (see their Table~6).
Our results seem to be in contradiction with the predictions of
asymmetric models by \citet{kasen07b} as presented in Figure~8 of
\citet{foley11a}. Other 2D models by \citet{kasen09} agree more with
the vanishing slopes of color versus \vsix, as shown in Figure~18 of
\citet{blondin11b}, although for those models the expansion velocities
are all larger than 11000 km s$^{-1}$.

\section{DISCUSSION AND CONCLUSIONS}
\label{sec:concl}

In this paper we have presented optical spectroscopic data of 93
low-redshift \sneia\ observed by the CSP between 2004 and 2009. The
data set amounts to 604 previously unpublished spectra
obtained on 502 different epochs in the range between $-12$
and $+155$ days with respect to $B$-band maximum light. To this we
have added 228 published spectra of 
some of the SNe, obtained from the SUSPECT database and from the CfA
sample \citep{blondin12} and the BSNIP sample \citep{silverman12a}. Based on
spectroscopic parameters such as line velocities and pseudo-equivalent
widths we have provided a quantified classification of \sndia\ subtypes
following the scheme of \citet{branch06}. We note, however, that the
separation between subclasses is somewhat arbitrary because there is
no clear discontinuity in the spectroscopic properties of \sneia\ when
we go from one group to another. With a study of spectroscopic parameters at
maximum light and their relationship with photometric quantities we
have attempted to understand the properties of \sneia\ and their
calibration as distance indicators. 

Among the Branch subtypes, Core Normal (CN) \sneia\ present the
largest degree of homogeneity in their optical spectrum at maximum
light, as was also found by \citet{blondin12} for the CfA sample. For
all subtypes, the dispersion increases toward both ends of the optical
range. At the blue end, the dispersion among Shallow Silicon (SS) and
Broad Line (BL) objects seems to be dominated by dust reddening. For
Cool (CL) SNe, the dispersion is mostly due to temperature
differences in the ejecta, which produces variations in the absorption 
features due to \ion{Ti}{2} between $\approx$4000 \AA\ and
$\approx$4500 \AA. At the extreme of this class (eCL) 
we find SN~1991bg-like objects with large \pwthree, intrinsically red
color at maximum light, and fast light-curve decline rate. These
properties allow us to distinguish eCL SNe from the rest of the CL
subclass which, as a whole shows no discontinuity in spectral
properties from CN SNe. The greater degree of association of SS SNe
with star-forming regions as compared with the other subtypes becomes
evident through the appearance of host-galaxy lines in the composite
spectrum of Figure~\ref{fig:spave}. 

At the red of the spectrum other interesting deviations appear related
with the strength of the \ion{Ca}{2}\,IR triplet and
\ion{O}{1}\,$\lambda$7774. The shape of the former feature reveals
high-velocity components present in most CN, BL and SS objects. CL SNe
show large \ion{Ca}{2}\,IR absorptions with no evidence of high-velocity
components. The strength of this absorption as measured by
\pweight\ tends to increase with \dm, which implies 
that it increases in the sequence of SS -- CN -- BL -- CL subtypes. At
a given \dm, however, the dispersion of \pweight\ can be large, driven
mostly by dominant high-velocity features. Interestingly, as
\pweight\ grows along the sequence described above, the other
\ion{Ca}{2} feature, \pwone, does not follow the same behavior but
rather stays nearly constant. We speculate that the difference in behavior
between both \ion{Ca}{2} lines may be due to temperature changes
affecting more the \ion{Ca}{2}\,IR triplet than the \ion{Ca}{2}\,H\&K
line. In analogy to the mechanism invoked by \citet{hachinger08}
to explain the dependence of the silicon ratio, \rsi, on \dm\, the
  \ion{Ca}{2}\,H\&K is likely saturated and thus, as opposed to the
  \ion{Ca}{2}\,IR triplet, does not grow with an increasing
  \ion{Ca}{2}\,$/$\,\ion{Ca}{3} abundance ratio as temperature
  decreases. The study of 
high-velocity features and their incidence among 
\sneia\ deserves further scrutiny. We emphasize that such study
requires spectra covering a long wavelength range with high S/N
around the \ion{Ca}{2} features.

In close agreement with \citet{blondin12} we find that 24\% of the
\sneia\ belong to the HV class---defined as having \vsix\,$>12000$ km
s$^{-1}$. When considering only ``normal'' \sneia---i.e., excluding
1991T- and 1991bg-like objects---the fraction of HV SNe in our sample
is 31\%. This percentage is close to that
of 35\% found in the sample of \citet{wang09}, and the difference
is further reduced if we adopt their dividing velocity of 11800 km
s$^{-1}$. We note that a study of the population of HV SNe in a
homogeneous sample, preferably from a ``blind'' search is required for
determining the actual fraction of these objects. The implications may
be important to our 
knowledge of the explosion physics considering the picture presented by
\citet{maeda10b}, where the differences in expansion velocities at
maximum light are explained as a viewing-angle effect of an asymmetric
explosion. The fraction of HV SNe would thus determine the average
solid angle subtended by the fast-expanding portion of the exploding WD.
 
As shown by \citet{wang09} and \citet{blondin12}, \vsix\ does not
correlate strongly with the pseudo-equivalent width of the same line,
unless the HV SNe are considered. Indeed, the Pearson correlation
coefficient grows from $\rho=0.50$ for non-HV SNe to $\rho=0.86$ for HV
objects. Most of the HV objects belong to the BL subtype 
for which the increase in \ion{Si}{2}
$\lambda$6355 and \ion{Si}{2} $\lambda$5972 velocity is accompanied by
an increase in 
\pwseven\ (the correlation coefficients are $\rho=0.95$ and
$\rho=0.80$ with each velocity, respectively). This behavior is {\em
  the opposite} for \pwsix. The strength of this line tends to {\em
  decrease} with increasing \vsix\ for HV or BL SNe (the correlation
coefficient being $\rho=-0.60$). Based on simple SYNOW spectral models,
\citet{branch09} explained this difference in behavior for the
\ion{Si}{2} $\lambda$5972 and $\lambda$6355 lines in terms of a difference in
the distribution of \ion{Si}{2} inside the ejecta. A shallower
distribution for BL SNe as compared with normal SNe can produce large
\ion{Si}{2} $\lambda$6355 absorptions while keeping the \ion{Si}{2}
$\lambda$5972 line
shallow. However, as the authors point out, several other ions may
affect the spectrum at the wavelength range of these features, which
may complicate the picture. 

The behavior of BL SNe also stands out in the relation between
velocity decline, \deltav, and light-curve decline rate, \dm. By
excluding the BL objects a correlation between both quantities can be
recovered with $\rho=0.79$ \citep[cf.][]{blondin12}. As already
pointed out by \citet{benetti05}, this indicates that SNe with
fast-declining light curves (their FAINT subtype) also decline faster
in the \ion{Si}{2} post-maximum velocity evolution, as compared with
the CN and SS objects (i.e., most of the LVG class of Benetti
et~al.). It is worth noting that BL SNe show, on average, larger negative
residuals in the absolute peak magnitude calibration versus decline
rate and color, as compared with other subtypes (see
Table~\ref{tab:phave}). This drives the dependence found between
residuals, $\Delta M_B$, and velocities from \ion{S}{2} and
\ion{Si}{2} (Figure~\ref{fig:vel256res}).  
 
We have detected two cases that indicate even further diversity in the
distribution of elements within the ejecta of \sneia. These are the SS
SN~2005M and the CN (but HV) SN~2006is (see
Section~\ref{sec:05M06is}). They show extreme \ion{Si}{2} velocities
at the low end of SS objects (SN~2005M), and at the high end of CN
objects (SN~2006is), accompanied by other light species such as \ion{O}{1} and
\ion{S}{2}. However, both objects show normal expansion velocities of
$\approx$11000 km s$^{-1}$ for the rest of the ions. SN~2006is
resembles the case of SN~2009ig that showed a strong, high-velocity
component of the \ion{Si}{2} $\lambda$6355 line before maximum
light \citep{foley12}. 

It is worth investigating whether models with departures from
spherical symmetry \citep[e.g.,][]{maeda10a} that have 
been suggested to explain the differences in velocity decline rates
\citep{maeda10b} can also provide an explanation to (a) the lack of growth
of the \ion{Si}{2} $\lambda$5972 feature among BL SNe, (b) the
relatively higher luminosity (negative residuals) of this subtype
after calibration by \dm\ and color, and (c) the peculiar \ion{Si}{2}
velocities of cases like SNe~2005M and 2006is. In addition to improving our
knowledge of the physical properties of \sneia, such studies may
help to increase the precision of distance estimations. 

Given the correlation of \pwtwo\ and \pwsix\ with light-curve decline rate, we
have replaced \dm\ by these \ew\ parameters in the calibration of peak
luminosities for \sneia. The resulting fits of peak absolute 
magnitudes versus \ew\ and color show similar precision as those
based on \dm. The rms scatter in the calibrated $B$-band peak absolute 
magnitude is as low as $0.13$ mag, with less than $\approx$$0.1$ mag of
estimated intrinsic scatter. Such dispersion is also similar to the
ones found by \citet{bailey09} and \citet{blondin11a} based on spectral
flux ratios. These results suggest that \sneia\ can be 
calibrated as distance indicators to within 6\%--8\% with just a
measurement of the color 
and a spectrum at maximum light. Moreover, the required S/N of the
spectrum needs not be very high since \ew\ measurements integrate over
a range of $\gtrsim$100 \AA. The use of \pwtwo\ in particular has the
advantage of being applicable to optical observations up to a redshift
of $z \approx 1$. 

The fits described above additionally provide information about the nature of 
extinction in the host galaxies, if the color dependence is interpreted
as the effect of reddening by dust. From the slopes of the color
dependence we have derived low values of the
total-to-selective absorption coefficient of $R_V \approx 2$, although
not as low as those found when \dm\ is used instead of
\ew\ \citep{folatelli10}. This result holds 
both when including or excluding intrinsically red SNe (those in the
CL group), indicating that the effect of dust reddening on luminosity
is similar to that of intrinsic reddening among CL SNe. 

\acknowledgments 

This research is supported by the World Premier International Research Center
Initiative (WPI Initiative), MEXT, Japan. G.~F. acknowledges financial
support by Grant-in-Aid for Scientific Research for Young Scientists
(23740175). This material is based upon work supported by NSF under  
grants AST--0306969, AST-0908886, AST--0607438, and AST-1008343. 
F.~F., J.~A. and G.~P. acknowledge support from FONDECYT through grants
3110042, 3110142 and 11090421. J.~A., F.~F, M.~H. and G.~P. acknowledge
support provided by the Millennium Center for Supernova Science through
grant P10--064-F (funded by “Programa Bicentenario de Ciencia y Tecnología
de CONICYT” and “Programa Iniciativa Científica Milenio de
MIDEPLAN”). M.~S. acknowledges the generous support provided by the
Danish Agency for Science and Technology and Innovation through a
Sapere Aude Level 2 grant. 


\clearpage
\LongTables

\clearpage
\begin{landscape}
\begin{deluxetable*}{lcccccccccccccc}
\tabletypesize{\tiny} 
\tablecolumns{15} 
\tablewidth{0pt} 
\tablecaption{Summary of observations, classification and photometric properties of the SN-Ia sample\label{tab:sne}} 
\tablehead{ 
\colhead{} & \colhead{} & \colhead{Epoch} & \colhead{SNID} & \colhead{Wang} & \colhead{Branch} & \colhead{} & \colhead{} & \colhead{$T_B(\mathrm{max})$} & \colhead{$m_B$} & \colhead{\dm} & \colhead{$E(B-V)$} & \colhead{$(B_{\mathrm{max}}-V_{\mathrm{max}})$} & \colhead{Low} & \colhead{$\Delta M_B$} \\ 
\colhead{SN} & \colhead{$N$} & \colhead{range} & \colhead{type} & \colhead{type} & \colhead{type} & \colhead{$z_{\mathrm{Helio}}$} & \colhead{$z_{\mathrm{CMB}}$} & \colhead{RJD} & \colhead{(mag)} & \colhead{(mag)} & \colhead{Gal (mag)} & \colhead{(mag)} & \colhead{Redd.} & \colhead{(mag)} \\ 
\colhead{(1)} & \colhead{(2)} & \colhead{(3)} & \colhead{(4)} & \colhead{(5)} & \colhead{(6)} & \colhead{(7)} & \colhead{(8)} & \colhead{(9)} & \colhead{(10)} & \colhead{(11)} & \colhead{(12)} & \colhead{(13)} & \colhead{(14)} & \colhead{(15)} 
} 
\startdata
2004dt & $5$ & $  +23.2 \;  +102.6$ &     normal &         HV &   BL &  0.01973 &  0.01883 &  53239.1 & $ 15.000$($035$) & $  1.235$($072$) &  0.023 & $ -0.046$($047$) &  \nodata & $  -0.39$($18$) \\ 
2004ef & $8$ & $   -8.0 \;   +37.5$ &     normal &         HV &   BL &  0.03099 &  0.02979 &  53264.0 & $ 16.828$($019$) & $  1.379$($061$) &  0.056 & \phs$  0.102$($027$) &  \nodata & $  -0.09$($12$) \\ 
2004eo & $3$ & $  -10.5 \;   +24.9$ &     normal &          N &   CL &  0.01570 &  0.01475 &  53278.2 & $ 15.067$($011$) & $  1.365$($063$) &  0.108 & \phs$  0.055$($020$) &        Y & $  -0.16$($17$) \\ 
2004ey & $4$ & $   -0.3 \;   +47.9$ &     normal &          N &   CN &  0.01579 &  0.01464 &  53303.9 & $ 14.713$($011$) & $  0.930$($013$) &  0.139 & $ -0.063$($022$) &        Y & \phs$   0.15$($16$) \\ 
2004gc & $4$ & $  +10.8 \;   +33.1$ &     normal &   $\cdots$ & \nodata &  0.03070 &  0.03074 &  53324.6 & $ 16.649$($028$) & $  1.088$($055$) &  0.206 & \phs$  0.150$($037$) &  \nodata & $  -0.25$($13$) \\ 
2004gs & $2$ & $   -3.1 \;    +2.7$ &     normal &          N &   CL &  0.02665 &  0.02752 &  53356.0 & $ 17.132$($017$) & $  1.613$($082$) &  0.031 & \phs$  0.158$($044$) &  \nodata & \phs$   0.07$($16$) \\ 
2004gu & $1$ & $   -4.3$ &     normal &        91T &   SS &  0.04586 &  0.04693 &  53363.4 & $ 17.478$($017$) & $  0.888$($299$) &  0.027 & \phs$  0.189$($021$) &  \nodata & $  -0.33$($23$) \\ 
2005A  & $3$ & $   +4.1 \;   +32.6$ &     normal &         HV &   BL &  0.01914 &  0.01835 &  53380.4 & $ 18.141$($151$) & $  1.222$($183$) &  0.030 & \phs$  1.019$($192$) &  \nodata & $  -0.10$($58$) \\
\enddata 
\tablecomments{Columns: (1) SN name; (2) Number of spectroscopic epochs; (3) Epoch range of previously unpublished spectra relative to $B$-band maximum light; (4) Spectral subtype based on SNID comparisons \citep{blondin07}; (5) Spectral subtype based on the scheme of \citet{wang09}: normal (N), SN~1991T-like (91T), SN~1991bg-like (91bg), high velocity (HV); (6) Spectral subtype based on \ew\ \citep{branch06}: core normal (CN), cool (CL), broad line (BL), shallow silicon (SS); (7) Heliocentric redshift from the NASA/IPAC Extragalactic Database (NED), except when otherwise noted; (8) Redshift in the 3K CMB frame converted from $z_{\mathrm{Helio}}$ using the application in NED; (9) Reduced Julian date (JD $-$ 2400000) of $B$-band maximum light; (10) $K$-corrected, apparent $B$-band peak magnitude, corrected for dust extinction in the Galaxy (uncertainties in thousandth of mag); (11) Observed decline rate, \dm\ (uncertainties in thousandth of mag); (12) Galactic reddening (NED); (13) Pseudo-color at maximum light (uncertainties in thousandth of mag); (14)``Y'' indicates the SN is considered to have suffered negligible host-galaxy reddening; (15) $B$-band absolute peak magnitude residual (see Section~\ref{sec:spphot}; uncertainties in hundredth of mag).\\ $\dagger$ Redshift computed from spectrum of the host galaxy obtained by the CSP.}

\end{deluxetable*} 
\clearpage
\end{landscape}

\clearpage
\begin{deluxetable*}{lccccccccc} 
\tabletypesize{\scriptsize} 
\tablecolumns{10} 
\tablewidth{0pt} 
\tablecaption{Previously unpublished spectroscopic observations \label{tab:spec}} 
\tablehead{ 
\colhead{UT Date} & \colhead{Reduced} & \colhead{Phase} & \colhead{Tel.} & \colhead{Inst.} & \colhead{Wavelength} & \colhead{Resol.} & \colhead{Exp.} & \colhead{Air-} & \colhead{Phot.} \\ 
\colhead{} & \colhead{JD} & \colhead{} & \colhead{} & \colhead{} & \colhead{Range (AA)} & \colhead{(AA)} & \colhead{(s)} & \colhead{mass} & \colhead{rms (mag)} \\ 
\colhead{(1)} & \colhead{(2)} & \colhead{(3)} & \colhead{(4)} & \colhead{(5)} & \colhead{(6)} & \colhead{(7)} & \colhead{(8)} & \colhead{(9)} & \colhead{(10)} 
} 
\startdata
\multicolumn{10}{c}{SN 2004dt} \\ 
2004-09-$14.210$$\ddagger$ & $ 53262.710$ & $   23.2$ & DUP & WF & $ 3800 -  9235$ & $  8.0$ & $  1800$ & $ 1.35$ & $ 0.063$ \\ 
2004-09-$19.334$$\ddagger$ & $ 53267.834$ & $   28.2$ & DUP & WF & $ 4530 -  9618$ & $  8.0$ & $  1800$ & $ 1.20$ & $ 0.109$ \\ 
2004-09-$20.216$$\ddagger$ & $ 53268.716$ & $   29.1$ & DUP & WF & $ 3800 -  9235$ & $  8.0$ & $  2400$ & $ 1.25$ & $ 0.059$ \\ 
2004-10-$24.269$ & $ 53302.769$ & $   62.5$ & CLA & LD & $ 3600 -  9000$ & $ 14.0$ & $   900$ & $ 1.30$ & $ 0.062$ \\ 
2004-12-$04.153$$\dagger$ & $ 53343.653$ & $  102.6$ & DUP & WF & $ 3800 -  9235$ & $  8.0$ & $  1800$ & $ 1.28$ & $ 0.216$ \\ 
\multicolumn{10}{c}{SN 2004ef} \\ 
2004-09-$07.162$$\dagger$ & $ 53255.662$ & $   -8.0$ & DUP & MS & $ 3780 -  7323$ & $  2.5$ & $  4500$ & $ 1.54$ & $ 0.288$ \\
\enddata 
\tablecomments{Columns: (1) UT date of the observation; 
 (2) Reduced Julian date of the observation (JD $-$ 2400000); 
 (3) Phase in rest-frame days since $B$-band maximum light;
 (4) Telescope code -- 3P6: ESO 3.6~m Telescope; 
     BAA: Las Campanas Magellan~I 6.5~m Baade Telescope; 
     CLA: Las Campanas Magellan~II 6.5~m Clay Telescope; 
     DUP: Las Campanas 2.5~m du Pont Telescope Telescope;
     GEM: Gemini-S 8.1~m Telescope; 
     HIL: Hiltner 2.4~m Telescope;
     MGH: McGraw-Hill 1.3~m Telescope;
     NTT: New Technology Telescope; 
     P20: Palomar 200~inch Hale Telescope; 
     T60: CTIO 1.5~m Telescope;
 (5) Instrument code -- 
     BC: Boller \& Chivens spectrograph;
     CS: Ritchey-Chretien spectrograph; 
     DB: Double Spectrograph;
     EF: ESO Faint Object Spectrograph and Camera (EFOSC-2);
     EM: ESO Multi-Mode Instrument (EMMI);
     GM: Gemini Multi-Object Spectrographs (GMOS);
     IM: Inamori Magellan Areal Camera and Spectrograph (IMACS),
     LD: Low Dispersion Survey Spectrograph (LDSS);
     MA: Magellan Echellette spectrograph (MagE);
     MS: Las Campanas Modular Spectrograh; 
     WF: Wide Field Reimaging CCD Camera (WFCCD);
 (6) Wavelength range covered;
 (7) Spectral resolution in \AA\ as estimated from arc-lamp lines;
 (8) Total exposure time;
 (9) Airmass at the middle of the observation;
 (10) RMS of the differences between synthetic and observed broadband magnitudes.\\
 $\dagger$ Spectrum corrected to match photometry (see Section~\ref{sec:obs}).\\ 
 $\ddagger$ Spectrum uncorrected for telluric absorption (see Section~\ref{sec:obs}).}

\end{deluxetable*} 

\clearpage
\begin{deluxetable*}{lcccccccc} 
\tabletypesize{\scriptsize} 
\tablecolumns{9} 
\tablewidth{0pt} 
\tablecaption{Values of velocity [km s$^{-1}$] at $B$-band maximum light, and \ion{Si}{2} $\lambda$6355 velocity decline [km s$^{-1}$].\label{tab:vels}} 
\tablehead{ 
\colhead{SN} & \colhead{v(\ion{Ca}{2} H\&K)} & \colhead{v(\ion{Si}{2} $\lambda$4130)} & \colhead{v(\ion{S}{2} $\lambda$5449)} & \colhead{v(\ion{S}{2} $\lambda$5622)} & \colhead{v(\ion{Si}{2} $\lambda$5972)} & \colhead{v(\ion{Si}{2} $\lambda$6355)} & \colhead{v(\ion{Ca}{2} IR)} & \colhead{\deltav} 
} 
\startdata
\multicolumn{9}{c}{CN} \\ 
2004ey  & $ 16203 \pm   1307$ & $  9640 \pm    381$ & $  9327 \pm    279$ & $  9238 \pm    191$ & $  9942 \pm    200$ & $ 11204 \pm    333$ & \nodata & \nodata \\ 
2005bo  & $ 11928 \pm    788$ & $  9527 \pm    153$ & $  8291 \pm    336$ & $  8527 \pm    118$ & $ 10065 \pm    202$ & $ 10857 \pm    291$ & \nodata & \nodata \\ 
2005el  & $ 12407 \pm     41$ & $  9668 \pm     46$ & $  8703 \pm     43$ & $  8966 \pm     29$ & $ 10329 \pm    182$ & $ 11003 \pm     51$ & \nodata & $  1328 \pm    294$ \\ 
2005hc  & \nodata & $ 10292 \pm    336$ & $  9071 \pm    587$ & $  9835 \pm    259$ & $ 11038 \pm    489$ & $ 11163 \pm    219$ & \nodata & \nodata \\ 
2005kc  & \nodata & $ 10089 \pm    352$ & $  8960 \pm    429$ & $  8916 \pm     93$ & $  9962 \pm    217$ & $ 10554 \pm    239$ & \nodata & $   707 \pm    270$ \\ 
2005ki  & \nodata & $ 10522 \pm    738$ & \nodata & \nodata & $ 10268 \pm      0$ & $  9811 \pm     92$ & \nodata & \nodata \\ 
2005ku  & \nodata & $ 11334 \pm    313$ & $  9793 \pm    842$ & $  9972 \pm    297$ & $ 11783 \pm   1592$ & $ 12215 \pm    697$ & \nodata & \nodata \\
\enddata 

\end{deluxetable*} 

\clearpage
\begin{landscape}
\begin{deluxetable}{lcccccccccccccccc}
  \tabletypesize{\scriptsize} 
  \tablecolumns{17} 
  \tablewidth{0pt} 
  \tablecaption{Slopes of expansion velocity evolution in units of [km
      s$^{-1}$ d$^{-1}$].\label{tab:velslp}} 
  \tablehead{ 
\colhead{\sneia} & \colhead{\ion{Ca}{2}} & \colhead{} & \colhead{\ion{Si}{2}} & \colhead{} & \colhead{\ion{S}{2}} & \colhead{} & \colhead{\ion{S}{2}} & \colhead{} & \colhead{\ion{Si}{2}} & \colhead{} & \colhead{\ion{Si}{2}} & \colhead{} & \colhead{\ion{Ca}{2}} & \colhead{} & \colhead{\ion{Si}{2}} & \colhead{} \\
\colhead{Subtype} & \colhead{H\&K} & \colhead{$n$\tablenotemark{a}} & \colhead{$\lambda$4130} & \colhead{$n$\tablenotemark{a}} & \colhead{$\lambda$5449} & \colhead{$n$\tablenotemark{a}} & \colhead{$\lambda$5622} & \colhead{$n$\tablenotemark{a}} & \colhead{$\lambda$5972} & \colhead{$n$\tablenotemark{a}} & \colhead{$\lambda$6355} & \colhead{$n$\tablenotemark{a}} & \colhead{IR} & \colhead{$n$\tablenotemark{a}} & \colhead{$\lambda$6355 (+20)\tablenotemark{b}} & \colhead{$n$\tablenotemark{a}} 
  }
  \startdata
  normal & $-170 \pm 111$ & 6 & $-85 \pm 16$ & 12 & $-168 \pm 10$ & 12 & $-176 \pm 11$ & 12 & $-43 \pm 26$ & 12 &  $-86 \pm 14$ & 13 &  $-144 \pm 141$ & 2 & \nodata & \nodata \\
  BL & \nodata & \nodata & \nodata & \nodata & $-255 \pm 132$ & 2 &  $-254 \pm 13$\tablenotemark{d} & 3 & $-218 \pm 99$ & 2 & $-250 \pm 36$\tablenotemark{d} & 3 & $-416 \pm 89$\tablenotemark{d} & 2 & \nodata & \nodata \\
  CL & $-65 \pm 91$ & 2 & $-192 \pm 96$ & 3 &  $-234 \pm 264$ & 3 & $-227 \pm 15$ & 3 & $-187 \pm 59$ & 6 & $-177 \pm 20$ & 6 & \phs$24 \pm 72$ & 3 & \nodata & \nodata \\
  SS & $-211 \pm 110$ & 2 & $-59 \pm 17$ & 4 & $-108 \pm 28$ & 6 & $-122 \pm 16$ & 6 & \phs$10 \pm 39$ & 4 & $-42 \pm 16$ & 6 & \nodata & \nodata & \nodata & \nodata \\
  normal+CL+SS & \nodata & \nodata & \nodata & \nodata & $-151 \pm 17$\tablenotemark{e} & 21 & $-169 \pm 11$\tablenotemark{e} & 21 & $-67 \pm 27$\tablenotemark{e} & 22 & $-96 \pm 14$\tablenotemark{e} & 25 & $-14 \pm 53$\tablenotemark{e} & 7 & \nodata & \nodata \\  
  All & $-156 \pm 65$\tablenotemark{c} & 11 & $-96 \pm 18$\tablenotemark{c} & 20 & $-155 \pm 18$ & 23 & $-177 \pm 11$ & 24 & $-77 \pm 26$ & 24 & $-112 \pm 17$ & 28 & $-21 \pm 61$ & 8 & $-77 \pm 16$\tablenotemark{c} & 21 \\
\enddata
\tablenotetext{a}{Number of SNe used in the average.}
\tablenotetext{b}{Value of the slope between +7 and +25 days.}
\tablenotetext{c}{Value applied to all \sneia.}
\tablenotetext{d}{Value applied to BL \sneia.}
\tablenotetext{e}{Value applied to normal, BL, and SS \sneia.}
\end{deluxetable}
\clearpage
\end{landscape}

\begin{deluxetable*}{lccc} 
  \tabletypesize{\small} 
  \tablecolumns{4} 
  \tablewidth{0pt} 
  \tablecaption{Pseudo-equivalent width feature definitions\label{tab:featw}} 
  \tablehead{ 
    \colhead{Feature} & \colhead{Feature} & \colhead{Blue-ward} & \colhead{Red-ward} \\
    \colhead{Name} & \colhead{ID\tablenotemark{a}} & \colhead{limit range (\AA)} & \colhead{limit range (\AA)} 
  }
  \startdata
  \pwone & \ion{Ca}{2}\,H\&K & 3500 -- 3800 & 3900 -- 4100 \\
  \pwtwo & \ion{Si}{2}\,$\lambda$4130 & 3900 -- 4000 & 4000 -- 4150 \\
  \pwthree & \ion{Mg}{2}, \ion{Fe}{2}\tablenotemark{b} & 3900 -- 4150 & 4450 -- 4700 \\
  \pwfour & \ion{Fe}{2}, \ion{Si}{2} & 4500 -- 4700 & 5050 -- 5550 \\
  \pwfive & \ion{S}{2}\,$\lambda$$\lambda$5449,5622 & 5150 -- 5300 & 5500 -- 5700 \\
  \pwsix & \ion{Si}{2}\,$\lambda$5972 & 5550 -- 5700 & 5800 -- 6000 \\
  \pwseven & \ion{Si}{2}\,$\lambda$6355 & 5800 -- 6000 & 6200 -- 6600 \\
  \pweight & \ion{Ca}{2}\,IR triplet & 7500 -- 8000 & 8200 -- 8900 \\
  \enddata
\tablenotetext{a}{Spectral lines that contribute most to the observed absorption
near maximum light.}
\tablenotetext{b}{For CL \sneia\ \pwthree\ is dominated by
  \ion{Ti}{2} lines.}
\end{deluxetable*}

\clearpage
\begin{deluxetable*}{lcccccccc} 
\tabletypesize{\scriptsize} 
\tablecolumns{9} 
\tablewidth{0pt} 
\tablecaption{Values of \ew\ [\AA] at $B$-band maximum light.\label{tab:ews}} 
\tablehead{ 
\colhead{} & \colhead{\ew1} & \colhead{\ew2} & \colhead{\ew3} & \colhead{\ew4} & \colhead{\ew5} & \colhead{\ew6} & \colhead{\ew7} & \colhead{\ew8} \\ 
\colhead{SN} & \colhead{(\ion{Ca}{2}\,H\&K)} & \colhead{(\ion{Si}{2}\,4130)} & \colhead{(\ion{Mg}{2})} & \colhead{(\ion{Fe}{2})} & \colhead{(\ion{S}{2}\,W)} & \colhead{(\ion{Si}{2}\,5972)} & \colhead{(\ion{Si}{2}\,6355)} & \colhead{(\ion{Ca}{2}\,IR)} 
} 
\startdata
\multicolumn{9}{c}{CN} \\ 
2004ey  & $ 107.2 \pm    5.3$ & $  13.6 \pm    1.5$ & $  96.8 \pm    1.5$ & $ 129.9 \pm    3.2$ & $  91.1 \pm    3.3$ & $  14.2 \pm    1.1$ & $  99.7 \pm    3.5$ & $  65.0 \pm   16.5$ \\ 
2005bo  & \nodata & $  16.8 \pm    1.6$ & $  85.2 \pm    3.9$ & $ 134.1 \pm    5.3$ & $  80.5 \pm    2.8$ & $  23.2 \pm    1.9$ & $  99.2 \pm    1.8$ & $ 158.8 \pm    5.9$ \\ 
2005el  & $ 110.7 \pm    1.8$ & $  20.6 \pm    0.5$ & $  82.4 \pm    5.8$ & $ 112.9 \pm    3.9$ & $  82.1 \pm    1.2$ & $  20.5 \pm    1.5$ & $  93.8 \pm    1.7$ & $ 130.1 \pm    5.2$ \\ 
2005hc  & \nodata & $  10.3 \pm    5.2$ & $ 102.0 \pm    5.2$ & $ 125.1 \pm    1.1$ & $  89.3 \pm    3.8$ & $  18.3 \pm    1.5$ & $  84.3 \pm    1.6$ & \nodata \\ 
2005kc  & \nodata & \nodata & $  75.7 \pm    2.4$ & $ 130.7 \pm    5.2$ & $  75.4 \pm    1.3$ & $  25.8 \pm    1.0$ & $  91.2 \pm    1.1$ & \nodata \\ 
2005ki  & $ 116.1 \pm    1.6$ & $  20.9 \pm    1.4$ & $  86.7 \pm    1.4$ & $ 110.8 \pm    1.1$ & $  81.4 \pm    0.6$ & $  22.3 \pm    2.0$ & $ 104.3 \pm    0.5$ & $  86.0 \pm    1.1$ \\ 
2005ku  & \nodata & $  11.5 \pm    1.7$ & $  79.7 \pm    8.1$ & $  91.0 \pm    3.7$ & $  54.6 \pm    1.4$ & $  12.1 \pm    1.3$ & $  83.0 \pm    3.0$ & \nodata \\
\enddata 

\end{deluxetable*} 

\clearpage
\begin{landscape}
\begin{deluxetable}{lcccccccccccccccc}
  \tabletypesize{\tiny} 
  \tablecolumns{17} 
  \tablewidth{0pt} 
  \tablecaption{Slopes of \ew\ evolution around maximum light in units of [\AA\ d$^{-1}$].\label{tab:ewslp}} 
  \tablehead{ 
\colhead{\sneia} & \colhead{\ew1} & \colhead{} & \colhead{\ew2} & \colhead{} & \colhead{\ew3} & \colhead{} & \colhead{\ew4} & \colhead{} & \colhead{\ew5} & \colhead{} & \colhead{\ew6} & \colhead{} & \colhead{\ew7} & \colhead{} & \colhead{\ew8} & \colhead{} \\
\colhead{Subtype} & \colhead{(\ion{Ca}{2}\,H\&K)} & \colhead{$n$\tablenotemark{a}} & \colhead{(\ion{Si}{2}\,4130)} & \colhead{$n$\tablenotemark{a}} & \colhead{(\ion{Mg}{2})} & \colhead{$n$\tablenotemark{a}} & \colhead{(\ion{Fe}{2})} & \colhead{$n$\tablenotemark{a}} & \colhead{(\ion{S}{2}\,W)} & \colhead{$n$\tablenotemark{a}} & \colhead{(\ion{Si}{2}\,5972)} & \colhead{$n$\tablenotemark{a}} & \colhead{(\ion{Si}{2}\,6355)} & \colhead{} & \colhead{(\ion{Ca}{2}\,IR)} & \colhead{$n$\tablenotemark{a}} 
  }
  \startdata
  normal & $-5.4 \pm 1.7$ & 4 & \phs$0.7 \pm 0.1$ & 11 & \phs$2.5 \pm 1.8$ & 12 & \phs$4.5 \pm 0.6$ & 13 & \phs$0.9 \pm 0.4$ & 13 & \phs$0.4 \pm 0.1$ & 13 & \phs$1.5 \pm 0.6$ & 13 & \phs$6.0 \pm 1.1$ & 11 \\
  BL & \nodata & \nodata & \nodata & \nodata & \phs$0.6 \pm 2.3$ & 3 & \phs$6.1 \pm 2.1$ & 3 & \phs$1.3 \pm 1.7$ & 3 & \phs$0.0 \pm 0.6$ & 3 & $-0.2 \pm 1.1$ & 3 & $8.2 \pm 3.8$ & 2 \\
  CL & \nodata & \nodata & $-0.6 \pm 0.3$ & 2 & \phs$9.9 \pm 0.5$\tablenotemark{c} & 5 & \phs$8.0 \pm 4.0$ & 5 & \phs$0.0 \pm 1.4$ & 4 & \phs$1.1 \pm 0.6$ & 5 & \phs$2.3 \pm 1.4$ & 5 & \phs$3.6 \pm 4.1$ & 4 \\
  SS & \nodata & \nodata & \phs$0.3 \pm 0.3$ & 5 & \phs$0.1 \pm 0.5$ & 7 & \phs$3.7 \pm 0.4$ & 7 & \phs$2.0 \pm 0.6$ & 7 & \phs$0.9 \pm 0.2$ & 7 & \phs$2.2 \pm 0.2$ & 6 & \phs$8.2 \pm 3.1$ & 7 \\
 normal+BL+SS & \nodata & \nodata & \nodata & \nodata & \phs$1.8 \pm 1.1$\tablenotemark{d} & 22 & \nodata & \nodata & \nodata & \nodata & \nodata & \nodata & \nodata & \nodata & \nodata & \nodata \\
  All\tablenotemark{b} & $-3.6 \pm 1.7$ & 7 & \phs$0.3 \pm 0.2$ & 19 & \phs$3.4 \pm 1.3$ & 27 & \phs$5.1 \pm 0.6$ & 28 & \phs$1.1 \pm 0.3$ & 27 & \phs$0.6 \pm 0.1$ & 28 & \phs$1.4 \pm 0.4$ & 27 & \phs$6.6 \pm 1.0$ & 24 \\
\enddata
\tablenotetext{a}{Number of SNe used in the average.}
\tablenotetext{b}{Values applied to all SNe, except in the case of \ew3 as indicated below.}
\tablenotetext{c}{Value applied to CL \sneia.}
\tablenotetext{d}{Value applied to normal, BL, and SS \sneia.}
\end{deluxetable}

\clearpage
\end{landscape}

\begin{deluxetable*}{lcccccccccc}  
  \tabletypesize{\scriptsize} 
  \tablecolumns{11} 
  \tablewidth{0pt} 
  \tablecaption{Mean photometric properties for different \sneia\ subtypes\label{tab:phave}} 
  \tablehead{ 
    \colhead{} & \colhead{$\langle$\dm$\rangle$} & \colhead{} & \colhead{$\langle(B_{\mathrm{max}}-V_{\mathrm{max}})\rangle$} & \colhead{} & \colhead{$\langle E(B-V)_{\mathrm{Host}}\rangle$} & \colhead{} & \colhead{$\langle M_B^0\rangle$} & \colhead{} & \colhead{$\langle \Delta M_B\rangle$} & \colhead{} \\
    \colhead{Sample} & \colhead{(mag)} & \colhead{$N$} & \colhead{(mag)} & \colhead{$N$} & \colhead{(mag)} & \colhead{$N$} & \colhead{(mag)} & \colhead{$N$} & \colhead{(mag)} & \colhead{$N$} \\
\colhead{(1)} & \colhead{(2)} & \colhead{(3)} & \colhead{(4)} & \colhead{(5)} & \colhead{(6)} & \colhead{(7)} & \colhead{(8)} & \colhead{(9)} & \colhead{(10)} & \colhead{(11)} 
  }
\startdata
Complete   & $  1.157\pm   0.072$ &  93 & $  0.141\pm   0.031$ &  93 & $  0.135\pm   0.024$ &  77 & $-19.308\pm   0.049$ &  68 & $  0.005\pm   0.029$ &  82 \\ 
CN         & $  1.092\pm   0.094$ &  25 & $  0.148\pm   0.060$ &  25 & $  0.133\pm   0.033$ &  25 & $-19.379\pm   0.056$ &  22 & $ -0.028\pm   0.043$ &  22 \\ 
BL         & $  1.070\pm   0.168$ &  22 & $  0.099\pm   0.060$ &  22 & $  0.153\pm   0.064$ &  22 & $-19.250\pm   0.101$ &  17 & $ -0.062\pm   0.046$ &  17 \\ 
SS         & $  0.888\pm   0.108$ &  10 & $  0.147\pm   0.046$ &  10 & $  0.168\pm   0.047$ &  10 & $-19.564\pm   0.139$ &  10 & $  0.027\pm   0.098$ &  10 \\ 
CL         & $  1.513\pm   0.123$ &  20 & $  0.224\pm   0.080$ &  20 & $  0.144\pm   0.063$ &   5 & $-19.048\pm   0.136$ &   5 & $  0.008\pm   0.060$ &  18 \\ 
N          & $  1.222\pm   0.076$ &  37 & $  0.145\pm   0.042$ &  37 & $  0.134\pm   0.027$ &  34 & $-19.248\pm   0.049$ &  29 & $  0.007\pm   0.032$ &  31 \\ 
HV         & $  1.005\pm   0.161$ &  19 & $  0.109\pm   0.068$ &  19 & $  0.163\pm   0.073$ &  18 & $-19.339\pm   0.115$ &  16 & $ -0.079\pm   0.057$ &  17 \\
\enddata 
\tablecomments{Weighted averages and standard deviations. \\ Columns: (1) Subsample used to average according to the classification schemes of Branch et~al.~and Wang et~al.~(see definitions in Section~\ref{sec:spdiv}); (2)--(3) Light-curve decline rate and number of SNe used in calculation; (4)--(5) Pseudo-color at maximum light and number of SNe used in calculation; (6)--(7) Host-galaxy color excess and number of SNe used in calculation; (8)--(9) Reddening-free $B$-band absolute peak magnitude and number of SNe used in calculation; (10)--(11) $B$-band Hubble residuals and number of SNe used in calculation.}  
\end{deluxetable*} 

\bigskip
\bigskip
\bigskip

\begin{deluxetable*}{lcccccccl}  
\tabletypesize{\scriptsize} 
\tablecolumns{9} 
\tablewidth{0pt} 
\tablecaption{Fits of peak magnitudes versus \ew\ and $(B-V)$ pseudocolor\label{tab:lumew}} 
\tablehead{ 
\colhead{Fit} & \colhead{} & \colhead{} & \colhead{} & \colhead{$R_V$} & \colhead{$\sigma_{\mathrm{SN}}$} & \colhead{WRMS} & \colhead{} & \colhead{} \\ 
\colhead{No.} & \colhead{$M_B(0)$} & \colhead{$a_B$} & \colhead{$\beta_B^{BV}$} & \colhead{(CCM+O)} & \colhead{[mag]} & \colhead{[mag]} & \colhead{$N_{\mathrm{SNe}}$} & \colhead{Sample} \\ 
\colhead{(1)} & \colhead{(2)} & \colhead{(3)} & \colhead{(4)} & \colhead{(5)} & \colhead{(6)} & \colhead{(7)} & \colhead{(8)} & \colhead{(9)}  
}
\startdata
\multicolumn{9}{c}{\pwtwo\ [$pW_0=15$ \AA]}\\
    1 & $ -19.07\pm   0.05$ & $ 0.0135\pm 0.0028$ & $  3.18\pm  0.14$ & $  1.85\pm  0.13$ & $  0.10$ & $  0.17$ &  55 & Complete \\ 
    2 & $ -19.07\pm   0.05$ & $ 0.0149\pm 0.0030$ & $  3.33\pm  0.26$ & $  1.98\pm  0.24$ & $  0.09$ & $  0.17$ &  40 & Non-HV \\ 
    3 & $ -19.09\pm   0.17$ & $ 0.0137\pm 0.0118$ & $  3.36\pm  0.33$ & $  2.01\pm  0.30$ & $  0.00$ & $  0.13$ &  20 & CN \\ 
    4 & $ -19.07\pm   0.07$ & $ 0.0154\pm 0.0035$ & $  3.30\pm  0.40$ & $  1.96\pm  0.36$ & $  0.08$ & $  0.16$ &  42 & $(B-V)<0.2$ \\ 
    5 & $ -19.07\pm   0.05$ & $ 0.0148\pm 0.0029$ & $  3.24\pm  0.27$ & $  1.90\pm  0.24$ & $  0.09$ & $  0.16$ &  37 & \dm\,$<1.7$ \\ 
\multicolumn{9}{c}{\pwsix\ [$pW_0=20$ \AA]}\\
    6 & $ -19.08\pm   0.04$ & $ 0.0112\pm 0.0018$ & $  3.19\pm  0.12$ & $  1.86\pm  0.11$ & $  0.12$ & $  0.19$ &  67 & Complete \\
\enddata 
\tablecomments{Fits of the type:
  $\mu=B_{\mathrm{max}}-M_B(0)-a_B\,[pW\,-\,pW_0]-\beta_B^{BV}\,
  (B-V).$ \\ Columns: (1) Fit identifier; (2) Peak absolute magnitude for
  $pW=pW_0$ and zero $(B-V)$ pseudocolor; (3)
  Luminosity-\ew\ slope; (4) Luminosity-color slope; (5) Corresponding
  parameter $R_V$ of the CCM+O reddening law; (6) Resulting intrinsic
  dispersion of SN data; (7) Weighted RMS of fit in magnitudes (see
  text); (8) Number of SNe used in fit; (9) Sample of SNe used in fit
  (see text).}  
\end{deluxetable*} 

\clearpage
\begin{deluxetable*}{lccccc}  
\tabletypesize{\small} 
\tablecolumns{6} 
\tablewidth{0pt} 
\tablecaption{Straight-line fits to $\Delta M_B$ versus expansion
  velocity at maximum light for \sneia\ with \dm\,$<1.7$
  mag.\label{tab:velres}}
\tablehead{ 
\colhead{Velocity} & \colhead{$a$} & \colhead{$b$} & \colhead{$\sigma_{\mathrm{SN}}$} & \colhead{WRMS} & \colhead{No. of} \\
\colhead{Feature} & \colhead{(mag)} & \colhead{(mag/$10^3$\,km\,s$^{-1}$)} & \colhead{(mag)} & \colhead{(mag)} & \colhead{SNe} 
}
\startdata 
\vone      & $  0.057$$(170)$ & $ -0.005$$(011)$ & $  0.106$ & $  0.178$ &  35 \\ 
\vtwo      & $  0.041$$(206)$ & $ -0.004$$(020)$ & $  0.052$ & $  0.147$ &  53 \\ 
\vthree    & $  0.396$$(189)$ & $ -0.046$$(021)$ & $  0.083$ & $  0.167$ &  49 \\ 
\vfour     & $  0.421$$(161)$ & $ -0.047$$(017)$ & $  0.067$ & $  0.157$ &  49 \\ 
\vfive     & $  0.699$$(211)$ & $ -0.067$$(020)$ & $  0.046$ & $  0.150$ &  55 \\ 
\vsix      & $  0.333$$(202)$ & $ -0.030$$(017)$ & $  0.074$ & $  0.161$ &  58 \\ 
\vseven    & $ -0.048$$(162)$ & $ -0.000$$(012)$ & $  0.094$ & $  0.162$ &  27 \\ 
\enddata 
\tablecomments{Fits of the type:
$\Delta M_B=a\,+\,b\,\times\,v$, with $v$ expressed in $10^3$\,km\,s$^{-1}$.
}
\end{deluxetable*} 

\bigskip

\begin{deluxetable*}{lcccccc}  
\tabletypesize{\small} 
\tablecolumns{7} 
\tablewidth{0pt} 
\tablecaption{Straight-line fits to $(B_{\mathrm{max}}-V_{\mathrm{max}})_0$ versus \ew\ for
  \sneia\ with \dm\,$<1.7$ mag.\label{tab:ewbv}} 
\tablehead{ 
\colhead{\ew\ Feature} & \colhead{$pW_0$} & \colhead{$a$} & \colhead{$b$} & \colhead{$\sigma_{\mathrm{SN}}$} & \colhead{WRMS} & \colhead{No. of} \\
\colhead{} & \colhead{(\AA)} & \colhead{(mag)} & \colhead{(mag/\AA)} & \colhead{(mag)} & \colhead{(mag)} & \colhead{SNe} 
}
\startdata 
\pwone     & 120 & $     0.008$$(083)$ & $    0.0000$$(0007)$ & $   0.044$ & $   0.049$ &   16 \\ 
\pwtwo     &  15 & $     0.002$$(033)$ & $    0.0034$$(0016)$ & $   0.033$ & $   0.039$ &   20 \\ 
\pwthree   & 120 & $     0.034$$(036)$ & $    0.0010$$(0003)$ & $   0.032$ & $   0.041$ &   21 \\ 
\pwfour    & 130 & $     0.012$$(078)$ & $    0.0012$$(0006)$ & $   0.037$ & $   0.044$ &   22 \\ 
\pwfive    &  90 & $     0.018$$(124)$ & $   -0.0009$$(0014)$ & $   0.043$ & $   0.050$ &   22 \\ 
\pwsix     &  20 & $     0.013$$(030)$ & $    0.0035$$(0013)$ & $   0.033$ & $   0.041$ &   22 \\ 
\pwseven   & 100 & $     0.023$$(069)$ & $    0.0012$$(0007)$ & $   0.038$ & $   0.046$ &   22 \\ 
\pweight   & 140 & $     0.020$$(030)$ & $    0.0006$$(0002)$ & $   0.033$ & $   0.040$ &   20 \\ 
\enddata 
\tablecomments{Fits of the type:
$(B_{\mathrm{max}}-V_{\mathrm{max}})_0=a\,+\,b\,\times\,\left[pW-pW_0\right].$
}
\end{deluxetable*} 

\end{document}